%% file: CFT-09-002_temp.tex
\pdfoutput=1
\documentclass[11pt,twoside,a4paper,pdftex,cmspaper,final,collab]{cms-tdr}

\begin{document}\cmsNoteHeader{CFT-09-002}
%%%%%%%%%%%%%%%%%%%%%%%%%%%%%%%%%%%%%%%%%%%%%%%%%%%%%%%%%%%%%%%%%%%%
%
%  Common definitions
%
%  N.B. use of \providecommand rather than \newcommand means
%       that a definition is ignored if already specified
%
%                                              L. Taylor 18 Feb 2005
%%%%%%%%%%%%%%%%%%%%%%%%%%%%%%%%%%%%%%%%%%%%%%%%%%%%%%%%%%%%%%%%%%%%

%%%%%%%%%%%%%%%%%%%%%%%%%%%%%%%%%%%%%%%%%%%%%%%%%%%%%%%%%%%%%%%%%%%%
%
% Hyphenations (only need to add here if you get a nasty word break)
%
\hyphenation{env-iron-men-tal}%    just an example
\hyphenation{had-ron-i-za-tion}
\hyphenation{cal-or-i-me-ter}
\hyphenation{de-vices}
%
% Hyphenations-end
%
% CVS info. These are modified by cvs at checkout time.
% The last version of these macros found before the maketitle will be the one on the front page,
% so only the main file is tracked.
% Edit by hand with care!
\RCS$Revision: 1.11 $
\RCS$Date: 2009/11/26 01:34:43 $
\RCS$Name:  $
%%%%%%%%%%%%% ptdr definitions %%%%%%%%%%%%%%%%%%%%%
\input{ptdr-definitions}
\newcommand{\degree}{\ensuremath{^\circ}}
\newcommand{\degC}{\ensuremath{^\circ\,\mathrm{C}}}
\newcommand{\tevc}{\ensuremath{\mathrm{TeV}/c}}
\newcommand{\gevc}{\ensuremath{\mathrm{GeV}/c}}
\newcommand{\mevc}{\ensuremath{\mathrm{MeV}/c}}
\newcommand{\chisq}{\ensuremath{\chi^2}\xspace}
\providecommand{\pt}{\ensuremath{p_T}}
\providecommand{\mum}{\ensuremath{\mathrm{\;\mu m}}}
\newcommand{\ckf}{Combinatorial Track Finder\xspace}
\newcommand{\rs}{Road Search\xspace}
\newcommand{\costf}{Cosmic Track Finder\xspace}
\newcommand{\tif}{Tracker Integration Facility\xspace}
\providecommand{\fixme}[1]{{\large\sffamily{\bfseries{}FIXME:} #1}}
\providecommand{\update}[1]{{\large\sffamily{\bfseries{}UPDATE:} #1}}
\providecommand{\comment}[1]{{\large\sffamily{\bfseries{}COMMENT:} #1}}
\newcommand{\chindf}{\ensuremath{\chi^2/\mbox{ndf}}}

%%%%%%%%%%%%%%%  Title page %%%%%%%%%%%%%%%%%%%%%%%%
\cmsNoteHeader{09-002}
\title{Commissioning and Performance of the CMS Silicon Strip Tracker
with Cosmic Ray Muons}
% Force line breaks with \\

%Author is always "The CMS Collaboration" for PAS, so author, etc will be ignored
\address[cern]{CERN}
\address[neu]{Northeastern University}
\author[neu]{George Alverson}\author[neu]{Lucas Taylor}\author[cern]{A. Cern Person}

% please supply the date in yyyy/mm/dd format. Today has been
% redefined to do so, but it should be fixed as of the final release date.
\date{\today}

% note that you cannot use \verb in the abstract text
\abstract{
During autumn 2008, the Silicon Strip Tracker was
operated with the full CMS experiment in a comprehensive test, in the presence of the 3.8~T magnetic
field produced by the CMS superconducting solenoid. Cosmic ray muons were
detected in the muon chambers and used to trigger the readout of all
CMS sub-detectors. About 15 million events with a muon in the tracker were collected.
The efficiency of hit and track reconstruction were measured to be
higher than 99\% and consistent with expectations from Monte Carlo
simulation.
This article details the commissioning and
performance of the Silicon Strip Tracker with cosmic ray muons. }

% these need to be filled in by hand and should (MUST) match the info
% in the TeX equivalents less the TeX markup
\hypersetup{%
pdfauthor={R. Bainbridge, V. Ciulli, J. Cole, L. Mirabito, T. Speer},%
pdftitle={Commissioning and Performance of the CMS Silicon Strip Tracker with Cosmic Ray Muons},%
pdfsubject={CMS},%
pdfkeywords={CMS, tracker, strip tracker, track reconstruction, CRAFT}}

\maketitle %maketitle comes after all the front information has been supplied

%%%%%%%%%%%%%%%%%%%%%%%%%%%%%%%%  Begin text %%%%%%%%%%%%%%%%%%%%%%%%%%%%%

\input{Introduction}

\input{commissioning}

\input{datasamples}
\input{localreco}
\input{tracking}

\input{Conclusions}
\input{Acknowledgments}

\bibliography{auto_generated}   % will be created by the tdr script.

\cleardoublepage\appendix\section{The CMS Collaboration \label{app:collab}}\begin{sloppypar}\hyphenpenalty=500\input{CFT-09-002-authorlist.tex}\end{sloppypar}
\end{document}

%% file: ptdr-definitions.tex
%%%%%%%%%%%%%%%%%%%%%%%%%%%%%%%%%%%%%%%%%%%%%%%%%%%%%%%%%%%%%%%%%%%%
%
%  Common definitions
%
%  N.B. use of \providecommand rather than \newcommand means
%       that a definition is ignored if already specified
%
%                                              L. Taylor 18 Feb 2005
%%%%%%%%%%%%%%%%%%%%%%%%%%%%%%%%%%%%%%%%%%%%%%%%%%%%%%%%%%%%%%%%%%%%

% Some shorthand
% turn off italics
\newcommand {\etal}{\mbox{et al.}\xspace} %et al. - no preceding comma
\newcommand {\ie}{\mbox{i.e.}\xspace}     %i.e.
\newcommand {\eg}{\mbox{e.g.}\xspace}     %e.g.
\newcommand {\etc}{\mbox{etc.}\xspace}     %etc.
\newcommand {\vs}{\mbox{\sl vs.}\xspace}      %vs.
\newcommand {\mdash}{\ensuremath{\mathrm{-}}} % for use within formulas

% some terms whose definition we may change
\newcommand {\Lone}{Level-1\xspace} % Level-1 or L1 ?
\newcommand {\Ltwo}{Level-2\xspace}
\newcommand {\Lthree}{Level-3\xspace}

% Some software programs (alphabetized)
\providecommand{\ACERMC} {\textsc{AcerMC}\xspace}
\providecommand{\ALPGEN} {{\textsc{alpgen}}\xspace}
\providecommand{\CHARYBDIS} {{\textsc{charybdis}}\xspace}
\providecommand{\CMKIN} {\textsc{cmkin}\xspace}
\providecommand{\CMSIM} {{\textsc{cmsim}}\xspace}
\providecommand{\CMSSW} {{\textsc{cmssw}}\xspace}
\providecommand{\COBRA} {{\textsc{cobra}}\xspace}
\providecommand{\COCOA} {{\textsc{cocoa}}\xspace}
\providecommand{\COMPHEP} {\textsc{CompHEP}\xspace}
\providecommand{\EVTGEN} {{\textsc{evtgen}}\xspace}
\providecommand{\FAMOS} {{\textsc{famos}}\xspace}
\providecommand{\GARCON} {\textsc{garcon}\xspace}
\providecommand{\GARFIELD} {{\textsc{garfield}}\xspace}
\providecommand{\GEANE} {{\textsc{geane}}\xspace}
\providecommand{\GEANTfour} {{\textsc{geant4}}\xspace}
\providecommand{\GEANTthree} {{\textsc{geant3}}\xspace}
\providecommand{\GEANT} {{\textsc{geant}}\xspace}
\providecommand{\HDECAY} {\textsc{hdecay}\xspace}
\providecommand{\HERWIG} {{\textsc{herwig}}\xspace}
\providecommand{\HIGLU} {{\textsc{higlu}}\xspace}
\providecommand{\HIJING} {{\textsc{hijing}}\xspace}
\providecommand{\IGUANA} {\textsc{iguana}\xspace}
\providecommand{\ISAJET} {{\textsc{isajet}}\xspace}
\providecommand{\ISAPYTHIA} {{\textsc{isapythia}}\xspace}
\providecommand{\ISASUGRA} {{\textsc{isasugra}}\xspace}
\providecommand{\ISASUSY} {{\textsc{isasusy}}\xspace}
\providecommand{\ISAWIG} {{\textsc{isawig}}\xspace}
\providecommand{\MADGRAPH} {\textsc{MadGraph}\xspace}
\providecommand{\MCATNLO} {\textsc{mc@nlo}\xspace}
\providecommand{\MCFM} {\textsc{mcfm}\xspace}
\providecommand{\MILLEPEDE} {{\textsc{millepede}}\xspace}
\providecommand{\ORCA} {{\textsc{orca}}\xspace}
\providecommand{\OSCAR} {{\textsc{oscar}}\xspace}
\providecommand{\PHOTOS} {\textsc{photos}\xspace}
\providecommand{\PROSPINO} {\textsc{prospino}\xspace}
\providecommand{\PYTHIA} {{\textsc{pythia}}\xspace}
\providecommand{\SHERPA} {{\textsc{sherpa}}\xspace}
\providecommand{\TAUOLA} {\textsc{tauola}\xspace}
\providecommand{\TOPREX} {\textsc{TopReX}\xspace}
\providecommand{\XDAQ} {{\textsc{xdaq}}\xspace}

%  Experiments
\newcommand {\DZERO}{D\O\xspace}     %etc.

% Measurements and units...

\newcommand{\de}{\ensuremath{^\circ}}
\newcommand{\ten}[1]{\ensuremath{\times \text{10}^\text{#1}}}
\newcommand{\unit}[1]{\ensuremath{\text{\,#1}}\xspace}
\newcommand{\mum}{\ensuremath{\,\mu\text{m}}\xspace}
\newcommand{\micron}{\ensuremath{\,\mu\text{m}}\xspace}
\newcommand{\cm}{\ensuremath{\,\text{cm}}\xspace}
\newcommand{\mm}{\ensuremath{\,\text{mm}}\xspace}
\newcommand{\mus}{\ensuremath{\,\mu\text{s}}\xspace}
\newcommand{\keV}{\ensuremath{\,\text{ke\hspace{-.08em}V}}\xspace}
\newcommand{\MeV}{\ensuremath{\,\text{Me\hspace{-.08em}V}}\xspace}
\newcommand{\GeV}{\ensuremath{\,\text{Ge\hspace{-.08em}V}}\xspace}
\newcommand{\TeV}{\ensuremath{\,\text{Te\hspace{-.08em}V}}\xspace}
\newcommand{\PeV}{\ensuremath{\,\text{Pe\hspace{-.08em}V}}\xspace}
\newcommand{\keVc}{\ensuremath{{\,\text{ke\hspace{-.08em}V\hspace{-0.16em}/\hspace{-0.08em}c}}}\xspace}
\newcommand{\MeVc}{\ensuremath{{\,\text{Me\hspace{-.08em}V\hspace{-0.16em}/\hspace{-0.08em}c}}}\xspace}
\newcommand{\GeVc}{\ensuremath{{\,\text{Ge\hspace{-.08em}V\hspace{-0.16em}/\hspace{-0.08em}c}}}\xspace}
\newcommand{\TeVc}{\ensuremath{{\,\text{Te\hspace{-.08em}V\hspace{-0.16em}/\hspace{-0.08em}c}}}\xspace}
\newcommand{\keVcc}{\ensuremath{{\,\text{ke\hspace{-.08em}V\hspace{-0.16em}/\hspace{-0.08em}c}^\text{2}}}\xspace}
\newcommand{\MeVcc}{\ensuremath{{\,\text{Me\hspace{-.08em}V\hspace{-0.16em}/\hspace{-0.08em}c}^\text{2}}}\xspace}
\newcommand{\GeVcc}{\ensuremath{{\,\text{Ge\hspace{-.08em}V\hspace{-0.16em}/\hspace{-0.08em}c}^\text{2}}}\xspace}
\newcommand{\TeVcc}{\ensuremath{{\,\text{Te\hspace{-.08em}V\hspace{-0.16em}/\hspace{-0.08em}c}^\text{2}}}\xspace}

\newcommand{\pbinv} {\mbox{\ensuremath{\,\text{pb}^\text{$-$1}}}\xspace}
\newcommand{\fbinv} {\mbox{\ensuremath{\,\text{fb}^\text{$-$1}}}\xspace}
\newcommand{\nbinv} {\mbox{\ensuremath{\,\text{nb}^\text{$-$1}}}\xspace}
\newcommand{\percms}{\ensuremath{\,\text{cm}^\text{$-$2}\,\text{s}^\text{$-$1}}\xspace}
\newcommand{\lumi}{\ensuremath{\mathcal{L}}\xspace}
\newcommand{\Lumi}{\ensuremath{\mathcal{L}}\xspace}%both upper and lower
%
% Need a convention here:
\newcommand{\LvLow}  {\ensuremath{\mathcal{L}=\text{10}^\text{32}\,\text{cm}^\text{$-$2}\,\text{s}^\text{$-$1}}\xspace}
\newcommand{\LLow}   {\ensuremath{\mathcal{L}=\text{10}^\text{33}\,\text{cm}^\text{$-$2}\,\text{s}^\text{$-$1}}\xspace}
\newcommand{\lowlumi}{\ensuremath{\mathcal{L}=\text{2}\times \text{10}^\text{33}\,\text{cm}^\text{$-$2}\,\text{s}^\text{$-$1}}\xspace}
\newcommand{\LMed}   {\ensuremath{\mathcal{L}=\text{2}\times \text{10}^\text{33}\,\text{cm}^\text{$-$2}\,\text{s}^\text{$-$1}}\xspace}
\newcommand{\LHigh}  {\ensuremath{\mathcal{L}=\text{10}^\text{34}\,\text{cm}^\text{$-$2}\,\text{s}^\text{$-$1}}\xspace}
\newcommand{\hilumi} {\ensuremath{\mathcal{L}=\text{10}^\text{34}\,\text{cm}^\text{$-$2}\,\text{s}^\text{$-$1}}\xspace}

% Some usual physics terms

\newcommand{\zp}{\ensuremath{\mathrm{Z}^\prime}\xspace}

% SM (still to be classified)

\newcommand{\kt}{\ensuremath{k_{\mathrm{T}}}\xspace}
\newcommand{\BC}{\ensuremath{{B_{\mathrm{c}}}}\xspace}
\newcommand{\bbarc}{\ensuremath{{\overline{b}c}}\xspace}
\newcommand{\bbbar}{\ensuremath{{b\overline{b}}}\xspace}
\newcommand{\ccbar}{\ensuremath{{c\overline{c}}}\xspace}
\newcommand{\JPsi}{\ensuremath{{J}/\psi}\xspace}
\newcommand{\bspsiphi}{\ensuremath{B_s \to \JPsi\, \phi}\xspace}
\newcommand{\AFB}{\ensuremath{A_\mathrm{FB}}\xspace}
\newcommand{\EE}{\ensuremath{e^+e^-}\xspace}
\newcommand{\MM}{\ensuremath{\mu^+\mu^-}\xspace}
\newcommand{\TT}{\ensuremath{\tau^+\tau^-}\xspace}
\newcommand{\wangle}{\ensuremath{\sin^{2}\theta_{\mathrm{eff}}^\mathrm{lept}(M^2_\mathrm{Z})}\xspace}
\newcommand{\ttbar}{\ensuremath{{t\overline{t}}}\xspace}
\newcommand{\stat}{\ensuremath{\,\text{(stat.)}}\xspace}
\newcommand{\syst}{\ensuremath{\,\text{(syst.)}}\xspace}
% these moved to similar defs
%\newcommand{\Etmiss}{\ensuremath{E_{\mathrm{T}\!{\rm miss}}}}
%\newcommand{\VEtmiss}{\ensuremath{{\vec E}_{\mathrm{T}\!{\rm miss}}}}

%%%  E-gamma definitions
\newcommand{\HGG}{\ensuremath{\mathrm{H}\to\gamma\gamma}}
\newcommand{\gev}{\GeV}
\newcommand{\GAMJET}{\ensuremath{\gamma + \mathrm{jet}}}
\newcommand{\PPTOJETS}{\ensuremath{\mathrm{pp}\to\mathrm{jets}}}
\newcommand{\PPTOGG}{\ensuremath{\mathrm{pp}\to\gamma\gamma}}
\newcommand{\PPTOGAMJET}{\ensuremath{\mathrm{pp}\to\gamma +
\mathrm{jet}
}}
\newcommand{\MH}{\ensuremath{\mathrm{M_{\mathrm{H}}}}}
\newcommand{\RNINE}{\ensuremath{\mathrm{R}_\mathrm{9}}}
\newcommand{\DR}{\ensuremath{\Delta\mathrm{R}}}

% Physics symbols ...

\newcommand{\PT}{\ensuremath{p_{\mathrm{T}}}\xspace}
\newcommand{\pt}{\ensuremath{p_{\mathrm{T}}}\xspace}
\newcommand{\ET}{\ensuremath{E_{\mathrm{T}}}\xspace}
\newcommand{\HT}{\ensuremath{H_{\mathrm{T}}}\xspace}
\newcommand{\et}{\ensuremath{E_{\mathrm{T}}}\xspace}
\newcommand{\Em}{\ensuremath{E\!\!\!/}\xspace}
\newcommand{\Pm}{\ensuremath{p\!\!\!/}\xspace}
\newcommand{\PTm}{\ensuremath{{p\!\!\!/}_{\mathrm{T}}}\xspace}
\newcommand{\ETm}{\ensuremath{E_{\mathrm{T}}^{\mathrm{miss}}}\xspace}
\newcommand{\MET}{\ensuremath{E_{\mathrm{T}}^{\mathrm{miss}}}\xspace}
\newcommand{\ETmiss}{\ensuremath{E_{\mathrm{T}}^{\mathrm{miss}}}\xspace}
\newcommand{\VEtmiss}{\ensuremath{{\vec E}_{\mathrm{T}}^{\mathrm{miss}}}\xspace}

%%%%%%
% From Albert
%

\newcommand{\ga}{\ensuremath{\gtrsim}}
\newcommand{\la}{\ensuremath{\lesssim}}
\newcommand{\swsq}{\ensuremath{\sin^2\theta_W}\xspace}
\newcommand{\cwsq}{\ensuremath{\cos^2\theta_W}\xspace}
\newcommand{\tanb}{\ensuremath{\tan\beta}\xspace}
\newcommand{\tanbsq}{\ensuremath{\tan^{2}\beta}\xspace}
\newcommand{\sidb}{\ensuremath{\sin 2\beta}\xspace}
\newcommand{\alpS}{\ensuremath{\alpha_S}\xspace}
\newcommand{\alpt}{\ensuremath{\tilde{\alpha}}\xspace}

\newcommand{\QL}{\ensuremath{Q_L}\xspace}
\newcommand{\sQ}{\ensuremath{\tilde{Q}}\xspace}
\newcommand{\sQL}{\ensuremath{\tilde{Q}_L}\xspace}
\newcommand{\ULC}{\ensuremath{U_L^C}\xspace}
\newcommand{\sUC}{\ensuremath{\tilde{U}^C}\xspace}
\newcommand{\sULC}{\ensuremath{\tilde{U}_L^C}\xspace}
\newcommand{\DLC}{\ensuremath{D_L^C}\xspace}
\newcommand{\sDC}{\ensuremath{\tilde{D}^C}\xspace}
\newcommand{\sDLC}{\ensuremath{\tilde{D}_L^C}\xspace}
\newcommand{\LL}{\ensuremath{L_L}\xspace}
\newcommand{\sL}{\ensuremath{\tilde{L}}\xspace}
\newcommand{\sLL}{\ensuremath{\tilde{L}_L}\xspace}
\newcommand{\ELC}{\ensuremath{E_L^C}\xspace}
\newcommand{\sEC}{\ensuremath{\tilde{E}^C}\xspace}
\newcommand{\sELC}{\ensuremath{\tilde{E}_L^C}\xspace}
\newcommand{\sEL}{\ensuremath{\tilde{E}_L}\xspace}
\newcommand{\sER}{\ensuremath{\tilde{E}_R}\xspace}
\newcommand{\sFer}{\ensuremath{\tilde{f}}\xspace}
\newcommand{\sQua}{\ensuremath{\tilde{q}}\xspace}
\newcommand{\sUp}{\ensuremath{\tilde{u}}\xspace}
\newcommand{\suL}{\ensuremath{\tilde{u}_L}\xspace}
\newcommand{\suR}{\ensuremath{\tilde{u}_R}\xspace}
\newcommand{\sDw}{\ensuremath{\tilde{d}}\xspace}
\newcommand{\sdL}{\ensuremath{\tilde{d}_L}\xspace}
\newcommand{\sdR}{\ensuremath{\tilde{d}_R}\xspace}
\newcommand{\sTop}{\ensuremath{\tilde{t}}\xspace}
\newcommand{\stL}{\ensuremath{\tilde{t}_L}\xspace}
\newcommand{\stR}{\ensuremath{\tilde{t}_R}\xspace}
\newcommand{\stone}{\ensuremath{\tilde{t}_1}\xspace}
\newcommand{\sttwo}{\ensuremath{\tilde{t}_2}\xspace}
\newcommand{\sBot}{\ensuremath{\tilde{b}}\xspace}
\newcommand{\sbL}{\ensuremath{\tilde{b}_L}\xspace}
\newcommand{\sbR}{\ensuremath{\tilde{b}_R}\xspace}
\newcommand{\sbone}{\ensuremath{\tilde{b}_1}\xspace}
\newcommand{\sbtwo}{\ensuremath{\tilde{b}_2}\xspace}
\newcommand{\sLep}{\ensuremath{\tilde{l}}\xspace}
\newcommand{\sLepC}{\ensuremath{\tilde{l}^C}\xspace}
\newcommand{\sEl}{\ensuremath{\tilde{e}}\xspace}
\newcommand{\sElC}{\ensuremath{\tilde{e}^C}\xspace}
\newcommand{\seL}{\ensuremath{\tilde{e}_L}\xspace}
\newcommand{\seR}{\ensuremath{\tilde{e}_R}\xspace}
\newcommand{\snL}{\ensuremath{\tilde{\nu}_L}\xspace}
\newcommand{\sMu}{\ensuremath{\tilde{\mu}}\xspace}
\newcommand{\sNu}{\ensuremath{\tilde{\nu}}\xspace}
\newcommand{\sTau}{\ensuremath{\tilde{\tau}}\xspace}
\newcommand{\Glu}{\ensuremath{g}\xspace}
\newcommand{\sGlu}{\ensuremath{\tilde{g}}\xspace}
\newcommand{\Wpm}{\ensuremath{W^{\pm}}\xspace}
\newcommand{\sWpm}{\ensuremath{\tilde{W}^{\pm}}\xspace}
\newcommand{\Wz}{\ensuremath{W^{0}}\xspace}
\newcommand{\sWz}{\ensuremath{\tilde{W}^{0}}\xspace}
\newcommand{\sWino}{\ensuremath{\tilde{W}}\xspace}
\newcommand{\Bz}{\ensuremath{B^{0}}\xspace}
\newcommand{\sBz}{\ensuremath{\tilde{B}^{0}}\xspace}
\newcommand{\sBino}{\ensuremath{\tilde{B}}\xspace}
\newcommand{\Zz}{\ensuremath{Z^{0}}\xspace}
\newcommand{\sZino}{\ensuremath{\tilde{Z}^{0}}\xspace}
\newcommand{\sGam}{\ensuremath{\tilde{\gamma}}\xspace}
\newcommand{\chiz}{\ensuremath{\tilde{\chi}^{0}}\xspace}
\newcommand{\chip}{\ensuremath{\tilde{\chi}^{+}}\xspace}
\newcommand{\chim}{\ensuremath{\tilde{\chi}^{-}}\xspace}
\newcommand{\chipm}{\ensuremath{\tilde{\chi}^{\pm}}\xspace}
\newcommand{\Hone}{\ensuremath{H_{d}}\xspace}
\newcommand{\sHone}{\ensuremath{\tilde{H}_{d}}\xspace}
\newcommand{\Htwo}{\ensuremath{H_{u}}\xspace}
\newcommand{\sHtwo}{\ensuremath{\tilde{H}_{u}}\xspace}
\newcommand{\sHig}{\ensuremath{\tilde{H}}\xspace}
\newcommand{\sHa}{\ensuremath{\tilde{H}_{a}}\xspace}
\newcommand{\sHb}{\ensuremath{\tilde{H}_{b}}\xspace}
\newcommand{\sHpm}{\ensuremath{\tilde{H}^{\pm}}\xspace}
\newcommand{\hz}{\ensuremath{h^{0}}\xspace}
\newcommand{\Hz}{\ensuremath{H^{0}}\xspace}
\newcommand{\Az}{\ensuremath{A^{0}}\xspace}
\newcommand{\Hpm}{\ensuremath{H^{\pm}}\xspace}
\newcommand{\sGra}{\ensuremath{\tilde{G}}\xspace}
\newcommand{\mtil}{\ensuremath{\tilde{m}}\xspace}
\newcommand{\rpv}{\ensuremath{\rlap{\kern.2em/}R}\xspace}
\newcommand{\LLE}{\ensuremath{LL\bar{E}}\xspace}
\newcommand{\LQD}{\ensuremath{LQ\bar{D}}\xspace}
\newcommand{\UDD}{\ensuremath{\overline{UDD}}\xspace}
\newcommand{\Lam}{\ensuremath{\lambda}\xspace}
\newcommand{\Lamp}{\ensuremath{\lambda'}\xspace}
\newcommand{\Lampp}{\ensuremath{\lambda''}\xspace}
\newcommand{\spinbd}[2]{\ensuremath{\bar{#1}_{\dot{#2}}}\xspace}

\newcommand{\MD}{\ensuremath{{M_\mathrm{D}}}\xspace}% ED mass
\newcommand{\Mpl}{\ensuremath{{M_\mathrm{Pl}}}\xspace}% Planck mass
\newcommand{\Rinv} {\ensuremath{{R}^{-1}}\xspace}

%%%%%%%%%%%%%%%%%%%%%%%%%%%%%%%%%%%%%%%%%%%%%%%%%%%%%%%%%%%%%%%%%%%%
%
% Hyphenations (only need to add here if you get a nasty word break)
%
\hyphenation{en-viron-men-tal}%    just an example

%% file: Introduction.tex
\section{Introduction}

The primary goal of the Compact Muon Solenoid (CMS) experiment~\cite{cms}
is to explore particle physics at the TeV energy scale exploiting the
proton-proton collisions delivered by the Large Hadron Collider
(LHC)~\cite{LHC}. 
The central tracking detector~\cite{cms} built for the CMS  
experiment is a unique instrument, in both size and complexity. It  
comprises two systems based on silicon sensor
technology: one employing silicon pixels and another using silicon
microstrips. The Pixel Detector surrounds the beampipe and contains
66~million detector channels~\cite{craftPixel}. The Pixel system is,
in turn, surrounded by the Silicon Strip Tracker (SST), which is the subject
of this paper.

\begin{figure}[b]
  \begin{center}
    \includegraphics[width=\textwidth]{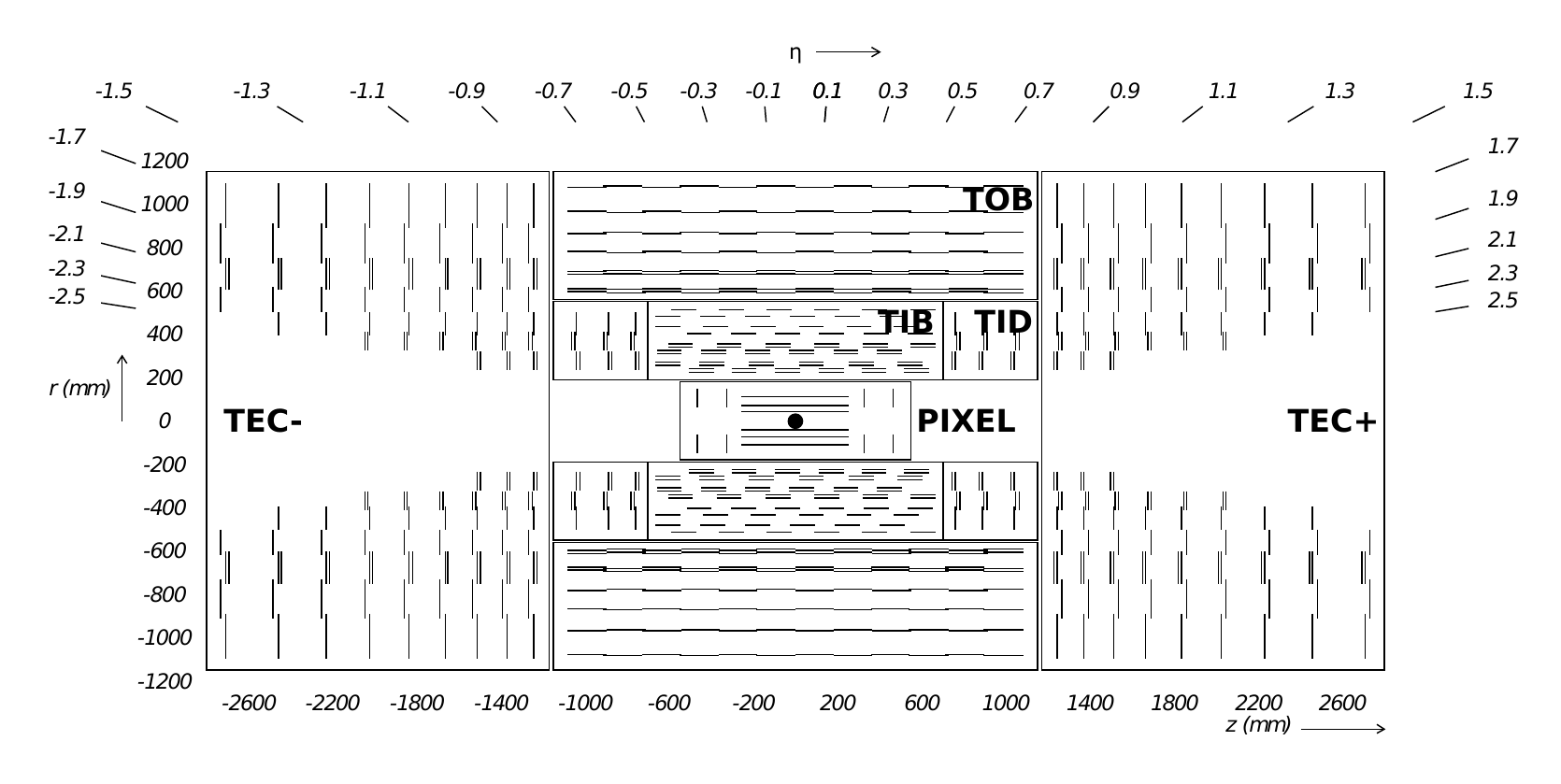}
    \caption{Schematic cross section of the CMS tracker. Each
      line represents a detector module. Double lines indicate
      double-sided modules which deliver stereo hits.}
    \label{fig:tk-layout}
  \end{center}
\end{figure}

The SST consists of four main subsystems, shown in
Fig.~\ref{fig:tk-layout}: the four-layer Tracker Inner Barrel (TIB),
the six-layer Tracker Outer Barrel (TOB) and, on each side of the
barrel region, the three-disk Tracker Inner Disks (TID), and the
nine-disk Tracker End Caps (TEC). Each TID disk is made of three rings
of modules, while TEC disks have seven rings. Overall, the tracker
cylinder is 5.5\,m long and 2.4\,m in diameter, with a total active
area of $198\, {\rm m}^2$, consisting of 15\,148 detector modules and
comprising 9.3 million detector channels.  Each detector module
consists of a carbon or graphite fibre frame, which supports the
silicon sensor and the associated front-end readout electronics.  Four
barrel layers and three rings in the end cap disks are equipped with
double-sided modules, each of which is constructed from two
single-sided modules mounted back-to-back with a stereo angle of
100\,mrad between the strips.
%This provides $Z$ information in the barrel and $r$ information in the end caps.
The silicon sensors are made up of single-sided $p^+$ strips on
$n$-bulk sensors with two different thicknesses: $320 \mum$ and
$500\mum$ in the inner four and outer six layers of the barrel,
respectively; $320\mum$ in the inner disks; and $320\mum$ and $500\mum$ in the inner four and outer
three rings of the end cap disks, respectively.  There are a total of
fifteen different types of sensors in the SST, which vary in terms of
strip length and pitch~\cite{sensors} to ensure that the single strip
occupancy is low
%and the two-hit resolution is good throughout the SST, 
even at full LHC luminosity.

The first experience of the SST operation and detector performance
study was gained in summer 2006, when a small fraction of the SST was
inserted into the CMS detector.  Cosmic ray muon data were recorded in
the presence of a solenoidal field up to the maximum design value of
4\,T. The results from this period of data-taking are described
elsewhere~\cite{mtccPaper}.  Construction of the full SST was
completed in 2007 and 15\% of the full SST was commissioned and
operated for several months prior to installation in the underground
CMS experimental hall.  The results of this period of stand-alone
operation, known as the Slice Test, are also described
elsewhere~\cite{TIF_Note, tifPaper}.

The installation of the SST within CMS was completed during 2008 and
the system underwent its first round of {\it in situ} commissioning
together with the other CMS sub-detectors during summer 2008.  The
first operation of the SST in a 3.8\,T magnetic field took place
during October-November 2008, when the CMS Collaboration conducted a
month-long data-taking exercise known as the Cosmic Run At Four Tesla
(CRAFT)~\cite{CRAFTGeneral}.  This exercise provided valuable
operational experience, as well as allowing, for the first time, a
full study of the SST performance after installation. First results
from the study are presented here.

This paper is laid out as follows.  The procedures used to commission
the SST and the results from the round of {\it in situ} commissioning
are presented and discussed in Section 2.  The final data samples from
CRAFT and the corresponding Monte Carlo simulations are described in
Section 3.  The performance results obtained from the CRAFT data
samples for hit and track reconstruction are presented in Sections 4
and 5, respectively.

%%%% Is this really needed to understand the paper???
%In all the results presented in this paper, the standard CMS reference system is used.
%This system has its origin in the centre of the detector,  the $z$ axis 
%is along the beam line in the anti-clockwise direction for
%an observer standing in the middle of the LHC ring. The $x$ axis points
%to the LHC centre and the $y$ axis points upward. The azimuthal 
%angle $\phi$ is measured starting from the $x$ axis toward the $y$ axis.
%The polar radius $r$ is defined as the distance
%from the $z$ axis in the transverse $(x,y)$ plane.

%% file: commissioning.tex
%% -----------------------------------------------------------------------------
%% -----------------------------------------------------------------------------
%% -----------------------------------------------------------------------------

\section{Commissioning the SST Control and Readout Systems}

In order to bring the SST detector into an operational state suitable
for data-taking, several commissioning procedures are required to
checkout, configure, calibrate, and synchronise the various hardware
components of the control and readout systems.  The majority of the
commissioning procedures are performed with the SST operating
independently of the rest of the CMS experiment.  Only the procedures
that concern synchronisation to an external trigger, described in
Section~\ref{sec:ext-synch}, require reconstructed particle
trajectories from cosmic ray muons or LHC pp collision data. The
commissioning of the SST aims to maximise the signal identification
efficiency for in-time particles and minimise pileup due to
out-of-time particles. The ultimate objective is to maximise the
tracking efficiency while minimising the number of tracks caused by
out-of-time signals from adjacent bunch crossings.

Section~\ref{sec:readout-system} provides an overview of the SST
control and readout systems. Section~\ref{sec:checkout} summarises the
checkout procedures used to determine the functional components of
these systems. Sections~\ref{sec:int-synch}-\ref{sec:ext-synch} review
the various commissioning procedures and their performances.

%% -----------------------------------------------------------------------------
%% -----------------------------------------------------------------------------
%% -----------------------------------------------------------------------------

\subsection{The control and readout systems}
\label{sec:readout-system}

The major components of the SST readout system~\cite{ttdr} are:
15\,148 front-end detector modules that host 76\,000
APV25~\cite{APV25} readout chips, an analogue optical link system
comprising 38\,000 individual fibres~\cite{LINKS}, and 440
off-detector analogue receiver boards, known as Front-End Drivers
(FED)~\cite{FED}. The SST control system~\cite{ieeeFred} is driven by
46 off-detector digital transceiver boards, known as Front-End
Controllers (FEC)~\cite{FEC}. The FECs distribute the LHC clock,
triggers and control signals to the front-end detector modules via
Communication and Control Units (CCU)~\cite{CCU}, which are hosted on
368 {\it control rings}. 
%The 40~MHz LHC clock provided by the Timing, Trigger, and Control
%(TTC) system~\cite{TTC}.

The APV25 readout chip samples, amplifies, buffers, and processes
signals from 128 detector channels at a frequency of 40~MHz. Fast
pulse shaping is therefore required to provide bunch crossing
identification and minimise pileup. This is difficult to achieve with
low noise and power levels, so the APV25 chip uses pre-amplifier and
shaper stages to produce a CR-RC pulse shape with a relatively slow
rise-time of 50~ns in an operating mode known as {\it peak}.  An
alternative mode, {\it deconvolution}, performs additional signal
processing to constrain the signal to a single bunch
crossing~\cite{Deconvolution} at the expense of a reduced
signal-to-noise ratio. Deconvolution is expected to be the standard
mode of operation. However,
%due to a lack of time to complete a final timing adjustment needed for
%operation in deconvolution mode, 
the results presented in this paper are based on data accumulated with
peak mode operation, unless stated otherwise.

\begin{figure}[t]
  \begin{centering}
    \includegraphics[width=0.54\textwidth]{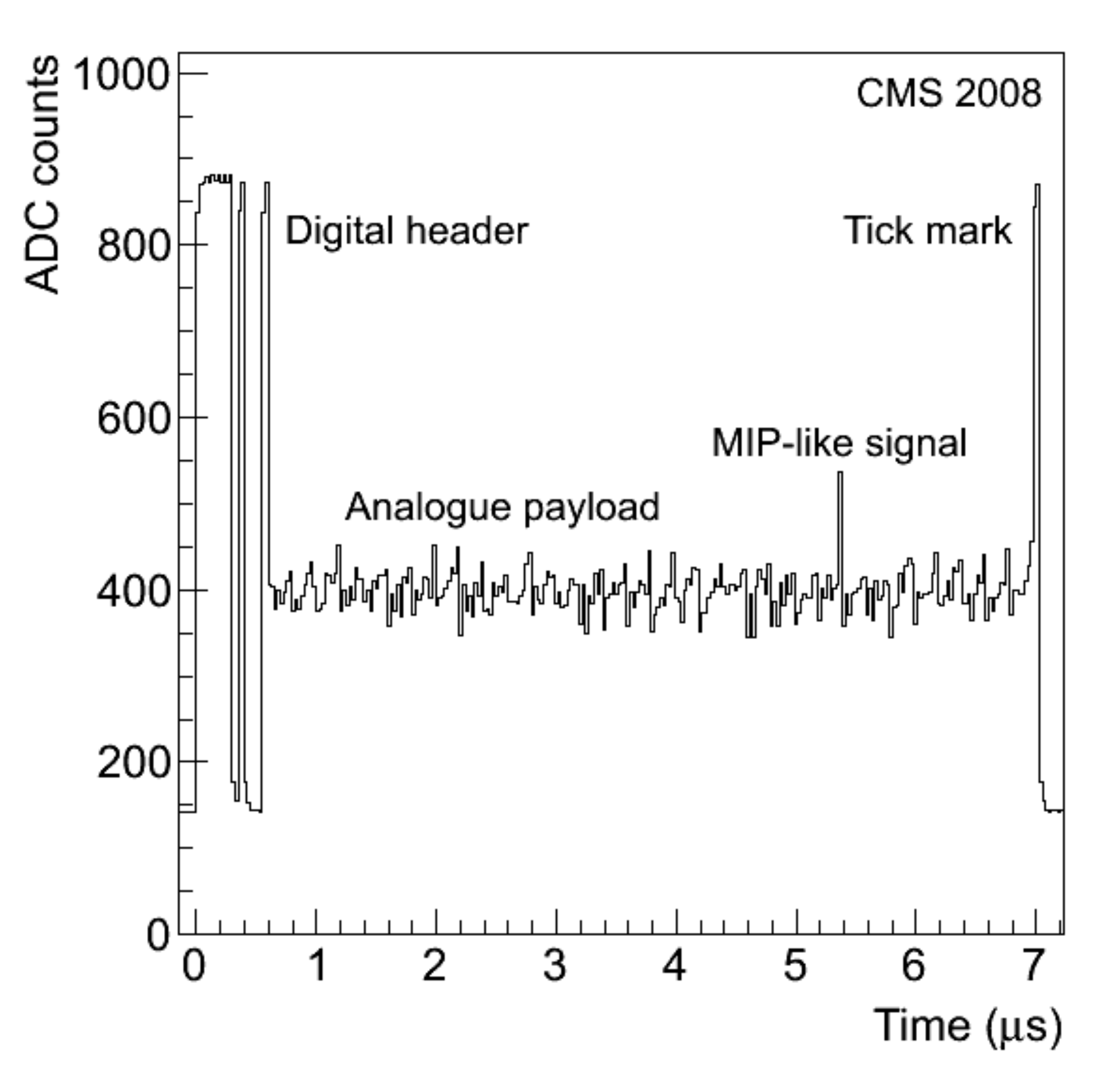}
    \includegraphics[width=0.44\textwidth]{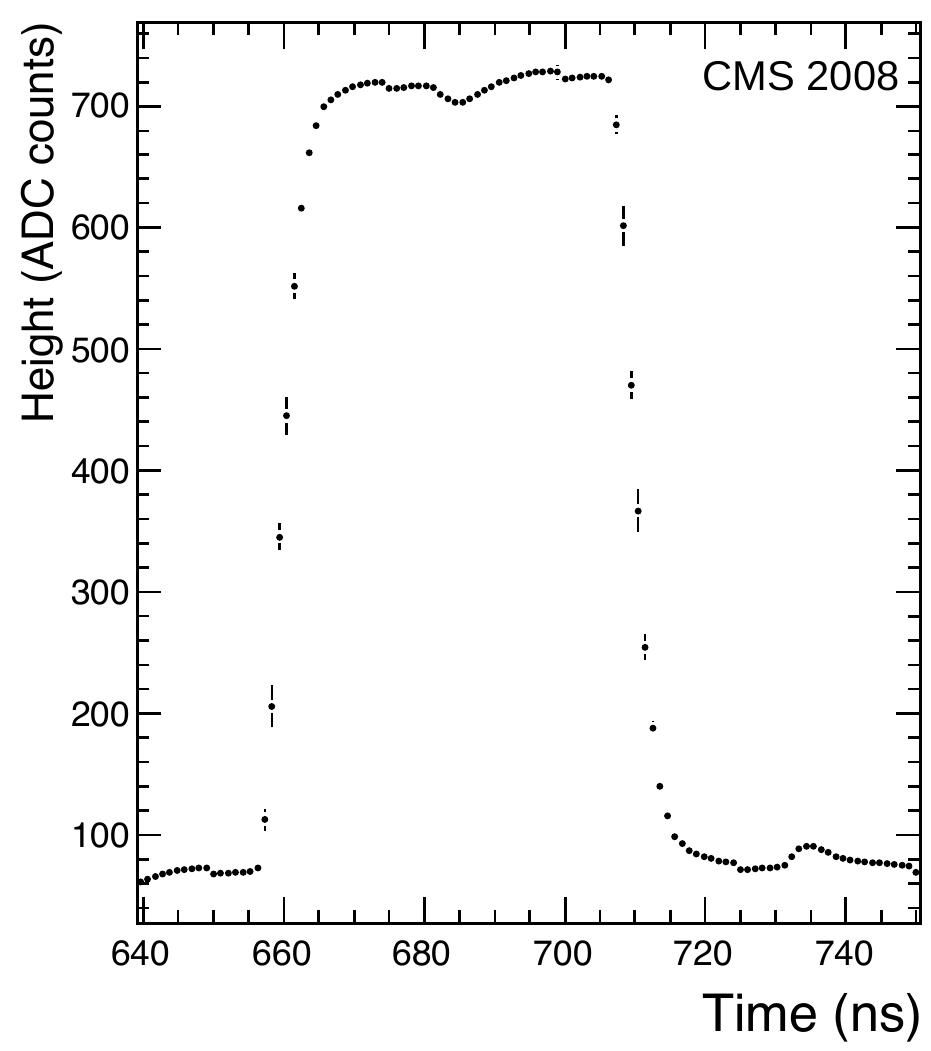}
    \par
  \end{centering}
  \caption{\label{fig:apv-data} (left) Two APV25 data frames multiplexed,
    containing a time stamp and the sensor pulse height
    information. (right) A feature of the APV25 data stream, known as
    a tick mark, that is heavily used by the checkout and
    commissioning procedures. The left and right figures have sampling
    intervals of 25~ns and 1.04~ns, respectively. }
\end{figure}

%On receipt of a Level-1 trigger, pulse height and bunch-crossing
%information from pairs of APV25 chips are generated and multiplexed
%onto a single line. The data are converted to optical signals and
%transmitted via optical links to the off-detector FED boards. The FEDs
%digitise, compress, and format the pulse height data from up to 96
%pairs of APV25 chips, before forwarding the resulting event fragments
%to the CMS data acquisition and high-level trigger
%systems~\cite{daqtdr}. 

Figure~\ref{fig:apv-data} (left) shows an example of the raw data
captured at 40~MHz by a single FED readout channel on receipt of a
trigger. The data contain frames from two APV25 chips that are
multiplexed (interleaved) together. A single frame comprises 12 bits
of binary information that encodes time and error information, known
as the digital header, followed by analogue pulse height data from 128
sensor strips. A trailing {\it tick mark} identifies the end of the
frame.
%The digital header encodes three header bits used to mark the
%beginning of a frame, eight address bits that provide a time stamp,
%and a single bit used to flag error conditions in the internal logic
%of the chip.
The structure observed in the pulse height data across the 128
channels is due to static offsets, known as {\it pedestals}, which are
unique to each detector channel. Small, time-varying {\it common mode}
shifts in the levels of all 128 channels are observed when
operating. Figure~\ref{fig:apv-data} (left) also shows an example of a
signal left by a minimum ionising particle. Signals are superimposed
on the pedestal and common mode levels, which must be subtracted
before the signal can be identified.

In the absence of a trigger, no data frames are output by the APV25
chip, but tick marks are produced every 70 clock
cycles. Figure~\ref{fig:apv-data} (right) shows the pulse shape of
multiplexed tick marks from two APV25 chips that are reconstructed
with an effective sampling frequency of 960~MHz. This tick mark
feature is used heavily in the checkout and commissioning procedures
detailed below.

The FEDs can format the pulse height data from the APV25 chips in
different ways. The first is Scope Mode (SM), which is simply a
capture of the raw data, as shown in Fig.~\ref{fig:apv-data}
(left). The second is Virgin Raw (VR), which removes all of the binary
information (digital header and tick marks) and simply provides the
digitised pulse height data from the sensors. Both modes provide
digital samples with a 10-bit range and are used when commissioning
the SST system and for debugging. The third and normal mode of
operation is Zero Suppressed (ZS). This uses Field Programmable Gate
Array (FPGA) chips to implement algorithms that perform pedestal
subtraction, common mode subtraction, and identification of channels
potentially containing signals above threshold. A threshold of five
times the detector channel noise is used for single channels, but a
threshold of only twice the channel noise is used for signals in
contiguous channels. The zero-suppressed data are output with an 8-bit
range.

%% -----------------------------------------------------------------------------
%% -----------------------------------------------------------------------------
%% -----------------------------------------------------------------------------

\subsection{Checkout of the detector components and cabling}
\label{sec:checkout}

The checkout procedures are used to identify: responsive and
functional devices in the control and readout systems; the cabling of
the readout electronics chain, from the front-end detector modules to
the off-detector FED boards; the cabling of the Low Voltage (LV) and
High Voltage (HV) buses of the power supply system~\cite{POWER}; and
the mapping of the detector modules to their geometrical position
within the SST superstructure. 
Automation is possible as each detector module hosts a Detector
Control Unit (DCU) chip~\cite{DCU}, which broadcasts a unique
indentifier via the control system. This identifier is used to tag
individual modules. 
%Automation is an important and necessary feature
%of the procedures, given the complexity of the SST control, readout,
%and power supply systems.
%Automation of these procedures is necessary due to the complexity of
%the system.

The cabling of the LV power supply system is established by
sequentially powering groups of detector modules and identifying
responsive devices via the control system. 
%Similarly, the HV cabling is determined by identifying changes in
%noise measurements for individual detector channels, as described in
%Section~\ref{sec:noise}, when the HV bias is applied or not to
%modules. Biasing a sensor decreases the strip capacitance and
%therefore the channel noise.
Similarly, the HV cabling is determined by applying HV to an
individual channel and identifying detector modules responding with a
decreased noise, due to reduced strip capacitance.

Each front-end detector module hosts a Linear Laser Driver (LLD)
chip~\cite{LLD}, which drives the optical links that transmit the
analogue signals to the off-detector FED boards. The cabling of the
readout electronics chain is established by configuring individual LLD
chips to produce unique patterns in the data stream of the connected
FED channels.

The final number of modules used in the CRAFT data-taking period
corresponds to 98.0\% of the total system. The most significant losses
were from one control ring in each of the TIB and TOB sub-systems. In
the TIB, this was due to a single faulty CCU. The remaining CCUs on
this ring have since been recovered using a built-in redundancy
feature of the control ring design. The fraction of operational
modules was increased to 98.6\% after data-taking, once problems
identified during checkout were investigated more fully.

%% -----------------------------------------------------------------------------
%% -----------------------------------------------------------------------------
%% -----------------------------------------------------------------------------

\subsection{Relative synchronisation of the front-end}
\label{sec:int-synch}

Relative synchronisation involves adjusting the phase of the LHC clock
delivered to the front-end so that the sampling times of all APV25
chips in the system are synchronous. Additionally, the signal sampling
time of the FED Analogue/Digital Converters (ADC) is appropriately
adjusted. This procedure accounts for differences in signal
propagation time in the control system due to differing cable
lengths. 
%The precision of the relative synchronisation procedure, described
%here, and the absolute synchronisation to an external trigger, as
%described in Section~\ref{sec:ext-synch}, is important because signal
%amplitude is attenuated by as much as 4$\%$ per nanosecond
%misalignment in time, because of the narrow pulse shape in
%deconvolution mode.
This synchronisation procedure is important because signal amplitude
is attenuated by as much as 4$\%$ per nanosecond mis-synchronisation
due to the narrow pulse shape in deconvolution mode.

Using the FED boards in Scope Mode, the tick mark pulse shape is
reconstructed with a 1.04~ns step width by varying the clock phase
using a Phase Locked Loop (PLL) chip~\cite{PLL} hosted by each
detector module, as shown in Fig.~\ref{fig:apv-data} (right). The
ideal sampling point is on the signal plateau, 15~ns after the rising
edge of the tick mark. The required delays are thus inferred from the
arrival times of the tick mark edges at the FED ADCs.
%Figure~\ref{fig:apv-timing} (left) shows the signal propagation times
%for each detector module in the TIB system as a function of its CCU
%position within the control ring. The observed structure reflects the
%differences in signal propagation times in the control system: the
%three bands are due to three different fibre lengths used to connect
%the FECs to the TTC system; the gradient across the bins is due to the
%different positions of the CCUs within their control rings; and the
%spread for each group within a single bin is due to the different
%location of detector modules relative to their
%CCUs. Figure~\ref{fig:apv-timing} (right) demonstrates the relative
%synchronisation of the TIB partition after performing the procedure;
%the RMS of the distribution is 0.72~ns and 99.9\% of APV25 chips are
%synchronised to within 2~ns of the median value. The largest deviation
%is 4~ns, which corresponds to a maximum signal attenuation of
%$\sim$16\% in deconvolution mode.
The pre-synchronisation timing spread of up to 160~ns is reduced to an
RMS of 0.72~ns, with the largest deviation of 4~ns corresponding to a
maximum signal attenuation of $\sim$16\% in deconvolution mode.

%\begin{figure}[t]
%  \begin{centering}
%    \includegraphics[width=0.48\textwidth]{Figures/delays}
%    \includegraphics[width=0.48\textwidth]{Figures/timing}
%    \par
%  \end{centering}
%  \caption{\label{fig:apv-timing} (left) Signal propagation times for
%    each detector module in the TIB system as a function of its CCU
%    position within the control ring. (right) Synchronisation of
%    all modules in the TIB after the relative synchronisation
%    procedure. }
%\end{figure}

%% -----------------------------------------------------------------------------
%% -----------------------------------------------------------------------------
%% -----------------------------------------------------------------------------

\subsection{Calibration of the readout system gain}
\label{sec:tick-mark-calib}

One of the largest contributions to gain variation in the readout
system is the distribution of laser output efficiencies caused by the
variation of laser-to-fibre alignment from sample to sample during
production of the transmitters. In addition some loss may have been
introduced at the three optical patch panels in the fibre
system. Changes in the LV power supply or environmental temperature can
also significantly affect the gain at the level of a FED readout
channel.

The calibration procedure aims to optimise the use of the available
dynamic range of the FED ADCs and also equalise the gain of the entire
readout system. This is achieved by tuning the bias and gain register
settings of the LLD chip for individual fibres. Four gain settings are
possible. The amplitude of the tick mark, which is assumed to be
roughly constant in time and across all APV25 chips within the system,
is used to measure the gain of each readout channel. The setting that
results in a tick mark amplitude closest to 640 ADC counts is chosen,
as this amplitude corresponds to the expected design gain of
0.8. After tuning the system, a spread of $\pm$20\% is observed, which
is expected because of the coarse granularity of the LLD gain settings.

The response of all detector channels can be further equalised during
offline reconstruction by correcting the signal magnitude by the
normalisation factor $f = 640~\mathrm{ADC~counts}~/ a_{\mathrm{tick
mark}}$, where $a_{\mathrm{tick mark}}$ is the tick mark amplitude in
ADC counts. The tick mark amplitude is a good indicator of the maximum
output of the APV25 chip, which corresponds to a charge deposit of
175\,000~$\mathrm{e}^{-}$. This method provides a calibration factor
of $274\pm14$~$\mathrm{e}^-$/ADC~count. The estimated systematic
uncertainty is 5\%, attributable to the sensitivity of the tick mark
amplitude to variations in the LV power supply and environmental
temperature~\cite{TIF_Note}.

\begin{figure}[t]
  \begin{centering}
    \includegraphics[height=0.40\textwidth]{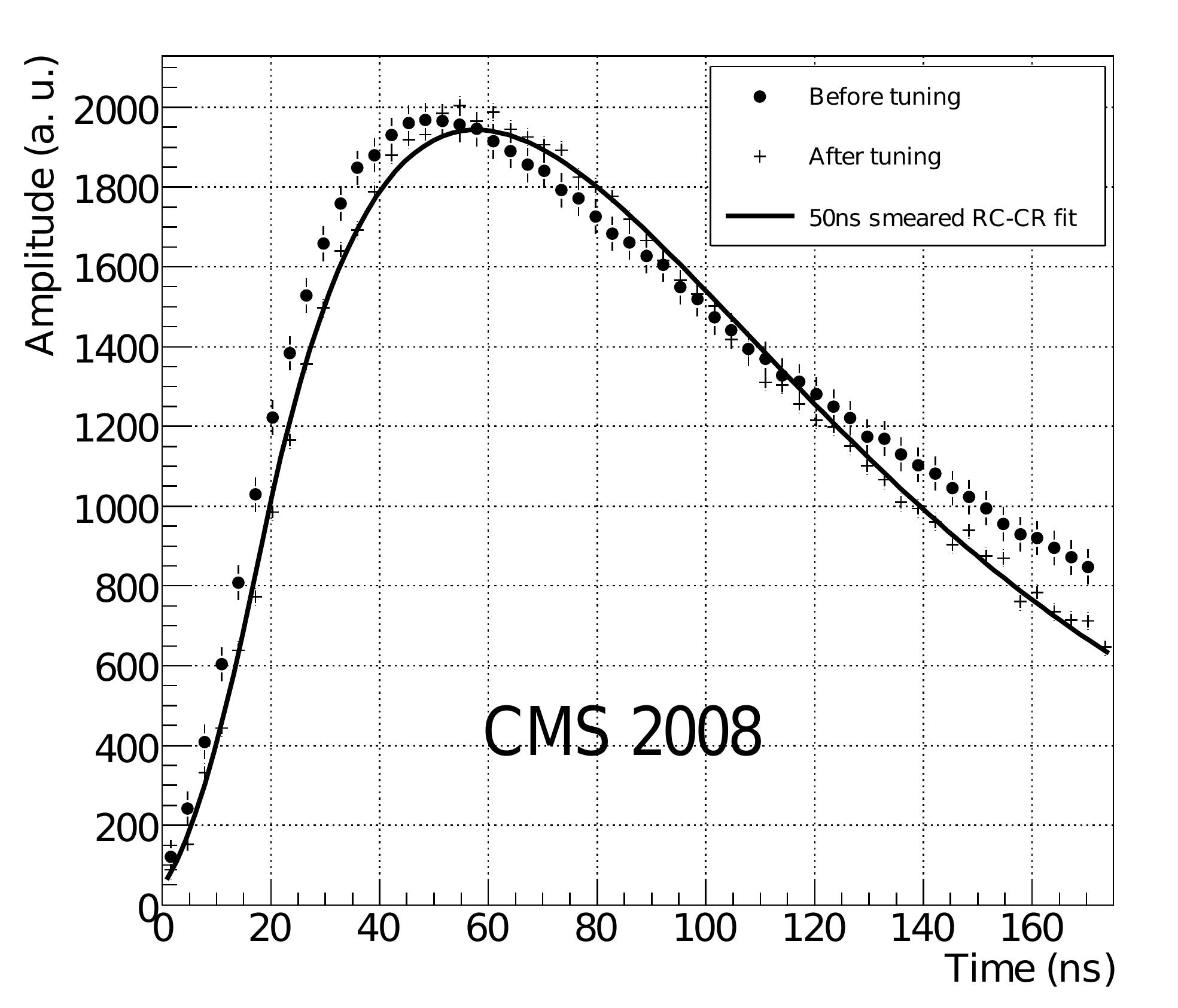}
    \includegraphics[height=0.40\textwidth]{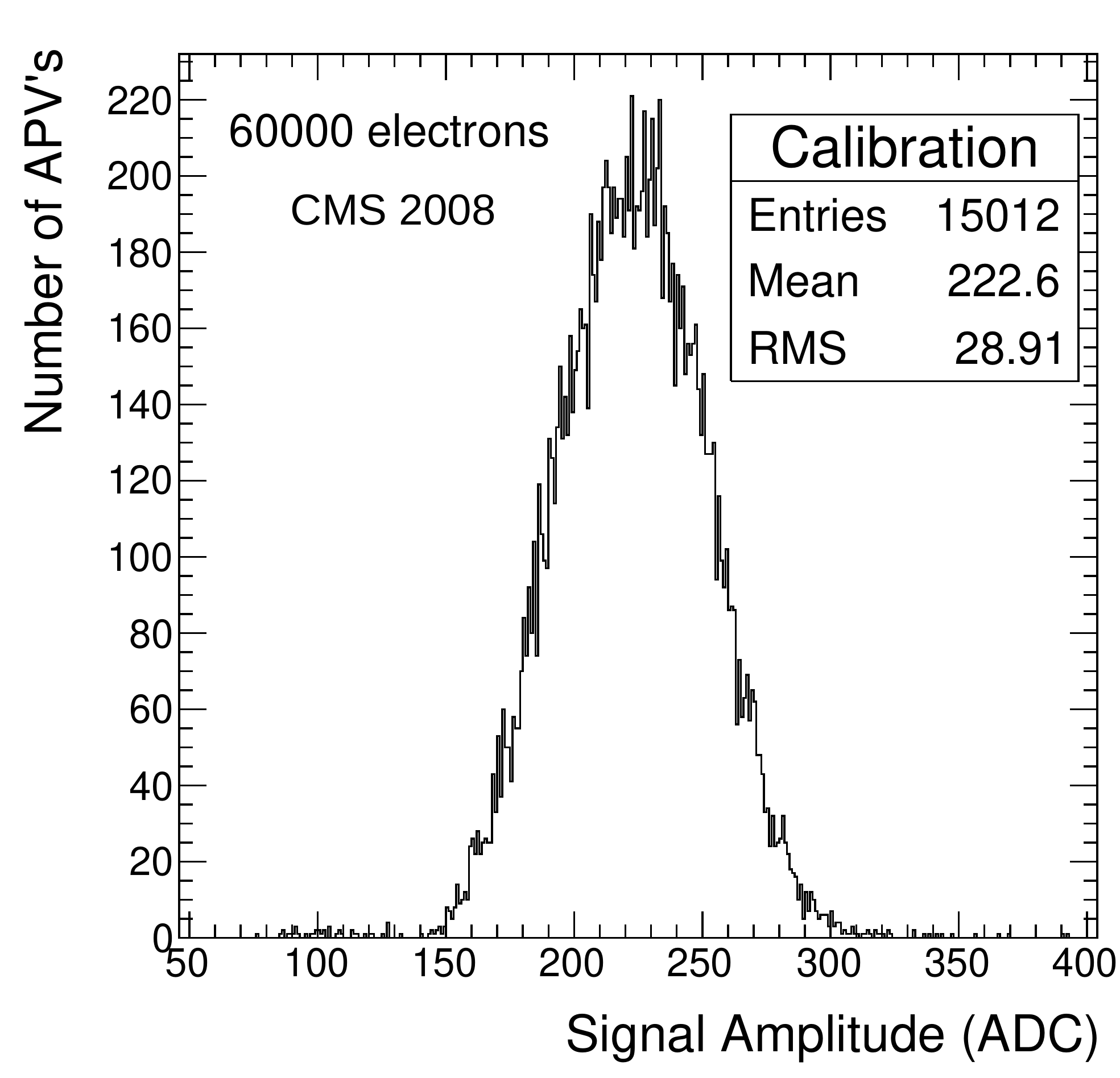}
    \par
  \end{centering}
  \caption{\label{fig:pulse-calib} (Left) An example of the CR-RC
    pulse shape of a single APV25 chip, before and after the pulse
    shape tuning procedure. (Right) Pulse height measurements using
    the on-chip calibration circuitry of APV25 chips in the TEC+. }
\end{figure}

%% -----------------------------------------------------------------------------
%% -----------------------------------------------------------------------------
%% -----------------------------------------------------------------------------

\subsection{Tuning of the APV25 front-end amplifier pulse shape}
\label{sec:pulse-shape}

The shape of the CR-RC pulse from the APV25 pre-amplifier and shaper
stages is dependent on the input capacitance, which depends on the
sensor geometry and evolves with total radiation dose. By default, all
APV25 chips are configured with pre-defined settings appropriate to
the sensor geometry, based on laboratory measurements~\cite{Timing}.
However, non-uniformities in the fabrication process result in a small
natural spread in the pulse shape parameters for a given input
capacitance.  This issue is important for performance in deconvolution
mode, which is sensitive to the CR-RC pulse shape. In order to
maximise the signal-to-noise ratio and confine the signal to a single
bunch crossing interval when operating in deconvolution mode, the rise
time of the CR-RC pulse shape must be tuned to 50~ns and the signal
amplitude at 125~ns after the signal maximum should be 36\% of the
maximum. This tuning also reduces the timing uncertainties associated
with the synchronisation procedures. Figure~\ref{fig:pulse-calib}
(left) demonstrates how the CR-RC pulse shape of an APV25, operating
in peak mode, can be improved by the procedure.

\begin{figure}[t]
  \begin{centering}
    \includegraphics[width=0.48\textwidth]{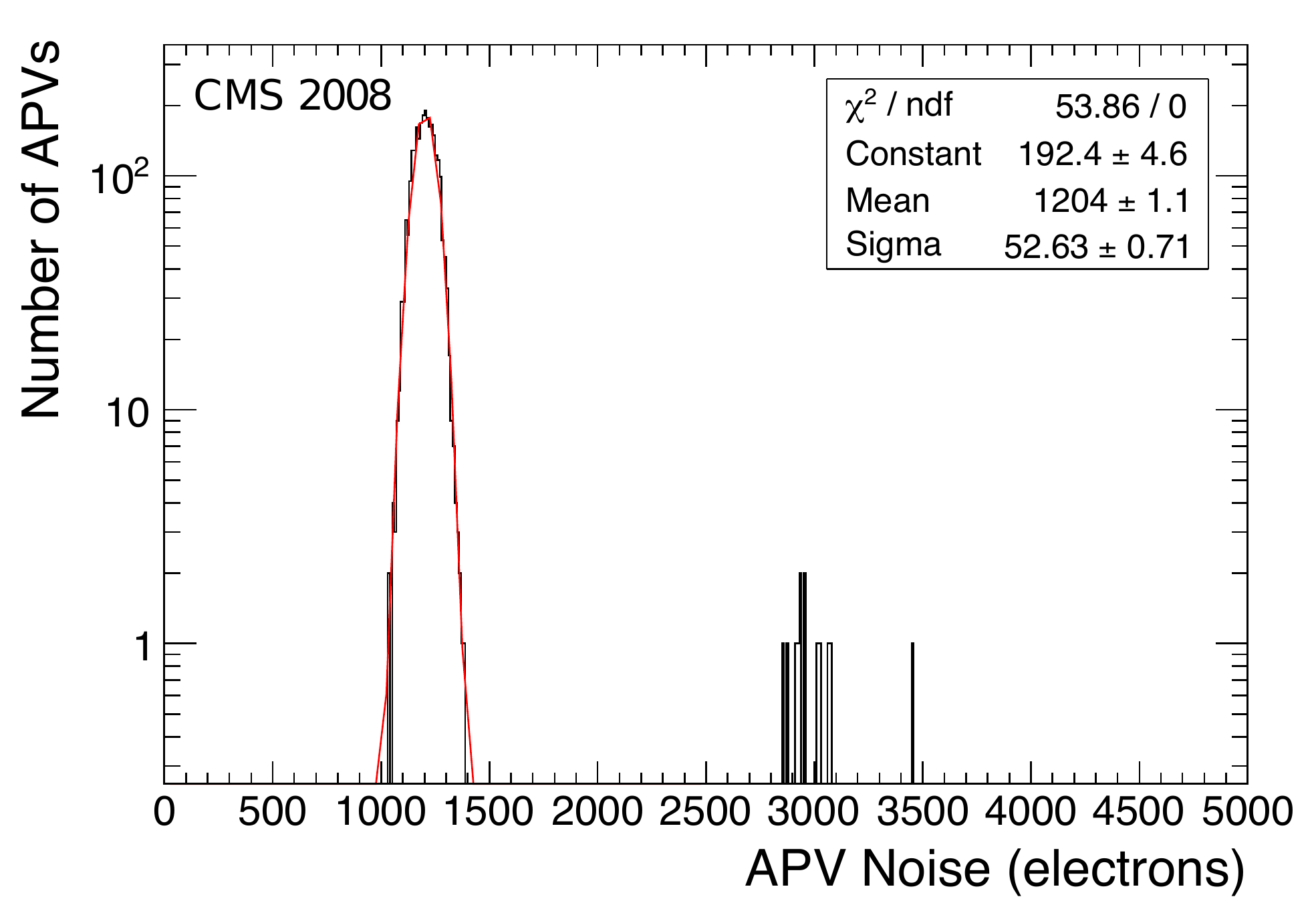}
    \includegraphics[width=0.48\textwidth]{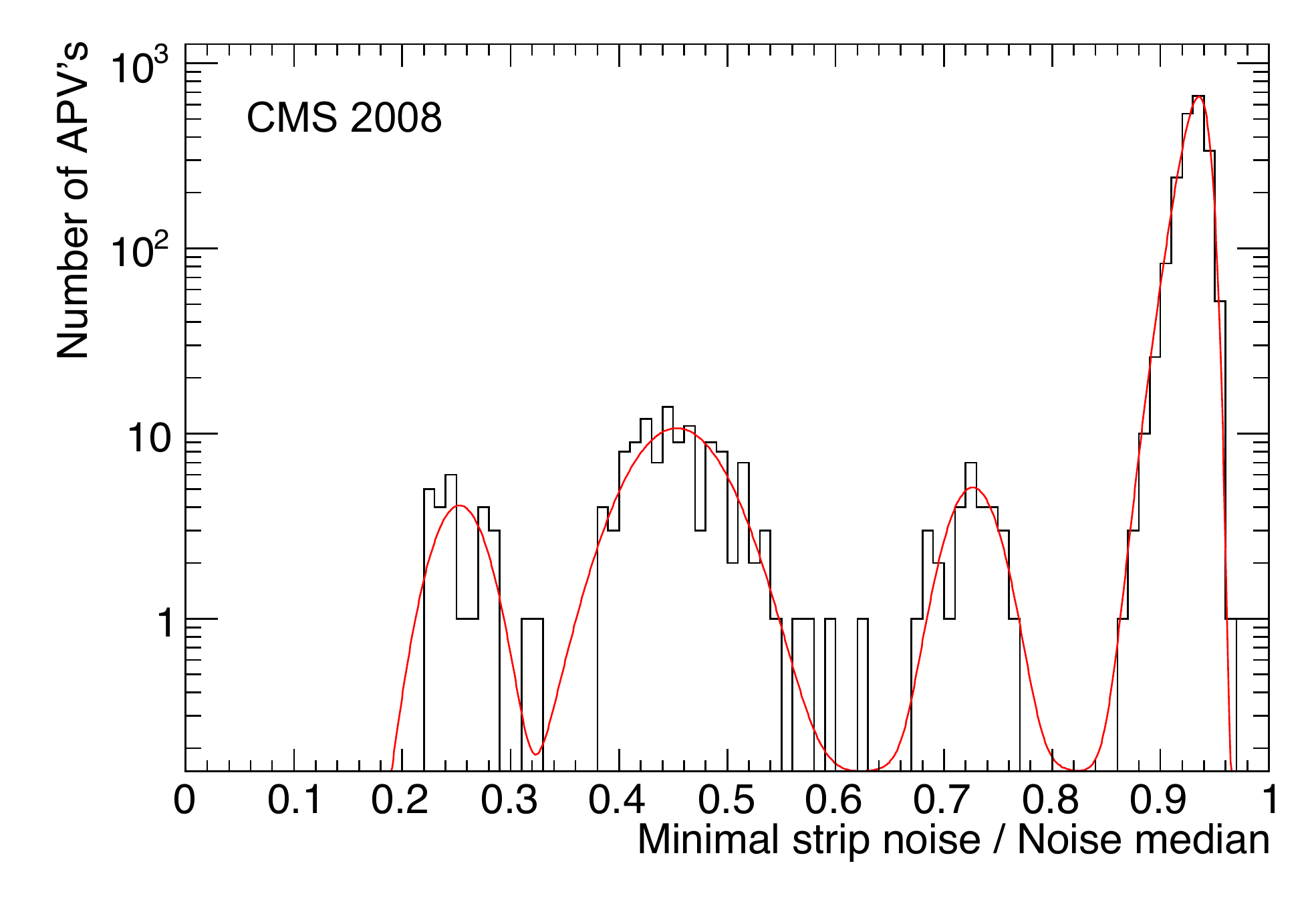}
    \par
  \end{centering}
  \caption{\label{fig:noise-analysis} (Left) Mean calibrated noise for
    individual APV25 chips on modules in the TOB single side layer 3. 
    (Right) The ratio of minimum noise to median noise per APV25
    chip. The distinct populations reflect the different noise sources
    within a module. }
\end{figure}

Figure~\ref{fig:pulse-calib} (right) shows the pulse height amplitude
(in ADC counts) observed for a charge injection of
60\,000~$\mathrm{e}^-$ using the on-chip calibration circuitry of the
APV25 chip. The charge injection provided by the calibration circuit
is known with a precision of 5\% and can be used to calibrate the
detector signal amplitude. A mean signal of 223~ADC~counts with a RMS
of 29~ADC~counts was observed, giving a calibration factor of
$269\pm13\, \mathrm{e}^-$/ADC~counts. This measurement is compatible
with the calibration based on tick mark amplitudes, described in
Section~\ref{sec:tick-mark-calib}.

%% -----------------------------------------------------------------------------
%% -----------------------------------------------------------------------------
%% -----------------------------------------------------------------------------

\subsection{Calibration of the detector channel pedestals and noise}
\label{sec:noise}

The mean level of the pedestals for the 128 channels of a given APV25
chip, known as the {\it baseline} level, can be adjusted to optimise
the signal linearity and the use of the available dynamic range of the
APV25. The baseline level for each APV25 chip is adjusted to sit at
approximately one third of the dynamic range.

Following this baseline adjustment, the pedestal and noise constants
for each individual detector channel must be measured, as these values
are used by the zero-suppression algorithms implemented in the FPGA
logic of the FEDs. 
Pedestals and noise are both measured using a random, low frequency
trigger ($\sim$10~Hz) in the absence of signal. Pedestals are first
calculated as the mean of the raw data in each detector channel from a
large event sample. They are subsequently subtracted from the raw data
values for each event. Common mode offsets are evaluated for each
APV25 chip per event by calculating the median of these
pedestal-subtracted data. The median value is then subtracted from
each channel. The noise for each detector channel is then defined to
be the standard deviation of the residual data levels, which can be
calibrated using the measurements described in
Sections~\ref{sec:tick-mark-calib}~and~\ref{sec:pulse-shape}.
Figure~\ref{fig:noise-analysis} (left) shows a distribution of the
mean noise measured per APV25 chip, for TOB single side layer 3. The
outliers correspond to APV25 chips from modules with unbiased sensors,
due to problems in the HV power supply.

\begin{table}
  \caption{Summary of the mean normalised noise for each type of
  sensor geometry. }
  \label{tab:Summary-of-noise}
  \vspace{1ex}
  \begin{centering}
    \begin{tabular}{|c|c|c|c|c|}
      \hline {Partition} & {Strip length (cm)} & {Total noise (
      $\mathrm{e}^-$)} & {Pitch adapter ( $\mathrm{e}^-$)} & {Bare APV
      ( $\mathrm{e}^-$)} \tabularnewline
      \hline 
      TEC Ring 1 &  8.52 & 757 & 421 & 245 \tabularnewline
      TEC Ring 2 &  8.82 & 791 & 434 & 265 \tabularnewline
      TEC Ring 3 & 11.07 & 832 & 450 & 250 \tabularnewline
      TEC Ring 4 & 11.52 & 843 & 437 & 257 \tabularnewline
      TEC Ring 5 & 14.44 & 1024 & 461 & 265 \tabularnewline
      TEC Ring 6 & 18.10 & 1097 & 513 & 270 \tabularnewline 
      TEC Ring 7 & 20.18 & 1146 & 510 & 258 \tabularnewline
      \hline 
      TOB Layers 1-4 & 18.32 & 1184 & 583 & 254 \tabularnewline 
      TOB Layers 5-6 & 18.32 & 1205 & 538 & 261 \tabularnewline
      \hline 
      TIB Layers 1-2 & 11.69 &  925 & 454 & 265 \tabularnewline
      TIB Layers 3-4 & 11.69 &  851 & 445 & 256 \tabularnewline
      \hline
    \end{tabular}
    \par\end{centering}
\end{table}

Modules with different sensor geometries are studied separately to
account for the different strip lengths and pitch adapter layouts that
affect the input capacitance. The mean normalised noise measured for
the different sensor geometries are summarised in
Table~\ref{tab:Summary-of-noise}. Fitting the mean noise versus
silicon strip length, the following parameterisation is obtained:
\[noise(\mathrm{e}^-)=(427\pm39)+(38.7\pm3.0)\times length(cm)\]
This is compatible with the measurement performed during the SST
integration period, prior to installation~\cite{cms}.

The individual sources of noise on the detector module can be
identified and measured by plotting the ratio of the minimum to the
median noise value for each APV25, as shown in
Fig.~\ref{fig:noise-analysis}~(right) and summarised in
Table~\ref{tab:Summary-of-noise}. The ratio takes advantage of the
fact that broken wire bonds on the detector modules effectively reduce
the input capacitance to individual channels of the APV25
chips. Broken wire bonds can occur between (in ascending order of
capacitance): the APV25 and pitch adapter; the pitch adapter and
silicon sensor; and sensors in two-sensor modules.  Fitting to the
first three populations, corresponding to the previous broken wire
configurations, provides an estimate of different noise
contributions. The fourth population corresponds to modules with no
broken wires. 

%% -----------------------------------------------------------------------------
%% -----------------------------------------------------------------------------
%% -----------------------------------------------------------------------------

\begin{figure}
  \begin{centering}
    \includegraphics[width=0.48\textwidth]{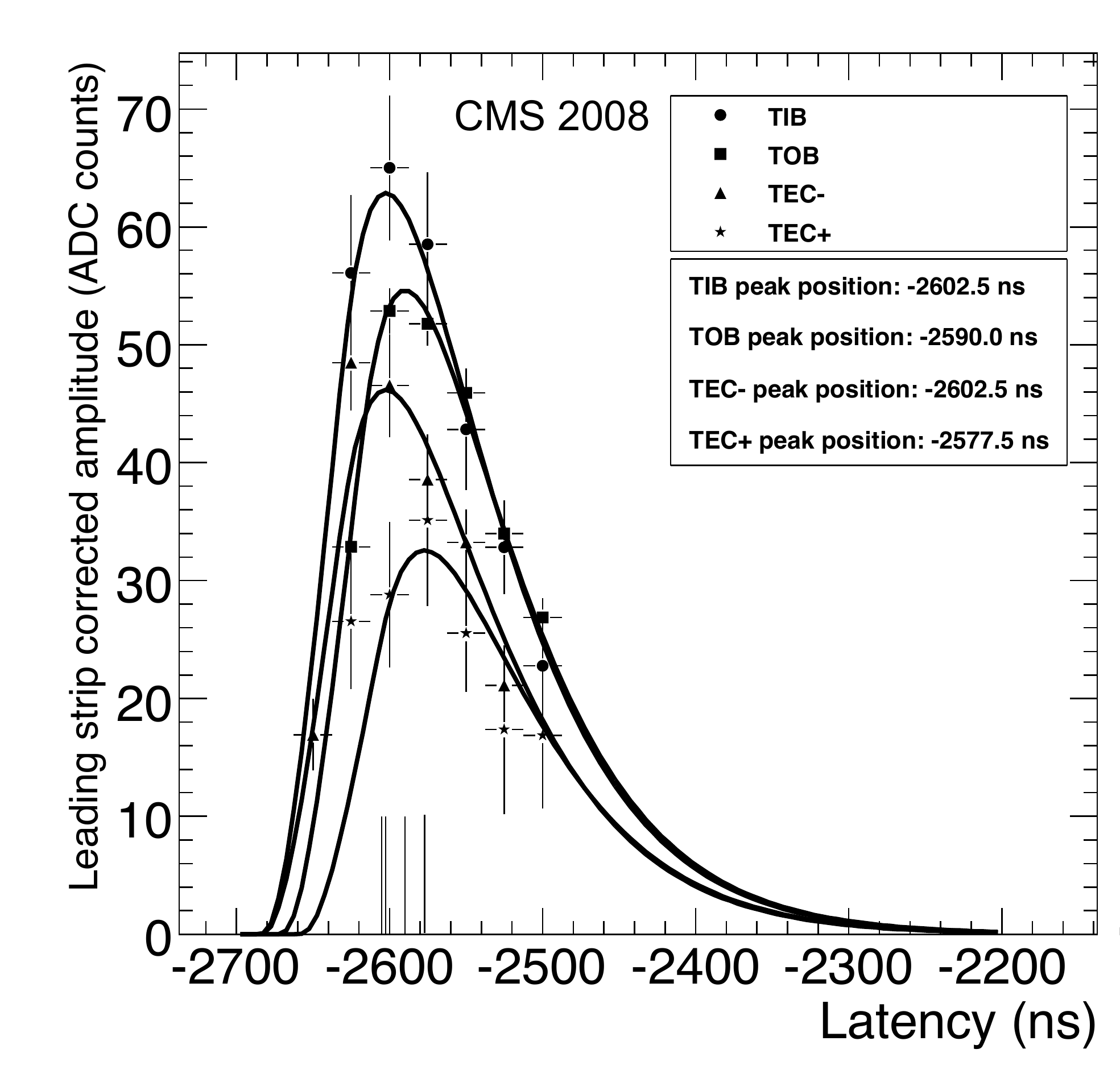}
    \includegraphics[width=0.48\textwidth]{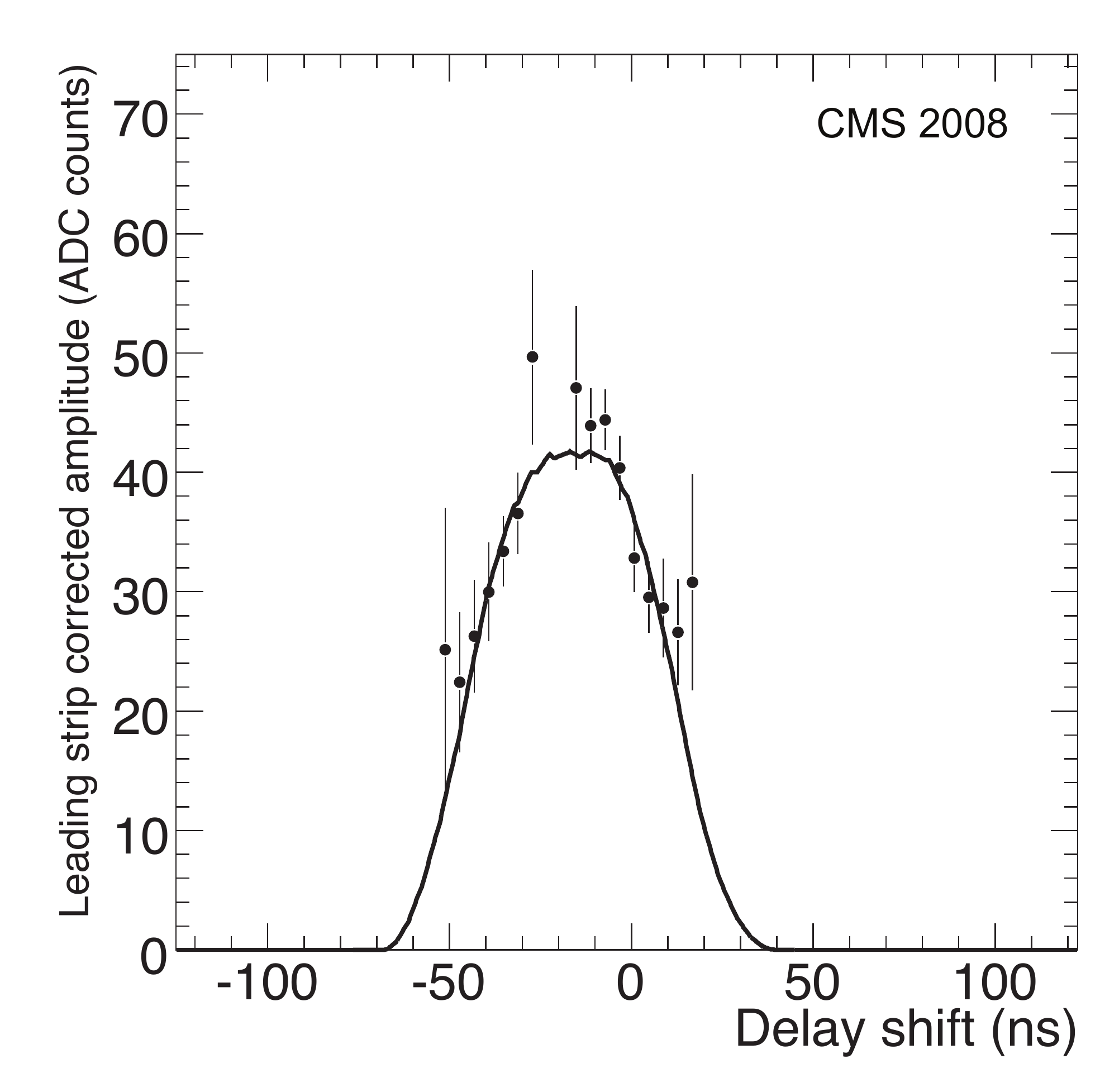}
    \par
  \end{centering}
  \caption{\label{fig:latency-scan} (Left) Mean signal of leading
    strip in clusters associated to tracks as a function of the
    latency (25 ns steps), for each of the four partitions. (Right) Fine delay scan
    for the TOB layer 3, in deconvolution. The mean position (-14.2\,ns)
    is including the mean time-of-flight of particles from
    the muon system to the silicon sensors (12\,ns).}
\end{figure}

\subsection{Absolute synchronisation to an external trigger}
\label{sec:ext-synch}

The last two commissioning procedures concern the synchronisation of
all modules in the SST with the Level-1 trigger of CMS. This was done
using a dedicated trigger provided by the Muon Drift Tube
sub-detector~\cite{MUON}, based on a coincidence between
centrally-located top and bottom chambers. The procedure requires
track reconstruction and the analysis was performed
offline~\cite{Timing}. Absolute synchronisation accounts for both the
delays introduced by the hardware configuration and the effects due to
the time-of-flight of particles.

The first of the two procedures is a coarse scan in time, in steps of
25~ns, by adjusting the latency between the trigger arrival and the
sampling time of the APV25 chip. The mean signal of the channel with
the largest signal amplitude ({\it leading strip}) in clusters
associated to reconstructed tracks was extracted as a function of the
latency. The signal magnitude was corrected for the track path length
through the active sensor volume, inferred from the track angle. The
latency measurement was performed for the tracker as a whole, but fine
adjustments for each partition were made relative to the TOB results:
TIB and TEC- were shifted by 12.5 ns and TEC+ by -12.5 ns, as shown by
the fits in Fig.~\ref{fig:latency-scan} (left). Time-of-flight is not
taken into account in this procedure, since the variations expected
across the detector ($\leq$10~ns with cosmic ray muons, 5~ns in
collisions) are lower than the target precision of 25~ns.

The last procedure comprises a fine tuning of the synchronisation.  It
involves skewing the clock delay in steps of 1~ns around the expected
optimal value for all modules of a given test layer, with the
configuration of all other modules in the SST unchanged with respect
to the value obtained from the coarse latency scan.  Clusters on the
test layer compatible with a reconstructed track are used to
reconstruct the pulse shape. Figure~\ref{fig:latency-scan} (right)
shows the resulting pulse shape from clusters found in modules of TOB
layer 3, acquired in deconvolution mode. With collision data, the
time-of-flight can be adjusted for each individual track, but this is
not the case for cosmic ray muons, for which the jitter from the
trigger cannot be subtracted.  The 14~ns shift observed is consistent
with the expected time-of-flight (12~ns) of cosmic ray muons from the
Muon Drift Tube chambers to the TOB layer 3.

From analysis of the latency and fine delay scans, correction factors
are computed to compensate the residual mis-synchronisation of each
partition. These corrections range from 1.0 to 1.06 with uncertainties
of ~0.03 and are used to correct the cluster charge in calibration and
$dE/dx$ studies, reported below. 

% \begin{table}
%   \caption{ Signal amplitude correction factors for each partition to account for
%     limitations in the synchronisation procedures used during CRAFT.
%     The uncertainty accounts for a $3\,\mathrm{ns}$ resolution on the
%     ideal sampling time and includes residual time-of-flight effects
%     in the tracker volume. }
%   \label{tab:syncCorrectionFactors}
%   \vspace{1ex}
%   \begin{centering}
%     \begin{tabular}{|c|c|}
%       \hline 
%       {Partition} & {Correction Factor} \tabularnewline
%       \hline 
%       TIB/TID & $1.018 \begin{array}{l} \scriptstyle{+0.012} \\ \scriptstyle{- 0.009} \end{array} $\\
%       TOB     & $1.0013 \begin{array}{l} \scriptstyle{+0.0065} \\ \scriptstyle{- 0.0012} \end{array} $\\
%       TEC+    & $1.058 \begin{array}{l} \scriptstyle{+0.032} \\ \scriptstyle{- 0.023} \end{array} $\\
%       TEC-    & $1.018 \begin{array}{l} \scriptstyle{+0.012} \\ \scriptstyle{- 0.009} \end{array} $\\
%       \hline
%     \end{tabular}
%     \par
%   \end{centering}
% \end{table}

%% -----------------------------------------------------------------------------
%% -----------------------------------------------------------------------------
%% -----------------------------------------------------------------------------

%% file: datasamples.tex
\section{Data Samples and Monte Carlo Simulations}
\label{sec:data-samples}
%\section{Data Sample}

In the following sections, the performance of the tracker will be analysed
using the data collected during CRAFT.
The event reconstruction and selection, data quality monitoring and data analysis were all performed
within the CMS software framework, known as CMSSW~\cite{ptdr}.
The data quality was monitored during both the online and offline
reconstruction~\cite{CRAFTWorkflow}.  The data were categorised and
the results of this categorisation procedure propagated to the CMS Dataset
Bookkeeping System~\cite{dbs}. Unless otherwise stated, only runs for
which the quality was certified as good, i.e., no problems were known
to affect the Trigger and Tracker performance, were used for the
analyses presented in this paper. 

The data-taking period can be split into three distinct intervals in time, based on magnetic field conditions
and tracker performance.  
%These intervals are summarised in Table~\ref{tab:ABC}.  
Each period has approximately uniform conditions.
In the first period, period A, part of the SST was not correctly
synchronised with the rest of the CMS detector.
%as discussed in Section~\ref{sec:ext-synch}.  
This
problem was fixed for data taken in subsequent periods.  The magnet was at its nominal field value
of 3.8~T during periods A and B, while period C corresponds to data
taken with the magnet switched off. 
%The temperature of the coolant was
%set to 11 $^{\circ}$C and was stable within 2 $^{\circ}$C for the
%whole period. The corresponding temperature of the sensors was
%measured to be between 21 and 27 $^{\circ}$C, depending on the type of
%module (single or double sided) and the number of APVs. 
Unless stated otherwise, the following results are based only on 
%2.2 million 
events from period B.
% For the results
% presented in the following Sections~4 and 5 only the data of period B
% are used, unless otherwise stated. 

% \begin{table}
% \caption{The different intervals into which the CRAFT data sample is divided for the studies presented in this paper.}
% \label{tab:ABC}
% \begin{center}
% \begin{tabular}{|c|c|c|l|} \hline
% Period &  Events ($\times 10^6$) & Magnetic & Comments \\
%        &                & Field    &          \\ \hline
% A      &  $10.9$        & On       & Synchronisation problem \\
% B      &  $2.2$         & On       &    \\
% C      &  $1.6$         & Off      &    \\ \hline
% \end{tabular}
% \end{center}
% \end{table}

For the studies presented in this paper, the events selected by the
Global Muon Trigger~\cite{CRAFTTrigger} were used. This data sample was
additionally filtered to include only events that contain at
least one reconstructed track in the tracker or that have a track
reconstructed in the muon chambers whose trajectory points back into
the SST barrel volume.
%The event numbers given in
%Table~\ref{tab:ABC} are quoted after filtering.

%The data quality was monitored during online and offline reconstruction and propagated to the CMS Dataset Book-keeping System~\cite{dbs} which 
%indexes event-data for the CMS Collaboration. Unless stated otherwise only good runs taken at B=3.8~T stable field are used for the analyses 
%presented in this paper. 
%%The track parameters are first determined without any additional cuts to get %an overview of the quality of reconstructed tracks through a 
%%comparison with %the MonteCarlo simulation (ref section tracking parameters?). Additional %quality cuts are then applied in order to achieve 
%%further analysis (ref %section efficiency, resolution?)
%\section{Monte Carlo Simulation}

Several analyses use a simulated sample of 21 million cosmic ray muons 
 to derive correction factors and compare results.
The sample was generated using the CMSCGEN package~\cite{CMSCGEN, CMSCGEN_2}.  
The detector was simulated using the standard program of CMSSW.
Modules known to be excluded from the read-out were masked in the simulation.
Besides this, the simulation was not optimised to the conditions of CRAFT. 
Nevertheless, the agreement with the data was sufficient for the purpose of
 the studies presented.

%The generated events were filtered before the detector simulation to
%select only those events with a cosmic ray muon in the fiducial tracker
%volume, i.e., a cylinder with a radius
%of 80~cm and length of 414~cm.  Muons outside this region can not be
%reconstructed in the SST, so this filter does not affect the results
%obtained on simulated data. 
%Furthermore, only muons with momentum at the
%surface greater than $4\,\gevc$ were kept, 
%as this corresponds to the minimum momentum required to penetrate through to the tracker.  
%The final simulation sample contains more than 21 million events.

%The simulation of the incident cosmic ray ray  muons is based on the generator CMSCGEN~\cite{CMSCGEN}. The parametrisation of the incident flux of cosmic ray
%rays has been developed for the L3 + Cosm`ics experiment~\cite{Adriani:2002uu} for energies beyond 10 GeV. The coverage of the spectrum at lower energies
%is modeled using data of the CAPRICEexperiment~\cite{CAPRICE}. 
%A filter has been specifically applied to select the events pointing to a cylinder modeling the tracker. The radius of this cylinder is set
%to 80 cm and the length is equal to 414 cm. This filter is used to simulate events which are crossing the tracker. The minimal momentum
%of the original muons is set to 4~\gevc, which corresponds to the minimal momentum needed to reach the tracker taking into account the
%energy loss through CMS. More than 21 millions of events have been generated and reconstructed using two different tracking algorithms
%(Sect.~\ref{sec:trackalgo}).

%% file: localreco.tex
\section{Performance of the Local Reconstruction}
\label{localreco}

In this section, the reconstruction at the level of the single detector module, is
presented.
The cosmic ray muon rate is small and events with more than one track are
rare. So with zero-suppression only a tiny fraction of the SST
channels are read out in any one event. These channels which pass
zero-suppression and therefore have non-zero ADC counts are known as
{\em digi}. Despite the zero suppression, digis may still only
consist of noise.
%As all the data taken during CRAFT were zero-suppressed,
%only a small fraction of the strips have non-zero ADC counts; these are known as {\em digis}.
%For each event, a list of digis in each module is used as input for the formation of clusters.  These clusters then, in turn,
%form the input for tracking algorithms.  

Clusters are formed from digs by means of a three threshold algorithm~\cite{ptdr}.  Clusters are seeded by digis which have a charge that is
at least three times greater than the corresponding channel noise.  For each seed, neighbouring strips are added if the strip charge 
is more than twice the strip noise.  A cluster is kept if its total charge is more than five times the cluster noise, defined as 
$\sigma_{\mathrm{cluster}} = \sqrt{ \sum_i \sigma_i^2 }$, where $\sigma_i$ is the noise from strip $i$, and the sum runs over all strips in the cluster.

In the following, the properties of both digis and clusters are studied and the performance of each SST subsystem is assessed.

\subsection{Occupancy}

The average number of digis per event and the occupancy are shown for
each SST subsystem in Table~\ref{tab:digirate}.  The strip occupancy
is computed after removing the masked modules (2.0 \%).  The average
occupancy in the SST is $4\times 10^{-4}$,
%to the 
%The average number of digis per event is shown in Table~\ref{tab:digirate} for each SST subsystem.
%The strip occupancy can be determined using these numbers, along with the number of readout channels, once the modules which have been masked in the commissioning are removed.
%These numbers may also be used to calculate the average occupancy for
%the complete SST, which has been determined to be  $4\times 10^{-4}$.
as expected from simulation and from the properties of
the zero suppression algorithm.
%(Section~\ref{sec:readout-system}).  
The digi occupancy is dominated by noise, but the cluster algorithm
reconstructs less than ten hits per event when there is no track within
the SST acceptance. 

\begin{table}
  \caption{Strip occupancies in the SST subsystems.  
    \label{tab:digirate}
  }
  \begin{center}
    \begin{tabular}{|l|c|c|c|c|} \hline
      & TIB & TOB & TID & TEC \\ \hline
      Average number of digis per event  &  720  & 1000 & 300 & 1700     \\
      Number of readout channels / 10$^6$ & 1.8 & 3.1 & 0.6 & 3.9        \\ 
      Strip occupancy from digis (\%) & 0.04 & 0.03 & 0.05 & 0.04        \\ \hline
      %strips masked  &     &     &     &     \\
      %APV25 masked  &  564   & 484 &  72 & 578    \\
      %APV25 total  &  13968 & 24192 & 4416 & 30208 \\  \hline
%      Total number of modules  & 2724 & 5208 & 816 & 6400 \\ 
%      Modules operated & 2616 & 5103 & 805 & 6283 \\
%      Number of excluded modules  & 108 (1)    &  105 (3)   &  11 (0)   &  117 (36)   \\
%      Modules remaining after offline analysis & 2615 & 5100 & 805 & 6247 \\
%      Particle occupancy from clusters (\%)& 0.04 & 0.04 & 0.04 & 0.05 \\ 
      Average number of clusters per event due to noise  &  1.0  & 2.0  & 0.3 & 3.0   \\  
    \hline
    \end{tabular}
  \end{center}
\end{table}

\subsection{Signal-to-noise ratio \label{sec:s2n} }
The signal-to-noise ratio is a benchmark for the performance of the SST.
It is particularly useful for studying the stability over
time.  
%The charge of the clusters generated by 
%cosmic ray muons as they pass through the sensitive volume of the
%sensor is proportional to the path length of the muons, so 
%the signal-to-noise ratio must be corrected for the traversed detector
%thickness obtained from the reconstructed   
%track.
In the signal-to-noise ratio, the cluster noise is divided by
$\sqrt{N_{\mathrm{strips}}}$, so that the resulting noise value is
approximately equal to the strip noise, independently of the size of the
cluster. 
\begin{figure}[hbtp]
  \begin{center}
    \includegraphics[width=0.48\linewidth]{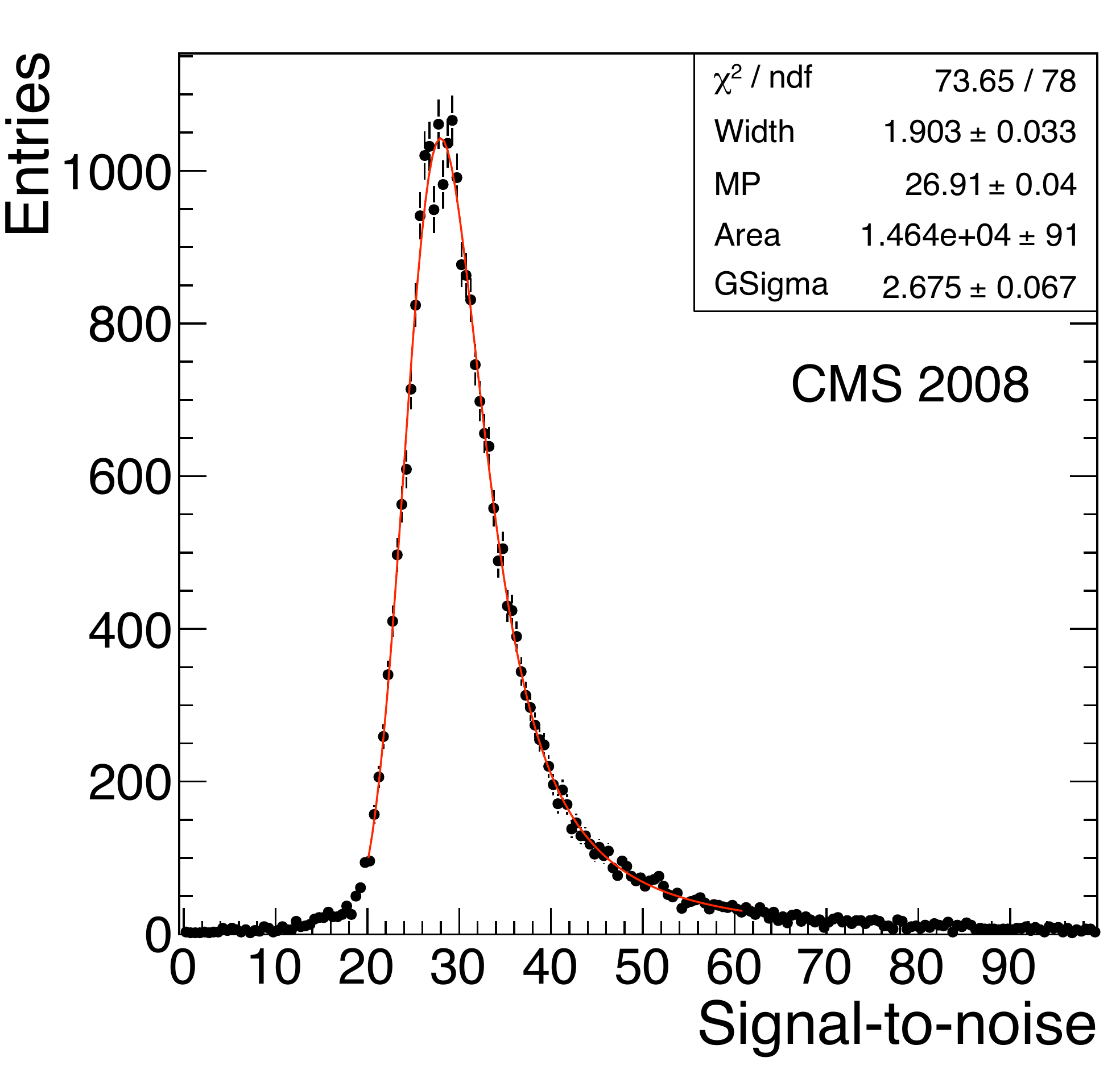}
    \includegraphics[width=0.48\linewidth]{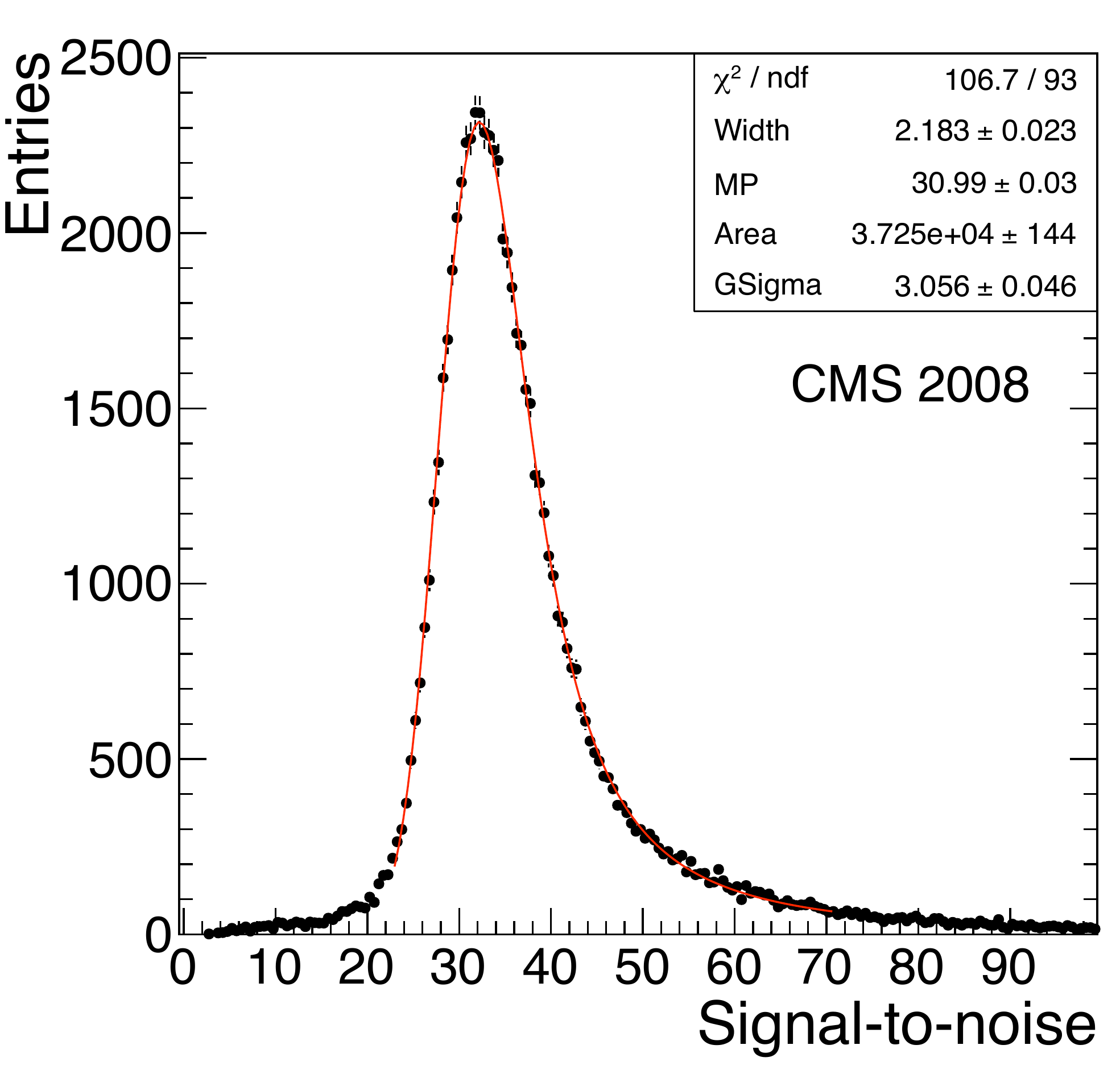}
%    \caption{Signal-to-noise ratio distributions of clusters associated to tracks in TIB (left) and TOB (right).}
    \caption{Signal-to-noise ratio distributions of clusters associated to tracks in TIB layer 1 (left) and TOB layer 5 (right).}
    \label{fig:CHG}
  \end{center}
\end{figure}
The path-length corrected signal-to-noise ratio distributions are
presented in Fig.~\ref{fig:CHG} for TIB layer 1 and TOB layer 5. 
The distributions have been fitted with a Landau function convoluted with a Gaussian function to determine the
most probable value for the signal-to-noise ratio. 
The result
is in the range 25-30 for thin modules and 31-36 for
thick ones, and within 5\% from the expected values.
% In
% Table~\ref{tab:ston} the fitted values of signal-to-noise ratio are
% compared to the expectations for all module types. The expected
% signal-to-noise ratio is calculated from noise values shown in
% Table~\ref{tab:Summary-of-noise} and the most probable signal for a
% minimum ionising particle, corresponding to about 22\,500 $e^-$ in
% thin sensors and 38\,000 $e^-$ in thick sensors. 
Thick sensors collect about a factor of $5/3$ more charge than the
thin sensors, but this does not simply scale up the signal-to-noise
ratio, as the noise is also larger for thick sensors, because of the
longer strips of these modules. 

\begin{figure}[hbtp]
  \begin{center}
    \includegraphics[width=0.7\linewidth]{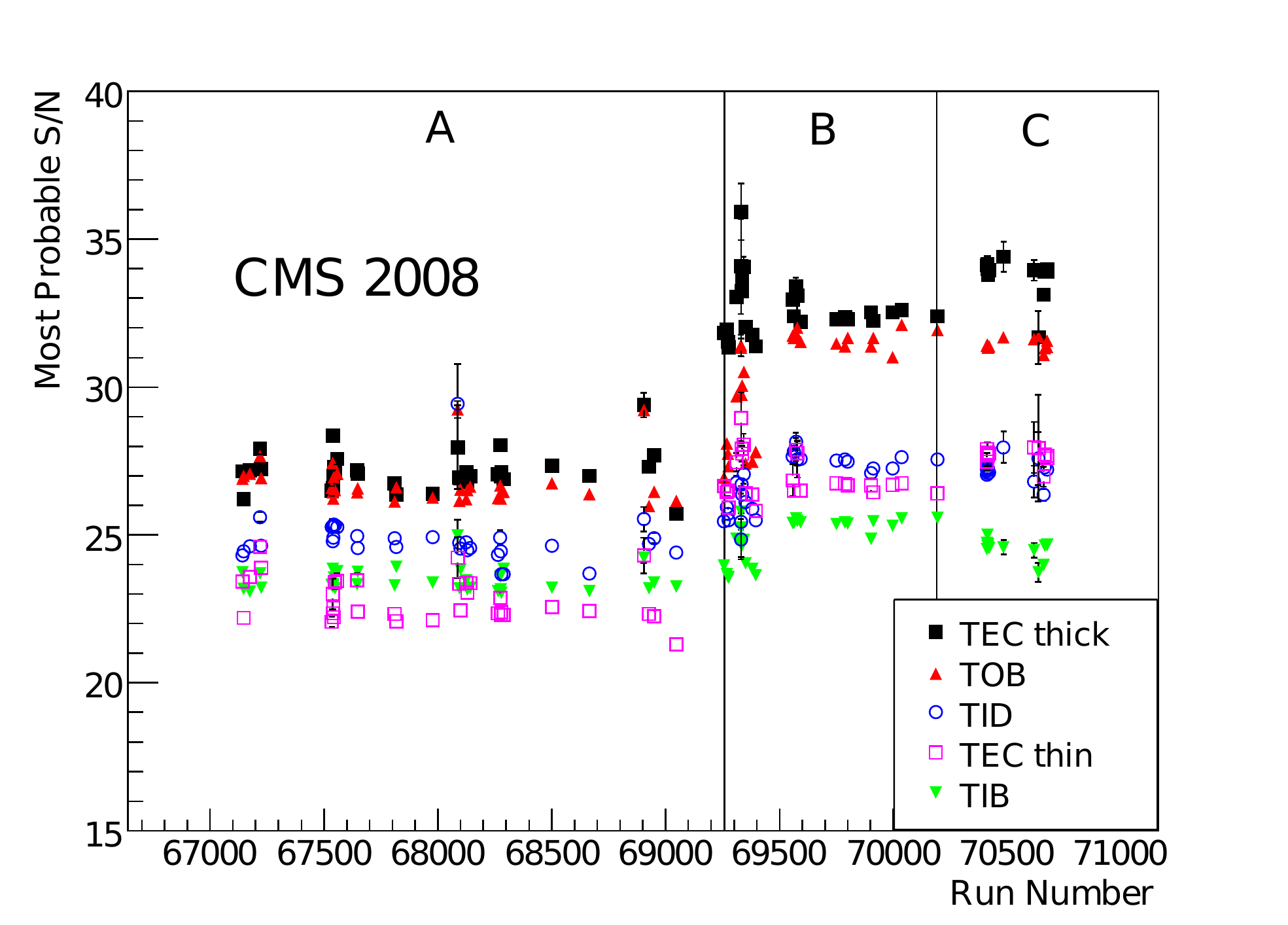}
    \caption{Signal-to-noise ratio versus the run number. The error
      bars represent the uncertainty associated with the Landau fit described in the text.}
    \label{fig:SNFIT}
  \end{center}
\end{figure}

The fit of the signal-to-noise ratio can also be performed on a run-by-run basis; Figure~\ref{fig:SNFIT}
shows the most probable value as a function of run number, allowing to
monitor the stability over a period of time.
Figure~\ref{fig:SNFIT} is divided into the three main data-taking
periods as discussed in Section~\ref{sec:data-samples}.  It can be seen that in period A
the signal-to-noise ratio was lower because muons were out-of-time in the modules not correctly
synchronised with the trigger. 
%than expected from previous measurements~\cite{TIF_Note}. 
Temporal variations of 5\% arise from
residual pedestal and timing mis-calibrations.

\subsection{Gain calibration \label{sec:gain}}
The charge released in the silicon sensors by the passage of a charged particle is processed by the readout electronics chain
described in Section~\ref{sec:readout-system}.  The ratio of ADC counts output after FED digitisation to the originally-released charge corresponds
to the gain of the electronics chain.
Particle identification using energy loss in the silicon detectors~\cite{pid-dedx} is known to be sensitive both to the absolute
calibration scale and to gain non-uniformities. It is therefore important that these non-uniformities be corrected for
and that the conversion factor between deposited energy and ADC counts is measured precisely.

\subsubsection{Inter-calibration of gain \label{sec:gainnorm}}
\begin{figure}
  \begin{center}
      \includegraphics[width=0.45\linewidth]{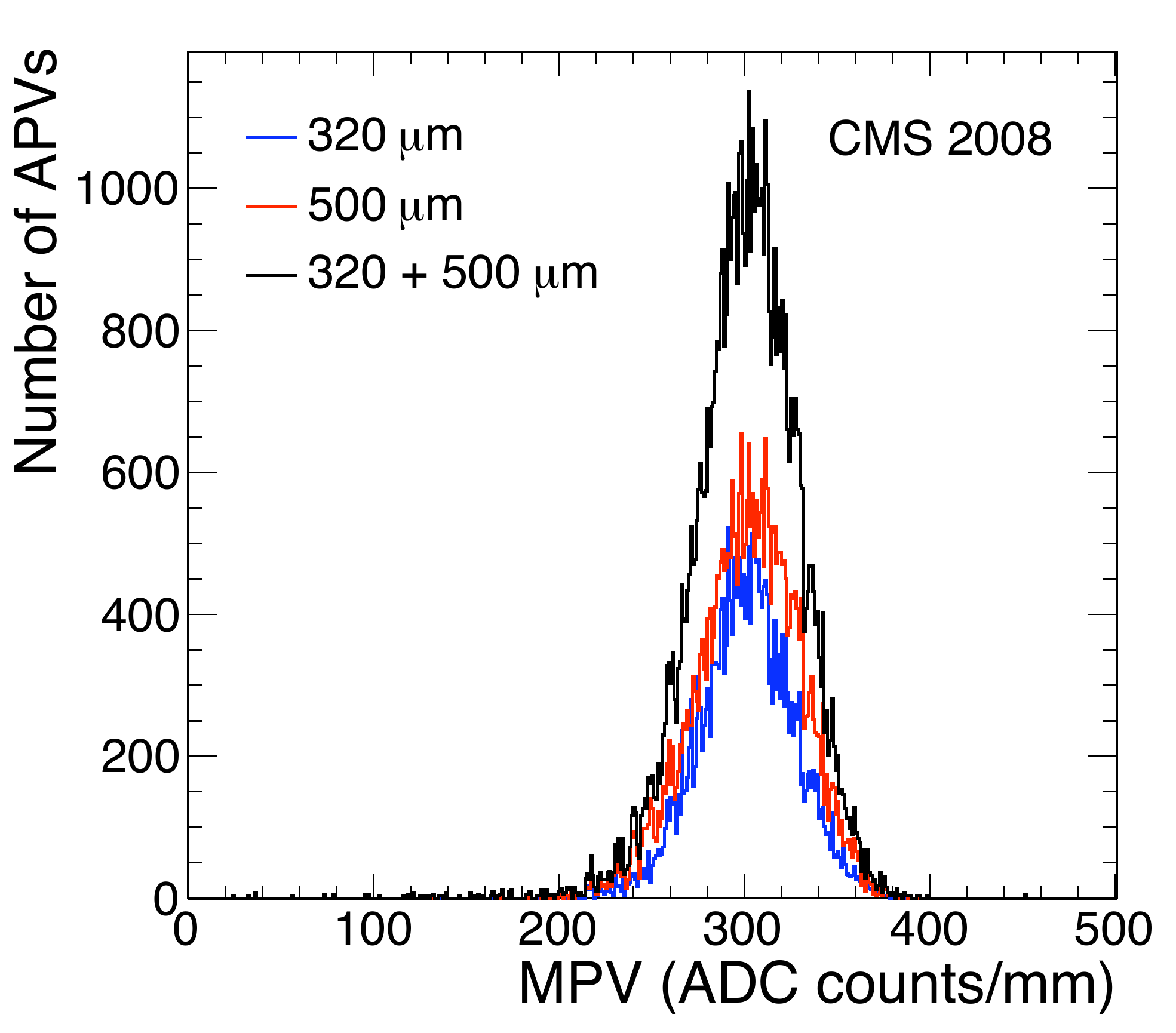}
    \caption{Most probable value of the cluster charge for different
      thicknesses before gain calibration.}
      \label{fig:gain}
  \end{center}
\end{figure}

The electronics gain can be made uniform throughout the SST simply by scaling the tick mark heights measured during calibration to an appropriate value. 
%chosen to be compatible with the direct measurements in Section~\ref{sec:tick-mark-calib}.
However, this procedure will not take into account gain changes due to
temperature variations and non-uniformities 
in the sensor response to a traversing particle, e.g., because of 
trigger synchronization, or because the sensor is not fully
depleted. For particle identification with energy loss,
non-uniformity must not exceed 2\%~\cite{pid-dedx}. 
This level of inter-calibration can be achieved only using the signals
produced by particles.  
The path length corrected charge of those clusters associated with
tracks was fitted with a separate Landau curve for each APV25 chip. 
Figure~\ref{fig:gain} shows the 
distribution of most probable values for APVs with at least 50
clusters, subdivided by sensor thickness.  The spread of these
distributions is around 10\%.

The most probable value of each distribution is then used to
compute the inter-calibration constants by normalising the signal to 300 ADC counts/mm -- the value expected for a minimum 
ionising particle with a calibration of 270 e$^-$/ADC~count
(Section~\ref{sec:pulse-shape}).
The inter-calibration constants determined in this manner were used in
the final reprocessing of the CRAFT data, resulting in a uniform
response.
%, as can be seen in Fig.~\ref{fig:gain} (right).   

\subsubsection{Absolute calibration using energy deposit information \label{sec:dedx}}
In addition to the inter-calibration constants, for particle
identification using energy loss, the ratio of deposited  
charge to ADC counts must be measured. 
The energy loss by particles traversing thin layers of silicon is
described by the Landau-Vavilov-Bichsel theory~\cite{bichsel}.
%, which differs significantly from the
%better-known Bethe-Bloch \linebreak predictions~\cite{Bethe,Bloch}.  
The most probable energy deposition per unit of length, $\Delta_p/x$, is described by the Bichsel function
and depends on both the silicon thickness and the particle momentum.  For muons, the function has a minimum at $0.5$ GeV/$c$ and then rises to reach a 
plateau for momenta greater than 10 $\gevc$.  The absolute gain calibration can be determined by fitting the Bichsel function predictions to the measured 
$\Delta_p/x$ values from the CRAFT data sample.

The quantity $\Delta_p/x$ is measured using the charge of clusters associated to tracks as a function of track momentum.  The resulting charge distributions
are fitted with a Landau convoluted with a Gaussian.  Only tracks with at least six hits and $\chi^2/\mbox{ndf}$ less than 10 are considered.  In addition,
only clusters with fewer than four strips are taken into account.  This last requirement is imposed in order to avoid mis-reconstructed clusters.

% In order to extract the absolute calibration factor, the response of all modules is first equalised using the tick mark amplitude, as described in
% Section~\ref{sec:tick-mark-calib}.  The gain normalisation described in the previous section is not used as it is based on the assumption that the
% average energy loss of a particle traversing the sensor does not vary between modules.  This is clearly not the case for thick and thin sensors.  In addition,
% the momentum spectrum of the incident cosmic ray muons will vary
% module-to-module, because of acceptance effects, which will change average
% energy loss.

Before the absolute calibration factor can be extracted from the
cluster charge data, two corrections must be applied.
Firstly, a correction is needed to take into account any charge loss in the zero-suppression process and during clustering.
This is determined using Monte Carlo simulations for each subsystem and for both thin and thick sensors in the end caps.  Secondly,
a correction is needed to handle the imperfect synchronisation between
the different subsystems.  Overall, the uncertainty due to these corrections is estimated to be about $1.5$\%.

\begin{figure}[ht]
  \begin{center}
    \includegraphics[width=0.5\linewidth]{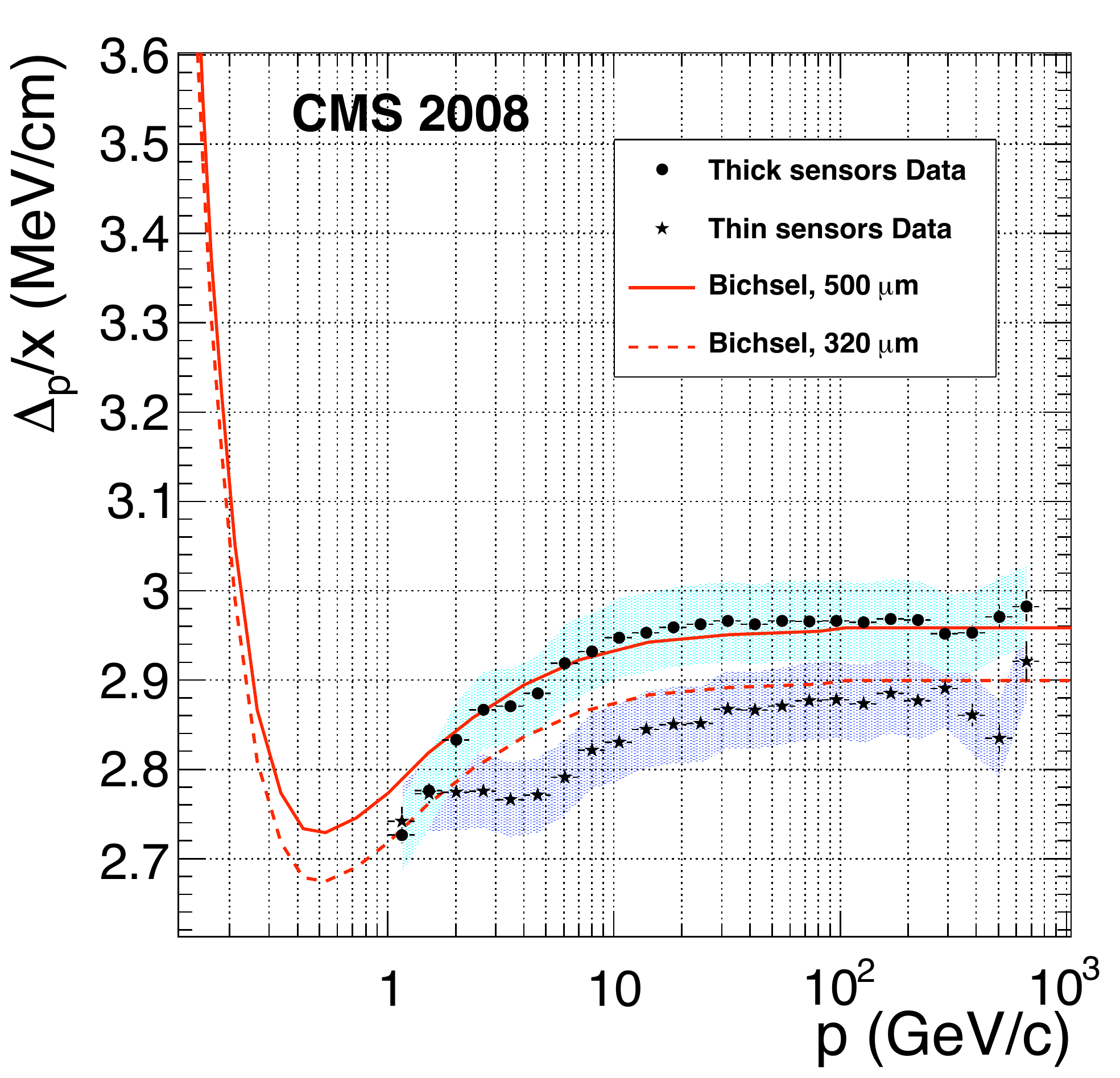}
    \caption{Most probable energy deposit per unit of length
      $\Delta_p/x$ as a function of track momentum, for thin and thick
      sensors. The shaded bands show the correlated systematic uncertainties on the measurements. 
The curves are the expectations from the Bichsel function~\cite{bichsel} as
explained in the text.}
    \label{fig:dEdx}
  \end{center}
\end{figure}

Figure~\ref{fig:dEdx} shows the most probable value of energy
deposition per unit length plotted as a function of the track momentum for both thin and thick sensors.  The error bars
reflect the uncertainty from the Landau fit, while the bands represent the fully-correlated systematic uncertainties from Monte Carlo corrections.
The small dip at 5 $\gevc$ arises from a temporary problem in the trigger provided
by a sector of the muon chambers, because of which this momentum region
was contaminated with out-of-time particles. 
The absolute calibration factor is determined separately for each subsystem and for both thin and thick sensors in TEC+ and TEC-.
The resulting values are given in Table~\ref{tab:dEdx}.  If a fit is
performed for all SST modules together, the absolute calibration
factor is found to be 262$\pm$ 3 e$^-$/ADC count, which is very similar to 
the result in the TOB alone, which dominates the data sample.  
However, thick and thin modules are
compatible and overall the result is in agreement with the value of
$269\pm13$ e$^-$/ADC count obtained from the pulse 
calibration described in Section~\ref{sec:pulse-shape}. 
\begin{table}
  \begin{center}
    \caption{\label{tab:dEdx} Absolute gain calibration measured from
      energy deposit per unit length, $\Delta_p/x$.}  
    \begin{tabular}{|l|c|c|c|c|c|c|} \hline
      Subsystem         & 
      TIB	        & 
      TOB  	        & 
      TEC+ thin         & 
      TEC+ thick	& 
      TEC- thin	        & 
      TEC- thick	\\ \hline
e$^-$/ADC count    & 
$262.3^{+2.5}_{-3.5}$ &
$261.5^{+0.5}_{-1.5}$ &  
$273^{+7}_{-9}$     &  
$270^{+7}_{-9}$     &  
$264^{+3}_{-4}$     &   
$261^{+3}_{-4}$     \\  
\hline      
\end{tabular}
\end{center} 
\end{table}  

\subsection{Lorentz angle measurement \label{sec:lorentz} }
\begin{figure}[h]
  \begin{center}
    \includegraphics[width=0.5\linewidth]{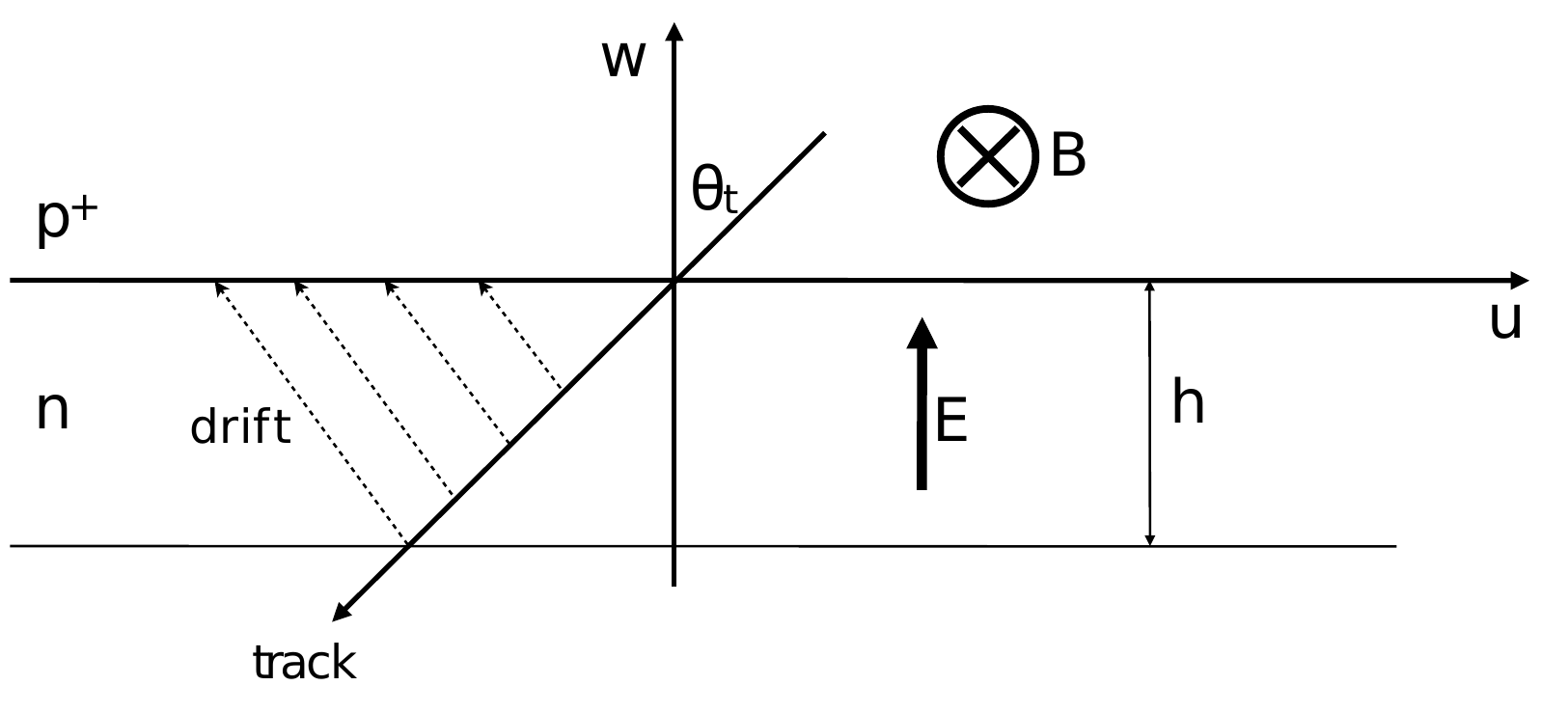}
    \caption{Lorentz drift in the microstrip sensors.}
    \label{fig:lorentz}
  \end{center}
\end{figure}

\begin{figure}[hbtp]
  \begin{centering}
    \includegraphics[width=0.5\linewidth]{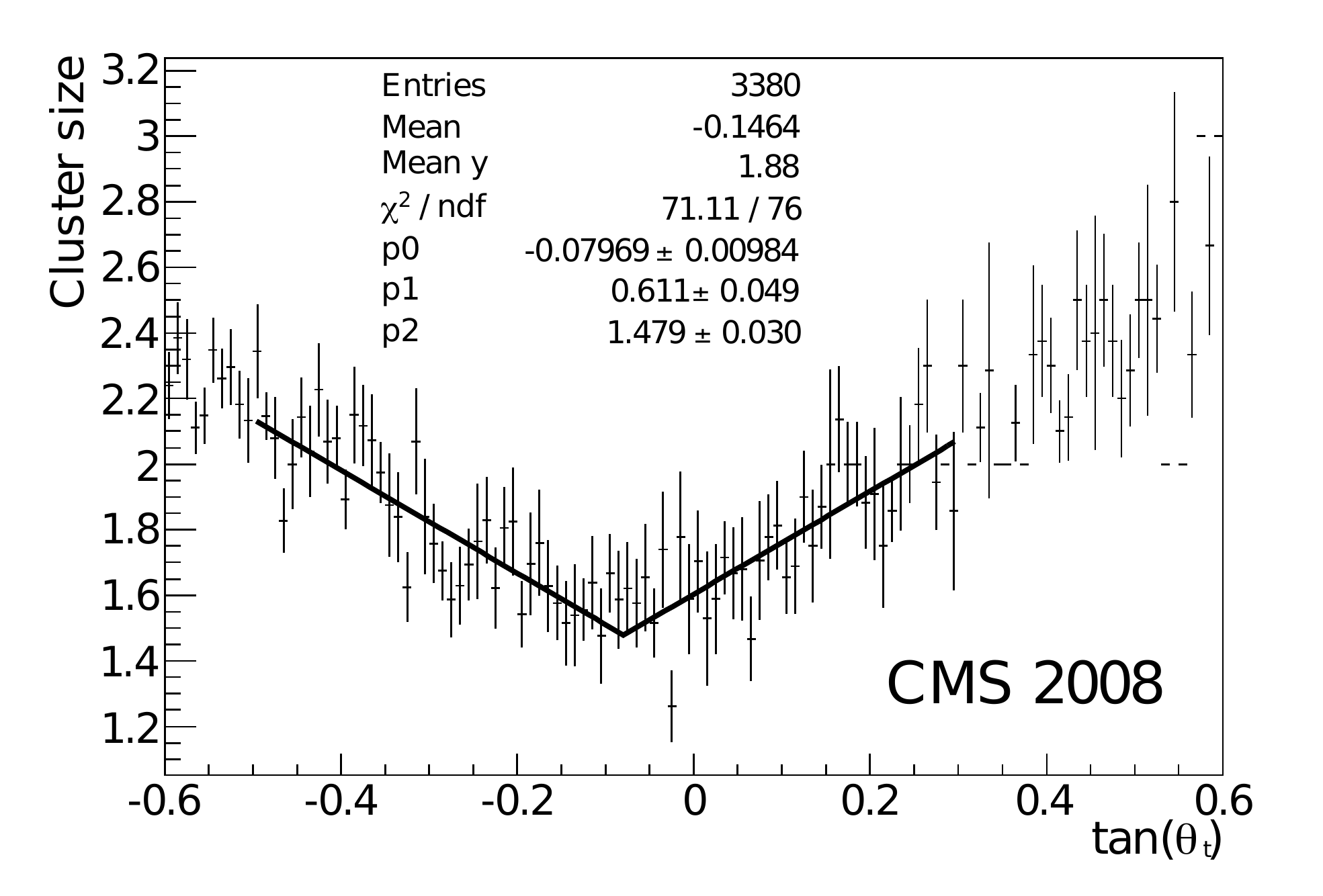}
    \caption{Cluster size versus incident angle in one module of TOB Layer 4.\label{fig:TK-loren}}
  \end{centering}
\end{figure}

In the silicon sensors, the electric field is perpendicular to the
strips. For normal incidence particles, typically only one strip is hit and the
cluster size increases with the angle of incidence.  In the presence
of a magnetic field, however, the drift direction is tilted by the
Lorentz angle, as illustrated in Fig.~\ref{fig:lorentz}. 
This is illustrated, for one module in layer 4 of TOB, in
Fig.~\ref{fig:TK-loren}, which shows a profile plot of cluster size
versus the tangent of the incidence angle. To extract the Lorentz
angle, this distribution is fitted to the function:
%Since the Lorentz angle is used during track reconstruction  
%to correct cluster positions, it is important to measure it, and, as  
%it may change as the silicon is irradiated, to have a method of  
%monitoring its long-term behaviour.
%The Lorentz angle can be directly measured by determining the angle at which the minimum cluster size is ob%served~\cite{LORENTZ}.  
%This is illustrated in Fig.~\ref{fig:TK-loren}, which shows a profile plot of cluster size versus the tangent of the incidence angle.
%This particular example is for one module in layer 4 of TOB with the
%magnetic field at its nominal value of 3.8 T.
\begin{displaymath}
  f(\theta_t)=\frac{h}{P}\cdot p_1 \cdot |\tan\theta_t - p_0| + p_2
\end{displaymath}
where $h$ is the detector thickness, $P$ is the pitch, and $p_0$, $p_1$ and $p_2$ are the fit parameters.  The parameter $p_0$ is,
in effect, $\tan \theta_L$, while $p_1$ represents the slope of the line divided by the 
ratio of thickness to pitch.
%while $p_1$ represents the product of the slope of the line and the ratio of thickness and pitch.
The third parameter, $p_2$, is the average cluster size at the minimum.

%Given that the Hall mobility is defined as $\mu_H = \tan\theta_{L}/B$, where $B$ is the magnetic field in Tesla, the result of the fitting
%procedure is practically a measurement of $\mu_H$.  
The Lorentz angle is measured for each individual module. 
%so the value
%for each layer can be determined simply as an average of the
%mobilities for eac%h module in the layer.  
The mean $\tan\theta_{L}$ is $0.07$ in TIB and $0.09$ in TOB, with
an RMS of $0.02$ and $0.01$, respectively. 
%The results of the
%layer-by-layer measurements are clearly consistent in TIB and in TOB,
A small difference between TIB and TOB %, though lower than observed, 
is expected because the hole mobility depends on the electric field, and
therefore, for the same applied voltage, on the thickness. 

The Lorentz angle correction applied to clusters during track reconstruction is relatively small -- of the order of 10 $\mu$m -- 
but it is still larger than the overall alignment precision~\cite{craftAlign}.  The alignment procedure can therefore provide a useful method of cross-checking
the Lorentz angle measurements.  In particular, it is useful to compare the residual distributions from data with and without the magnetic
field applied.  Results from the tracker alignment procedure confirm the measurements presented here~\cite{craftAlign}.

% \begin{table}
%     \caption{Mean and RMS values of Hall mobility $\mu_H$ determined
%       for each layer in the TIB and TOB.}
%   \label{tab:lorentzangle}
%   \vspace{1ex}
%   \begin{centering}
%     \begin{tabular}{|c|c|c|}
%       \hline
%       \multicolumn{3}{|c|}{$\mu_H$ (T$^{-1}$)} \\
%       \hline 
%       Layer & Mean  & RMS  \\
%       \hline
%       TIB 1& 0.0184 &0.005\\  
%       TIB 2& 0.0166 &0.004\\     
%       TIB 3& 0.0182 &0.004\\ 
%       TIB 4& 0.0186 &0.004\\ 
%       TOB 1& 0.0212 &0.004\\ 
%       TOB 2& 0.0227 &0.003\\     
%       TOB 3& 0.0223 &0.005\\ 
%       TOB 4& 0.0234 &0.003\\ 
%       TOB 5& 0.0235 &0.002\\ 
%       TOB 6& 0.0237 &0.002\\
%       \hline
%     \end{tabular}
%     \par\end{centering}
% \end{table}

\subsection{Hit efficiency}
The hit efficiency is the probability to find a cluster in a given silicon sensor that has been traversed by a charged particle.
In order to calculate the hit efficiency, track seeding, finding, and reconstruction must be performed.  The results presented here
have been determined using the \ckf for cosmic ray muons events (see Section~\ref{sec:trackalgo} for further details), excluding
the clusters in the layer of the SST for which the hit efficiency is to be determined.  The efficiency for a given module in this layer 
is then calculated by finding tracks that pass through that module and determining if a cluster is, in fact, present.

A single run from the CRAFT dataset has been used in order to assure that the number of excluded
modules did not change.  A very long run was chosen to ensure that the track statistics were sufficient.  There were between 16\,400 and 104\,800 
tracks per barrel layer and between 1700 and 6500 per end cap layer.  The analysis was limited to events that contained only one track, which 
was required to have a minimum of eight hits and no more than four missing hits.  To ensure that the muon has actually passed 
through the module under study, the location of the extrapolation of
the track trajectory on the module surface was required to be no closer to the sensor edge than five times the position uncertainty of the extrapolated point.

\begin{figure}[bhtp]
  \begin{center}
    \includegraphics[width=0.48\linewidth]{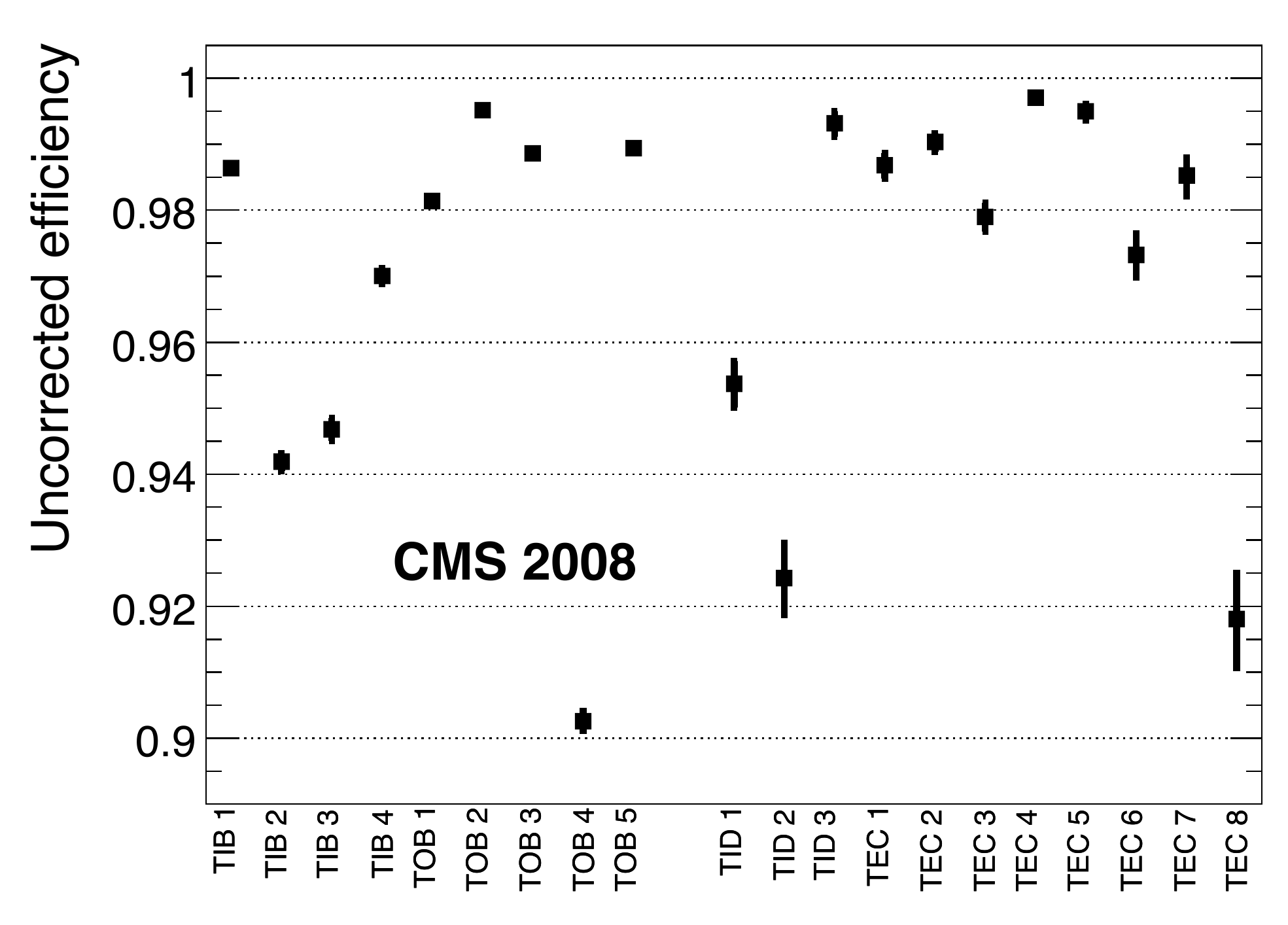}
    \includegraphics[width=0.48\linewidth]{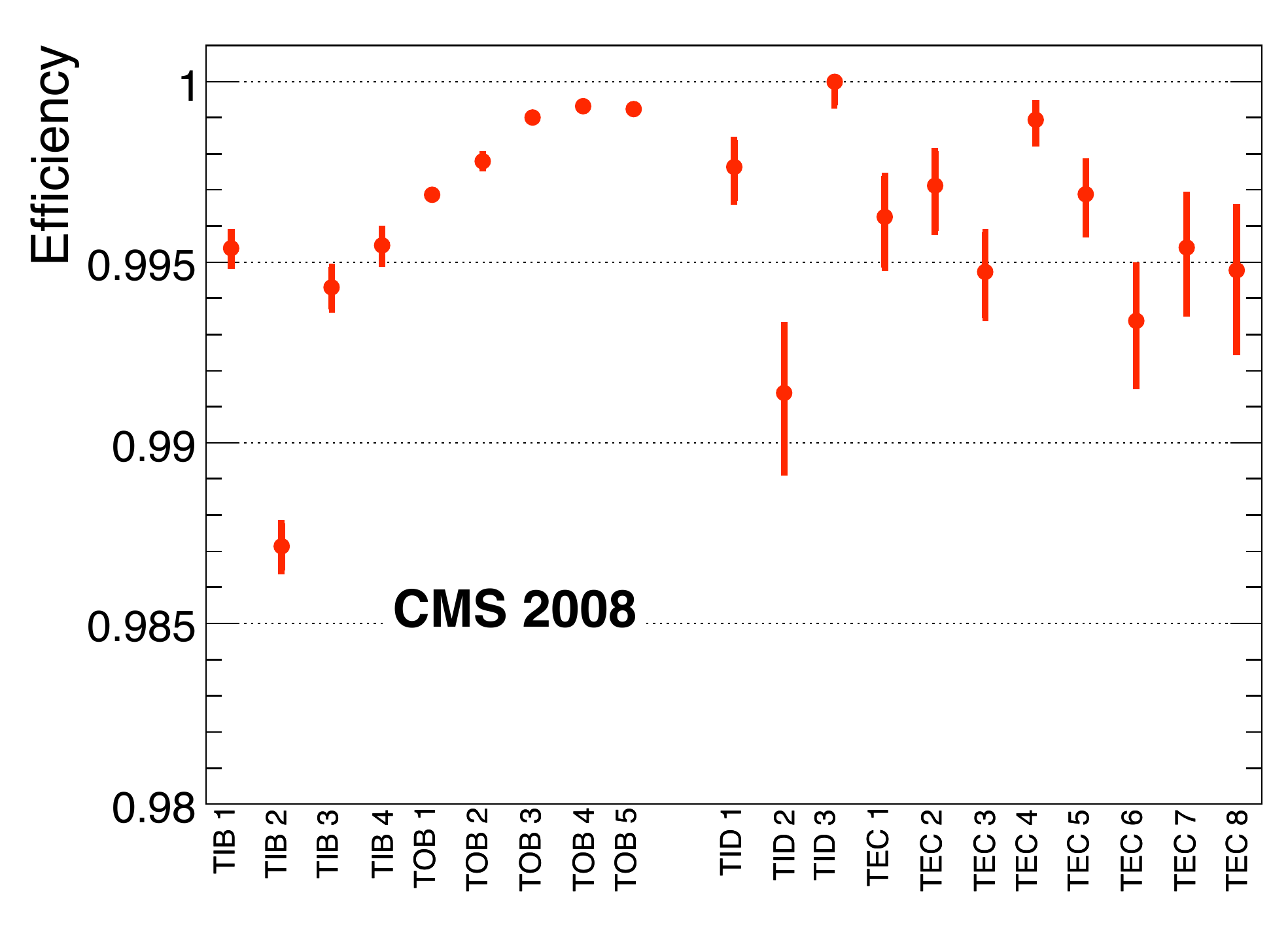}
    \caption{Average module hit efficiency per layer/disk, 
      without any correction for disconnected or otherwise exclude modules 
      (left) and after applying such corrections (right). The efficiency 
      cannot be measured in the outermost 
      layers of TOB (layer 6) or TEC (layer 9) without modifying the
      track reconstruction algorithm, because the track
      reconstruction requires the presence of a hit in the outermost
      layer or disk, depending on the track trajectory.} 
    \label{fig:hiteff}
  \end{center}
\end{figure}

The efficiency results per SST layer are shown in Fig.~\ref{fig:hiteff}.  These measurements, which include all SST modules, are
compatible with the expected overall percentage of excluded modules.
If the modules that were
excluded because of known problems were ignored in the efficiency calculation, the resulting efficiency would be greater than
99\% for most layers. No more than about $0.001$ of the inefficiency arises from isolated dead strips~\cite{tifPaper}, 
which are not taken into account in the efficiency calculation for
Fig.~\ref{fig:hiteff} (right).  The
rest is attributed to modules that were problematic only for
a short period of time and were therefore not identified by the other
procedures described in this paper.  Subsequent improvements, such as
detailed logging of modules affected by temporary power supply
problems (HV trips etc.), will improve the bookkeeping of inefficiency for future data-taking. 
% or from modules that had not clearly been identified as problematic.

%% file: tracking.tex
\providecommand{\cov} {95\% coverage\xspace}
%\newpage

\section{Track Reconstruction}

In this section, the performance of the track reconstruction using the full tracker, including the pixel detector, is presented.
Details of the commissioning and the performance of the hit reconstruction in the pixel detector can be found elsewhere~\cite{craftPixel}.
%Detailed description of the commissioning of the pixel detector and the performance of the local reconstruction is given in Ref.~\cite{craftPixel}.

\subsection{Track reconstruction algorithms\label{sec:trackalgo}}

The two main algorithms used to reconstruct tracks from cosmic ray muons in CRAFT data are the \ckf (CTF) and the Cosmic Track Finder (CosmicTF).
The \ckf is the standard track reconstruction algorithm intended for
use with proton-proton collisions and the main focus of the present study; for these runs, it has been specially
re-configured to handle the different topology of cosmic muon events.  The second algorithm was devised specifically for the reconstruction
of single track cosmic ray muon events. 
Since it is meant as a cross-check of the \ckf, it has not been tuned to the same level of performance.
A full description of these algorithms can be found elsewhere~\cite{tifPaper}.

There have been two significant changes in the \ckf since its first use in the Slice Test, both relating to the seed finding phase.
The Slice Test was performed without the presence of a magnetic field and with only limited angular coverage.  Now that the full tracker is 
available, seed finding in the barrel uses TOB layers only and both hit triplets and pairs are generated.  In the end caps, hits in adjacent disks are
used to form hit pairs.
The presence of the 3.8\,T magnetic field means that for hit-triplet seeds, the curvature of the helix yields an initial estimate of the momentum.  For 
hit pairs seeds, an initial estimate of $2\,\gevc$ is used, which corresponds to the most probable value.
The detector has been aligned with the methods described in Reference~\cite{craftAlign}.

\subsection{Track reconstruction results}

The number of tracks reconstructed by the two algorithms in the data from Period B, without applying any additional track quality
criteria, except those applied during the track reconstruction itself,
are 2.2 million 
%2\,196\,949 
using the \ckf and 
2.7 million 
%2\,785\,396 
by the \costf.

%To ensure that no known detector problem would affect the analysis of the performance of the track reconstruction, only the runs in Period~B 
%(c.f.\ Table~\ref{tab:ABC}) are used.  Period~C is not used, since the magnetic field was off during this period.
%Without applying any additional track quality criteria except those applied during the track reconstruction itself,
%The number of tracks reconstructed by the two algorithms in Period~B, without applying any additional track quality criteria except those applied
% during the track reconstruction itself,  2'196'949 tracks were reconstructed by  the \ckf and  2'785'396 tracks by the \costf. 
%%  is given in Table~\ref{tb:trkNbr}.
%%\begin{table}[tb]
%%\caption{\label{tb:trkNbr} %.
%%Number of tracks reconstructed by each algorithm in Period B.}
%%\begin{center}
%%\begin{tabular}{|l||c|c|c|}
%%\hline
%%Algorithm & \ckf & \costf  \\
%%\hline
%%Tracks & 2'196'949 &  2'785'396 \\
%%\hline
%%\end{tabular}
%%\end{center}
%%\end{table}

\begin{figure}[th]
\begin{center}
 \centerline{
 \includegraphics[width=7.5cm]{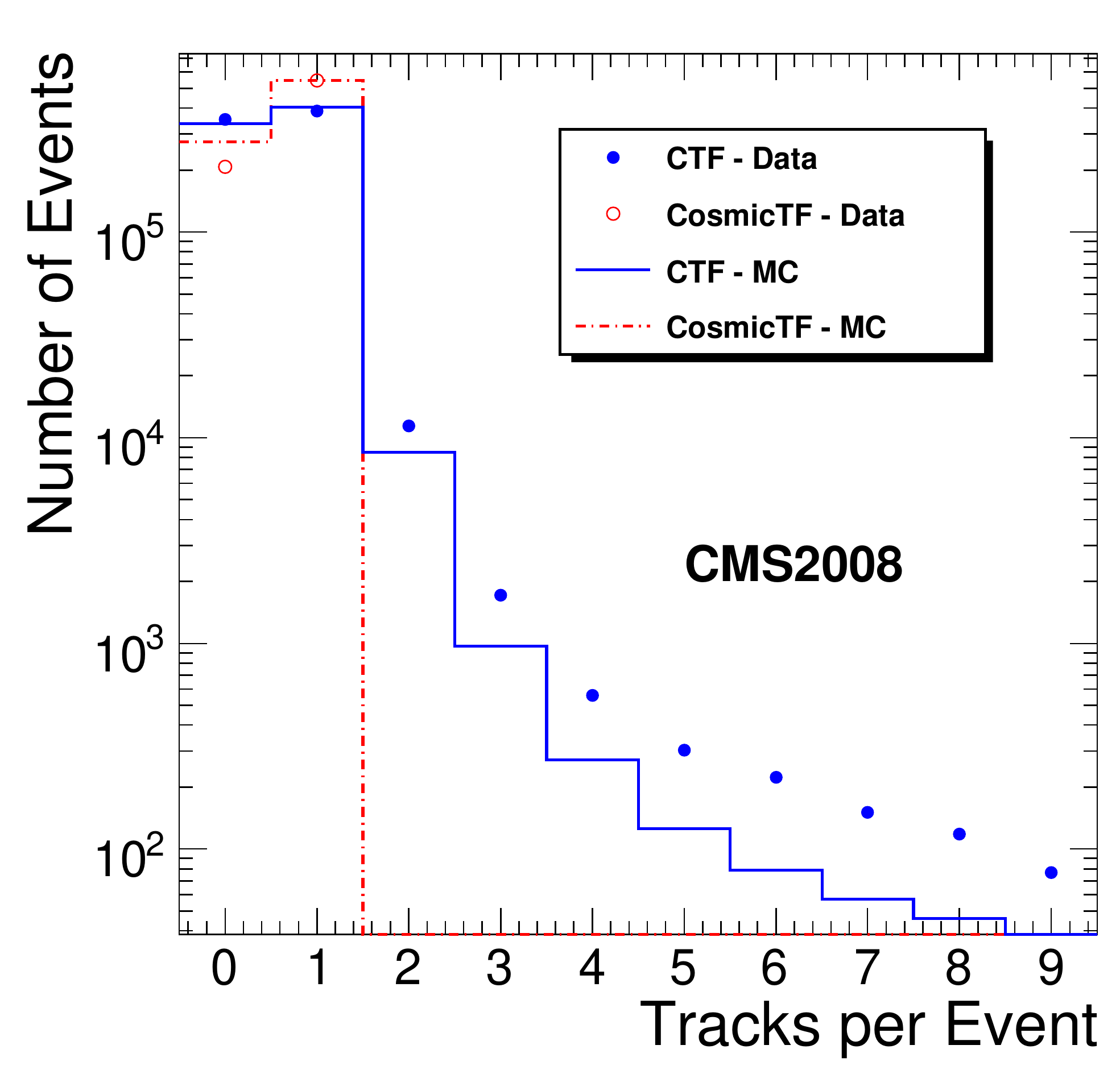}
 }

\caption{ %.
Distribution of the number of tracks reconstructed per event with the two different algorithms.
For each algorithm, the total number of simulated Monte Carlo tracks are normalised to the number of observed tracks. 
}
\label{fig:trkNbr}
\end{center}
\end{figure}

\begin{figure}[th]
\begin{center}
 \centerline{
 \includegraphics[width=5cm]{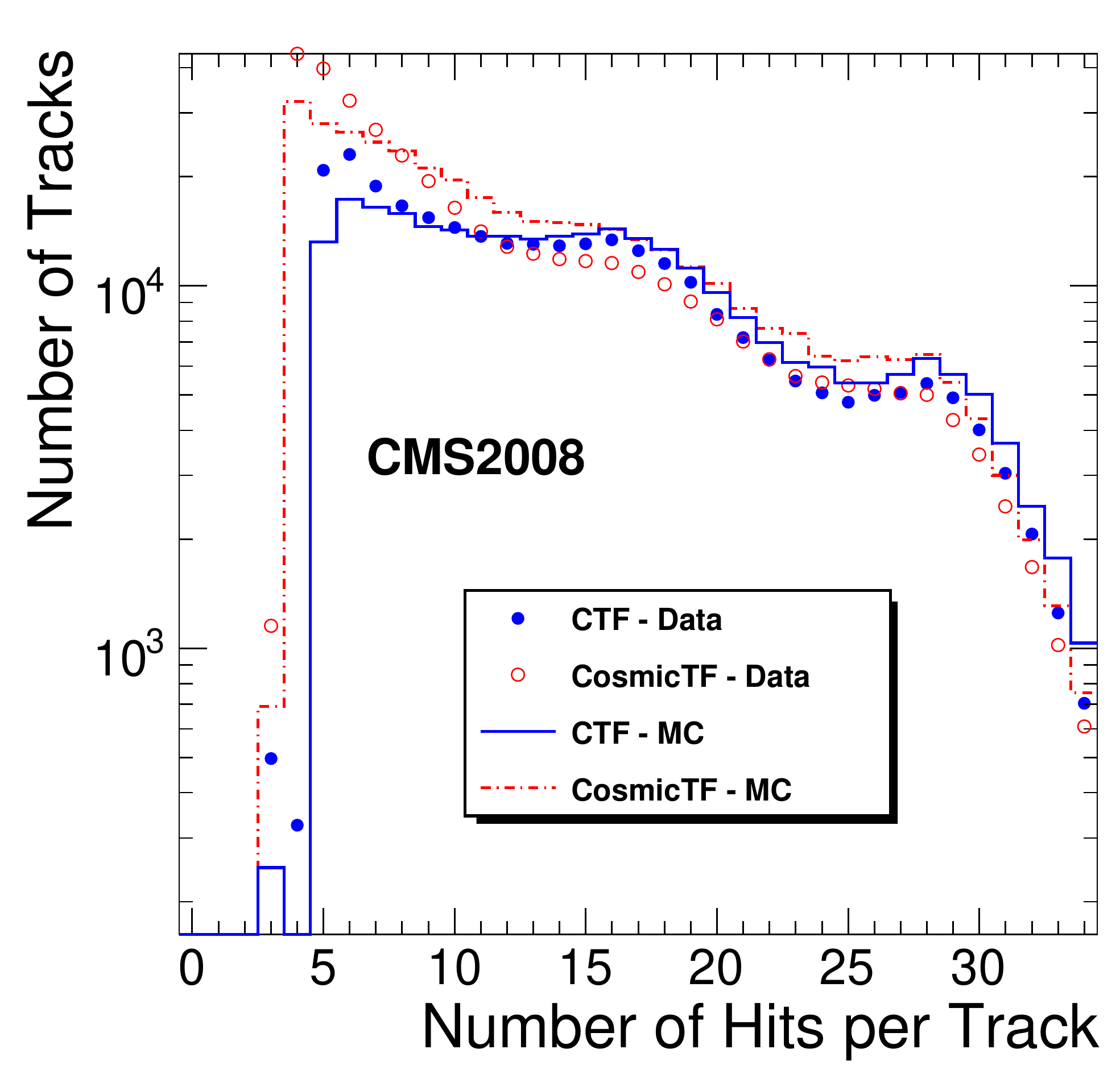}
 \includegraphics[width=5cm]{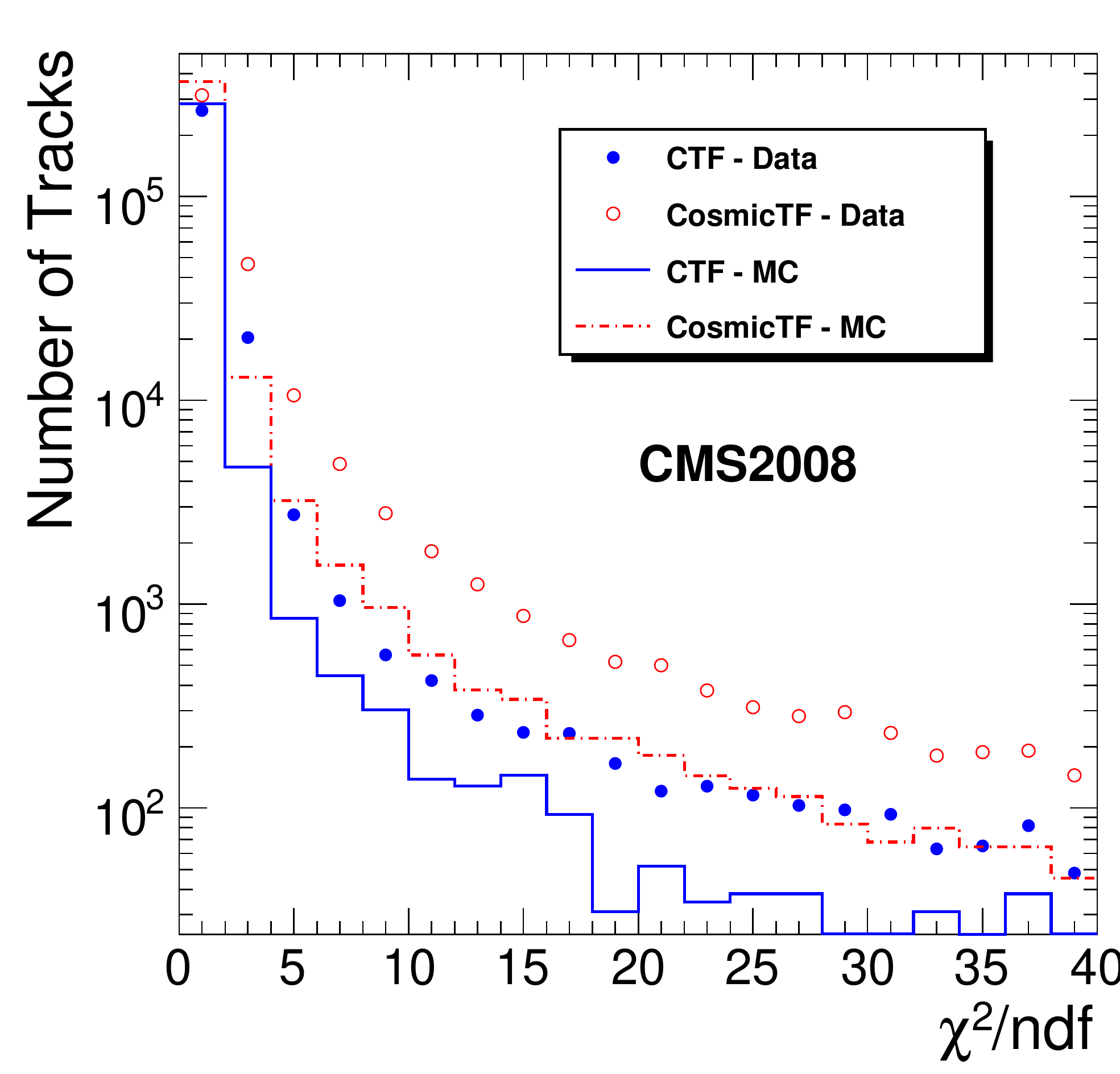}
 \includegraphics[width=5cm]{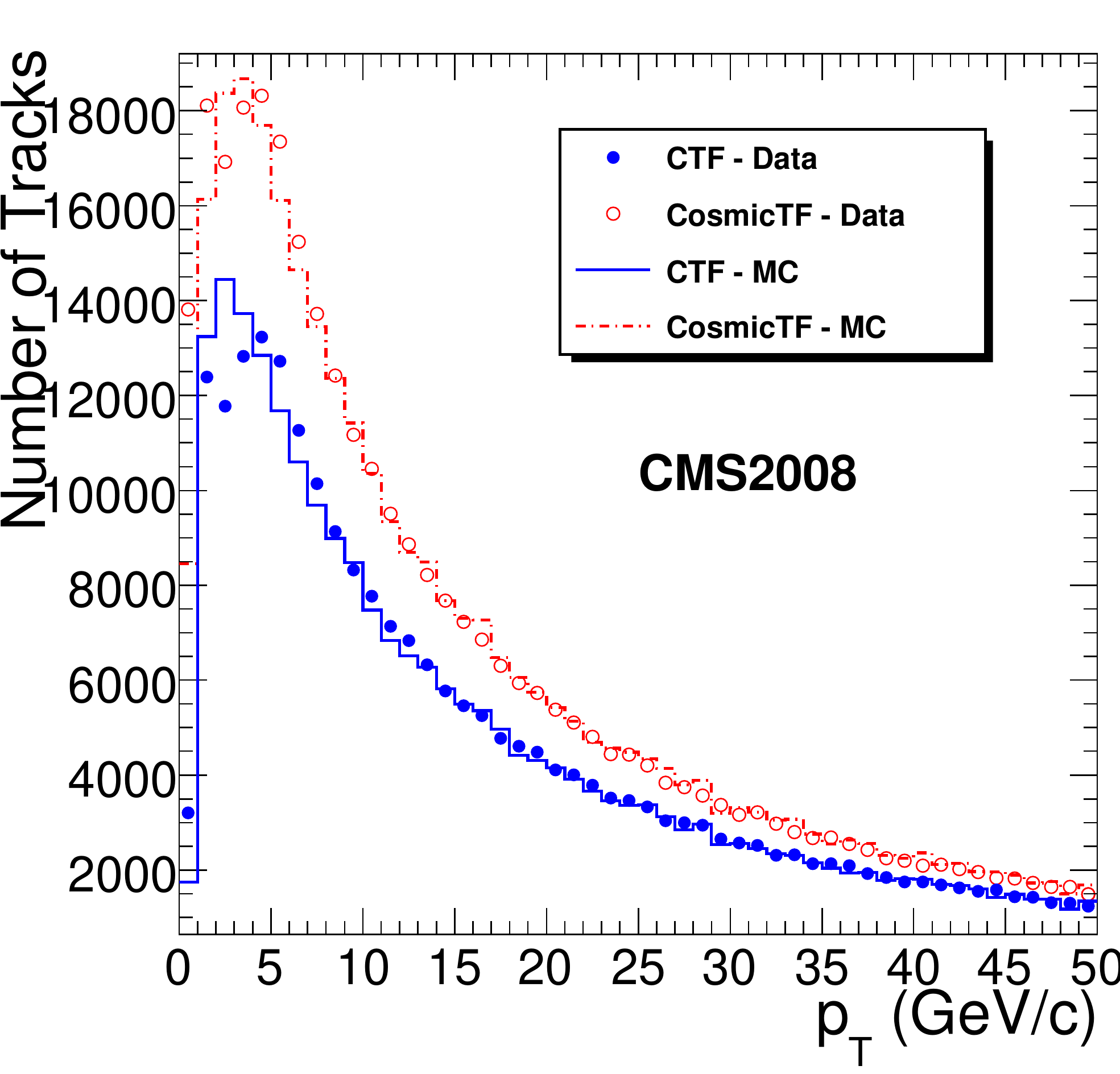}
 }

\caption{ %.
Distributions of several track-related variables for the two different algorithms in  single track events:
the number of hits per track (left), 
\chindf\ (middle) and the transverse momentum (right).
Note that for the \chindf\ distribution, a log-scale is used for the y-axis.
For each algorithm, the total number of simulated Monte Carlo tracks are normalised to the number of observed tracks. 
}
\label{fig:trkValidation}
\end{center}
\end{figure}

The number of reconstructed tracks per event is shown in Fig.~\ref{fig:trkNbr}, and
Fig.~\ref{fig:trkValidation} shows the distributions of a number of track-related quantities 
compared between a subset of the data and Monte Carlo simulation. 
The large number of events without reconstructed tracks is mainly due 
to muons outside of the fiducial volume for which fewer than five hits are reconstructed in the tracker.
% needed to be reconstruced by the \ckf.
% to the acceptance of the tracker for the muons selected with the procedure described in Section~\ref{sec:data-samples}.
% Indeed, in the Monte Carlo simulation, a third of the muons 

It can be seen that reasonable agreement is found between the data and the Monte Carlo simulation, although there
are some discrepancies that require further investigation.
These are thought to be due to the reconstruction of showers by the
track reconstruction algorithms.  
The \ckf is capable of reconstructing more than one track per event, but as
 it has not been optimised to reconstruct showers, multi-track events tend
 to contain a number of fake or badly reconstructed tracks.
These are mostly low momentum tracks with a small number of hits and large
 \chisq values, and the fake rate is estimated to be around 1\%.
For this reason, only single track events are used in the rest of the results presented in this paper, and the distributions shown in Fig.~\ref{fig:trkValidation} are only for single track events.
Small discrepancies remain for tracks with fewer hits and low momentum.
These could be due to detector noise and limitations in the simulation in describing the low
momentum range of cosmic ray muons, such as the position of the concrete plug
covering the shaft.
The simulation assumed that the CMS access shaft was always
closed by a thick concrete plug, while, during the data-taking period, it
was also opened or half-opened.
The absence of the concrete plug causes more low momentum muons to reach
 the tracker~\cite {MUON2}.
The noise is responsible for fake hits added to genuine tracks and, occasionally, fake
 tracks, which contribute to the discrepancies in the \chisq distribution.

% Discrepancies are larger for tracks with fewer hits and low momentum thereforeit could be either some noise in the detector which is not correctly described in the simulation or some problem in the simulation of the low momentum range of cosmics. This causes some fake tracks, and adds fake hits to genuine tracks, thereby increasing their chi2. Furthermore, residual misalignment of the tracker is probably also increasing slightly the chi2 (The MC is with an ideal detector).

%  These tracks cause the discrepancies
% observed between data and Monte Carlo simulations in all the distributions. 

By design the Cosmic Track Finder reconstructs only one track.
The difference between the number of tracks reconstructed
by the two algorithms is mainly due to the minimum number of hits required during the pattern recognition phase.  In the \ckf a minimum of
five hits are required, while only four are required in the case of the \costf.  It can be seen that
a small number of tracks have fewer hits than
these minimum requirements.  This is due to the fact that hits deemed to be outliers can still be removed in the track fitting phase.
It can also be seen from Fig.~\ref{fig:trkValidation} that there is a significant number of tracks with a high number of hits, indicating
that tracks can be followed through the whole tracker and be reconstructed with hits in both the upper and lower hemispheres.

\subsection{Track reconstruction efficiency}

The track reconstruction efficiency for the two algorithms described above has been measured using two different methods.
First, the efficiencies were measured by searching for a reconstructed track and matching it to a muon reconstructed only in
the muon chambers.  In the second method, the efficiency was measured using data just from the tracker, by reconstructing
tracks independently in the upper and lower hemispheres of the tracker.
In addition, the likely performance of the \ckf in proton-proton collisions was estimated by running the algorithm with the
appropriate settings and measuring the efficiency by comparing the two
segments of traversing cosmic ray muons, i.e.\  the second method.
%The track reconstruction efficiency of the cosmic ray reconstruction detailed above has been measured with two different methods.
%First, the efficiencies of the reconstruction algorithms have been measured by searching for a reconstructed track and matching it to a muon reconstructed
% in the muon chambers only.  In a second method, the efficiency is measured using only data from the
% tracker, where tracks are reconstructed independently in the upper and lower hemisphere of the tracker.
%Finally, to ensure the performance of the \ckf in future proton-proton collisions, it has been run with similar settings as will be used there,
%and its efficiency has been measured by comparing the two segments of traversing cosmic rays.

\subsubsection{Track reconstruction efficiency using muons reconstructed by the muon chambers}
\label{sec:tkEffMu}

In the first method, the track reconstruction efficiency is measured with respect to muons reconstructed using 
information from the muon chambers, and required to point within the geometrical acceptance of the tracker.  This ensures that the muons have been identified independently of the tracker.
The muons are first reconstructed by the muon chambers, combining segments of muon tracks reconstructed in the top and bottom hemispheres of the muon detectors
in a global fit.  These reference muons are required to have at least 52 hits in the muon chambers, which corresponds to having hits in at least five Drift Tube chambers.
Combining segments from the two hemispheres removes muons which are
absorbed by the CMS steel yoke before reaching the tracker.
It also improves the track direction reconstruction, which is needed for the propagation through the detector.  

The efficiency is estimated with respect to reference muons with a topology similar to that expected in proton-proton collisions.
This is achieved by requiring that the point of closest approach of the extrapolated muon to the centre of the detector is less than 30~cm in both
the transverse and longitudinal directions.  The absolute value of the pseudorapidity, $|\eta|$, is required to be less than $1$ and the azimuthal angle is required to
be in the range $0.5<|\phi|<2.5$, effectively restricting the tracks to the barrel.
These cuts also ensure that the tracks cross most of the layers of the tracker and cross most modules
perpendicularly.  
The efficiency is then measured by searching for a corresponding track reconstructed in the tracker.

\begin{figure}[tbh]
\begin{center}
 \centerline{
 \includegraphics[width=7.5cm]{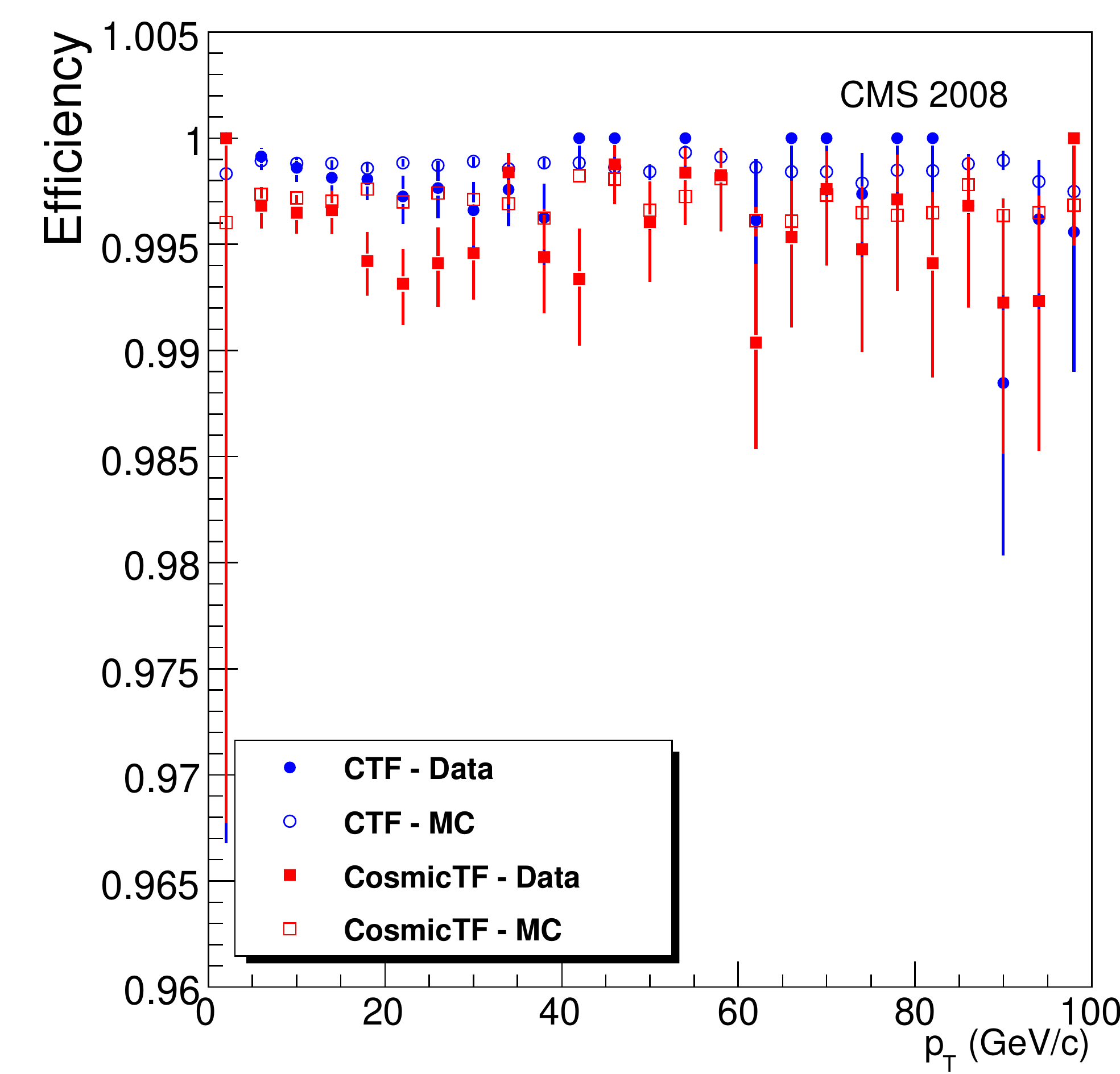}
 \includegraphics[width=7.5cm]{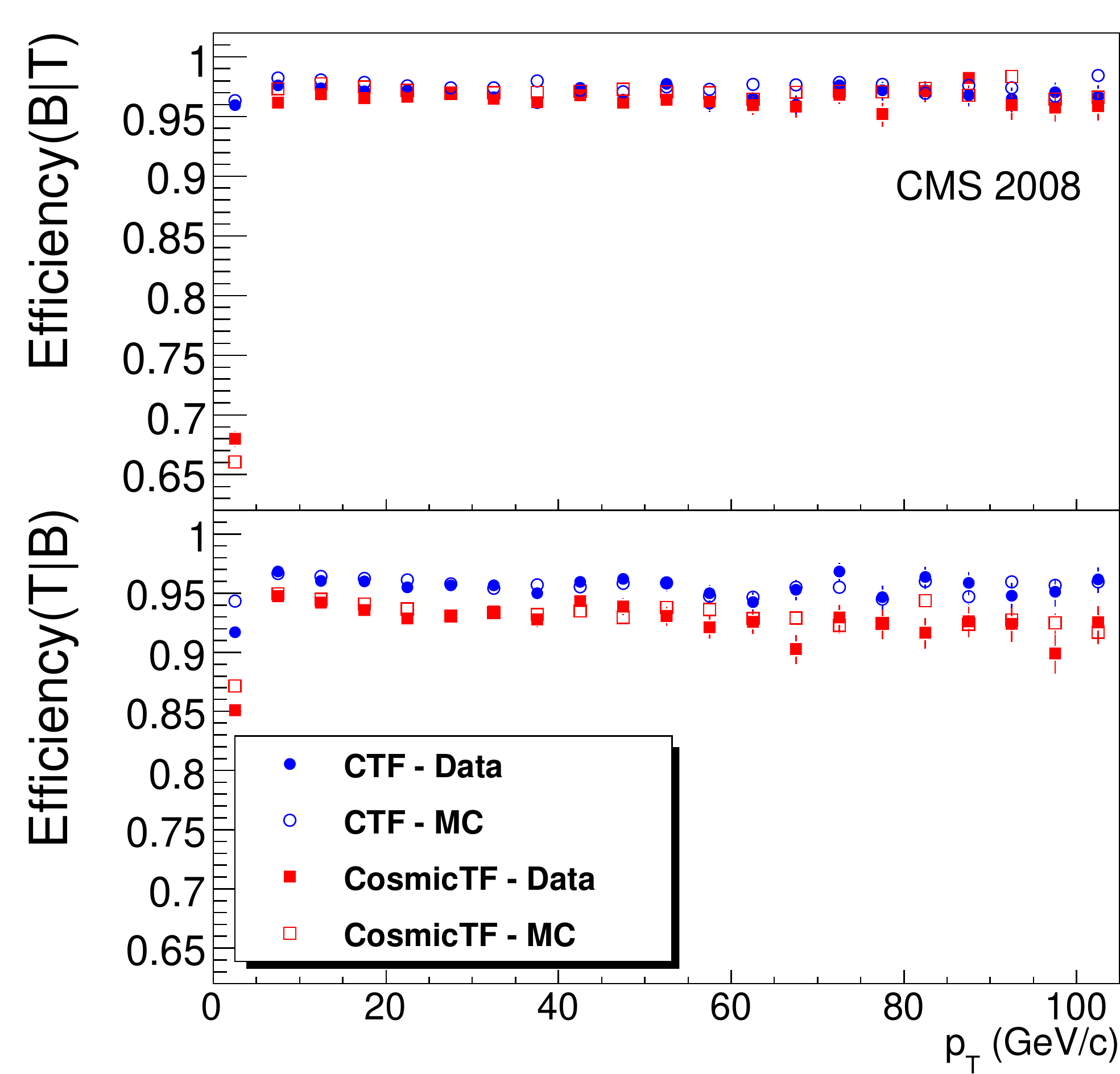}
 }
\caption{ %.
Track reconstruction efficiency as a function of the measured transverse
momentum of the reference track, as measured with the track-muon matching
method (left) and the Top/Bottom comparison method (right). 
\label{fig:trkEff}
}
\end{center}
\end{figure}

The efficiencies measured in the data and in the Monte Carlo simulation are compared in Fig.~\ref{fig:trkEff}~(left) and
summarised in Table~\ref{tab:stdEffLHC}.
The efficiencies are higher than 99\% for both data and Monte Carlo
 simulation and for the two tracking algorithms.  
The difference between data and Monte Carlo observed  around $20\,\gevc$
for the \costf, while statistically significant, is
 small and has not been pursued further, since this algorithm will not be used in
 proton-proton collisions. 
The overall differences between data and Monte Carlo simulation are found to be smaller
 than 0.5\%.  

%The track reconstruction efficiency is most likely over-estimated as no quality cuts
%have been applied to the tracks to remove fakes.

%The track reconstruction efficiencies are then estimated with respect to reference muons with a topology similar to that expected in proton-proton collisions.
%The point of closest approach of the extrapolated muon to the centre of the detector is thus required to be less than 30~cm in both the transverse and
% longitudinal directions, its pseudorapidity $|\eta|$ is required to be below 1, and its polar angle is required to be in the range  $0.5<|\phi|<2.5$.
%This ensures that tracks cross most of the layers of the tracker and cross most modules perpendicularly. %\fixme{nbr of tracks?}
%The efficiencies measured in the data and compared to those obtained from the Monte Carlo simulation are shown in Figure~\ref{fig:trkEff}~(left) and
% summarised in Table~\ref{tab:stdEffLHC}.  The efficiencies are higher than 99\% for both data and Monte Carlo simulation and
% for the two tracking algorithms, and the differences between data and Monte Carlo simulation are found to be smaller than 1\%.
%The track reconstruction efficiency is, in that case, probably over-estimated since no quality cuts on tracks are applied to remove fakes.

\begin{table} [th]
    \caption{Track reconstruction efficiencies for the two algorithms in Data and in Monte Carlo simulation, measured with the muon-matching method.}
    \label{tab:stdEffLHC}
  \begin{center}
\begin{tabular}{|l|c|c|c|c|} \hline
	& \multicolumn{2}{c|}{CTF}
		& \multicolumn{2}{c|}{CosmicTF}
	\\ \cline{2-5}
		 & Data & MC & Data & MC  \\ \hline
%Efficiency [\%] & 99.776 $\pm$ 0.023 &99.8752 $\pm$ 0.0047 & 99.474 $\pm$ 0.035&    99.7195 $\pm$ 0.0071    \\
Efficiency (\%) & 99.78 $\pm$ 0.02 &99.88 $\pm$ 0.01 & 99.47 $\pm$ 0.04&    99.72 $\pm$ 0.01    \\
      \hline
    \end{tabular}
  \end{center}
\end{table}

\subsubsection{Track reconstruction efficiency using tracker data only}

In the second method, the efficiency is measured using data from the tracker; no muon chamber information is included.
%This removes any possible bias arising from the reconstruction and the selection of the muons or from matching muon chamber and
%tracker tracks.  
This method has been used in previous cosmic ray muon data-taking exercises, when the efficiency was evaluated using
track segments reconstructed separately in the TIB and TOB~\cite{tifPaper}.
As cosmic ray muons pass through the tracker from top to bottom, the tracker was divided into two hemispheres along the $y=0$ horizontal plane for this
study.  The tracks were reconstructed independently in the two hemispheres.  Tracks reconstructed in the upper hemisphere are referred to as 
{\em top tracks} and those reconstructed in the lower hemisphere as {\em bottom tracks}.
Tracks in one hemisphere are used as references to measure the efficiency in the other hemisphere. Two such measurements are performed:  $\epsilon (T|B)$, where, given a bottom track, a matching top track
is sought and vice versa ($\epsilon (B|T)$).  The matching is performed by requiring that the two opposite-half tracks have pseudorapidities that
satisfy $\vert \Delta\eta\vert<0.5$.

Only events containing a single track with a topology similar to that expected in proton-proton collisions are analysed and
the same track requirements that were applied in Section~\ref{sec:tkEffMu} are used.  To reconstruct the two track legs independently, only
seeds with hits in the top or bottom hemisphere are selected and, before the final track fit, the hits in the other hemisphere are removed from
the track.  
%The procedure was applied to both of the reconstruction algorithms.  
After track segment reconstruction, a track is only retained for
further analysis if it contains at least 7 hits and satisfies the requirement $\chindf > 10 $.  Furthermore, to ensure that a matching track can 
be reconstructed, the extrapolation of the reference track into the other hemisphere is required to cross at least five layers.

The efficiencies measured using this method are shown in Fig.~\ref{fig:trkEff}~(right) and Table~\ref{tb:trkEffTB}.
The difference seen for low momentum tracks for the \costf is small, and has not been pursued further.
The lower efficiency for top tracks is primarily caused by a large inactive area in the upper half of TOB layer 4, which would otherwise
be used to build track seeds.  This will not be an issue for the track reconstruction that will be used in proton-proton collisions as
in this case, tracks are seeded principally in the pixel detector with the tracking then proceeding towards the outer layers of the SST.
The efficiencies measured in the Monte Carlo simulation are consistent with those measured in the data to within $1\%$.

%The lower efficiency for top tracks is mainly due to a large inactive area in the upper half of the fourth TOB layer, which is used to build the seeds.
%%% JEC - reference back to hit map in local reco section
% This area can be readily seen in Figure~\ref{fig:hittkmap}. 
%Since the track reconstruction algorithm that will be used during genuine proton-proton collisions is proceeding from the 
% inside of the tracker to the outside, from seeds constructed mainly in the pixel detector, this would not have any effect in collisions (c.f.\ Section \ref{lhc_trk}).
%The efficiencies measured in the Monte Carlo simulation match those measured in the data within $1\%$, once the detector inefficiencies are
%accounted for in the simulation.

\begin{table}[hbt]
\caption{\label{tb:trkEffTB} %.
Overall track reconstruction efficiency measured with the top/bottom comparison method.}
\begin{center}
\begin{tabular}{|l|c|c|c|c|} \hline
	& \multicolumn{2}{c|}{CTF}
		& \multicolumn{2}{c|}{CosmicTF}
	\\ \cline{2-5}
		 & Data & MC & Data & MC  \\ \hline
$\epsilon (B|T)$ (\%)	&97.03$\pm$0.07  &97.56$\pm$0.04 & 94.01$\pm$0.10  &93.41$\pm$0.06 \\
$\epsilon (T|B)$ (\%)	&95.61$\pm$0.08  &95.79$\pm$0.05 & 92.65$\pm$0.11  &93.19$\pm$0.07 \\ \hline
\end{tabular}
\end{center}
\end{table}

\subsubsection{Inside-out tracking method}\label{lhc_trk}

Finally, to evaluate the algorithm that will be used during proton-proton collisions, the efficiency of
the \ckf with the appropriate settings is measured.  The reconstruction process~\cite{ptdr} starts in the centre of the 
tracker and proceeds to the outside, using seeds constructed primarily in the pixel detector.  The default \ckf
is optimised to reconstruct tracks that originate near the interaction
point.  By contrast, very few cosmic ray muons will pass through
this region. In order to take this into account,
only tracks for which the point of closest approach to the centre of the detector is less than 4~cm in the transverse
direction and 25~cm in the longitudinal direction are used, effectively crossing the three barrel layers of the pixel detector.

The tracks are reconstructed from a seed made with hit pairs from any combination of the innermost three layers of the SST; the nominal beam spot
is used as an additional constraint in the transverse plane to provide the initial estimate of the track parameters.  This is a legitimate
approximation as long as the transverse impact parameter of the tracks is much smaller than the radius of the innermost detector layer used.
%The tracks are reconstructed from a seed made with pairs of hits from any combination of the innermost three layers of the silicon strip tracker; the
% nominal beam spot is used as a third constraint in the transverse plane to provide the initial estimate of the track parameters, a legitimate
% approximation as long as the transverse impact parameter of the tracks is much smaller than the radius of the innermost detector layer used.
%% Such seeding algorithm differs from the one that will be used for primary LHC collision tracks only in the choice of the tracker layers
%% used for collecting hits and in the size of the beam spot region; in that case the silicon pixel layers will be used, searching for tracks
%% with a transverse impact parameter smaller than  O(100um).  [in footnote?] When reconstructing proton proton collision this
%% reconstruction will be complemented by other more dedicated seeding algorithms to find tracks produced at larger impact transverse
%% parameter, which are beyond the scope of this analysis.
Hits in the silicon pixel detector are not used in this analysis in the seed finding phase, as this imposes too strong a constraint on the tracks
to come from the nominal beam spot.  
They are, however, identified in the pattern recognition phase and added to the track.

% , as an ``outside-in'' search for extra hits is 
% always performed after the original ``inside-out'' search.  A small bias could, in principle, be introduced by the fact that the algorithm assumes
% that muons lose energy when moving outwards, while with cosmic rays it will actually be the opposite in the top half of the detector.
% In practice, this bias has been verified to be negligible, as the tracks are required to have a transverse momentum above $10\,\gevc$.

%Hits in the silicon pixel detector are not used in the seed finding phase, since this imposes a strong constraint for the tracks to originate from the nominal beam spot.
%However, they are later collected in the pattern recognition phase because an outside-in search for extra hits is always performed after the inside-out search.
%The pattern recognition used after the seeding step is in all respects identical to the one used for collision tracks.
%A small bias could in principle be introduced by the fact the algorithm assumes that muons lose energy when moving outwards, while it is the
% opposite for cosmic muons in the top half of the detector. This bias has been verified to be negligible, since the tracks are required
% to have a transverse momentum above $10\,\gevc$.
%% Anyway in practice such bias is negligible for the muons considered in this
%% study, that are required to have pt > 10 GeV; this was explicitly verified by cross checking the results with those obtained modifying
%% the track reconstruction algorithm to account correctly the energy loss for cosmics.

The reconstruction efficiencies are estimated with respect to a reference track in one hemisphere of the tracker.  A compatible seed and track
is sought in the other hemisphere within a cone of radius $\Delta R < 1.0$ (where $\Delta R = \sqrt {\Delta \eta ^2 + \Delta \phi^2}$) opposite
to the reference track.  The cone size is kept very large compared to the angular resolution so that the matching procedure cannot bias the
efficiency measurements.  To avoid multi-track events, a track is not used as a reference if there is another track in the same hemisphere
or within the matching cone.  Fake tracks created by noisy hits are rejected by requiring that the reference tracks have at least 10 hits.
%The reconstruction efficiencies are then estimated with respect to a reference track, searching for a compatible seed and track in a cone of
% $\Delta R<1.0$ (where $\Delta R = \sqrt {\Delta \eta ^2 + \Delta \phi^2}$) opposite to the reference track.
%The cone is kept very large compared to the angular resolution so that no negative biases on the measured reconstruction efficiency can come from the
% matching procedure. To avoid multi-track events, a track is not used as reference if there is
% another track in the same hemisphere and within the aforementioned matching cone. 
%To reject fake tracks built from noisy hits that are by chance compatible with a helix, reference tracks are required to have at least 10 hits. 
%%\fixme
%%The very few tracks with more than 25 hits have also been excluded, since they  could arise from the rare cases where the full cosmic ray is
%% reconstructed as a single track in some narrow phase space regions. This could introduce biases in the efficiencies which are irrelevant for the
%% collision tracks.

\begin{figure}[tb]
\begin{center}
 \centerline{
 \includegraphics[width=5cm]{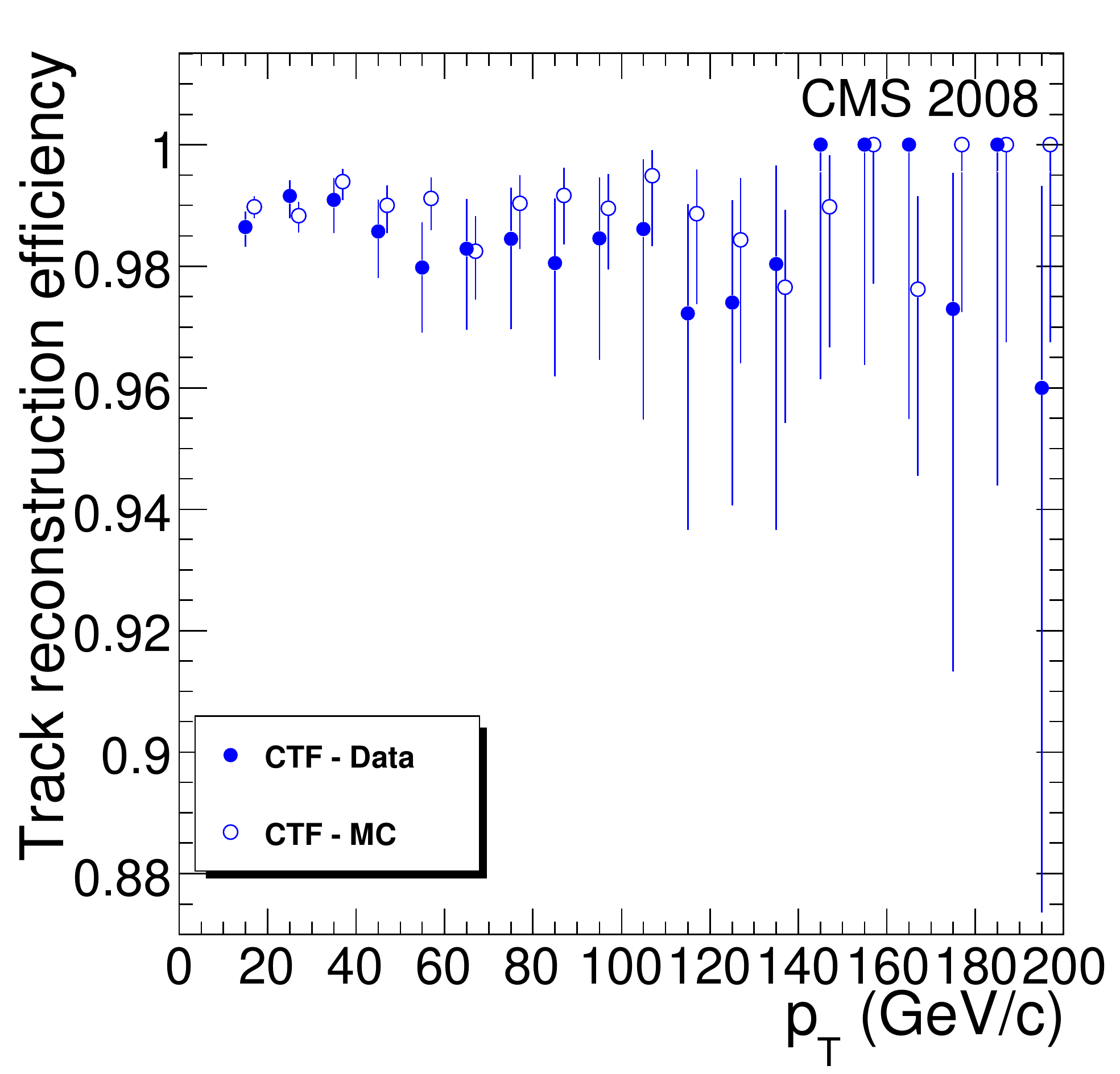}
 \includegraphics[width=5cm]{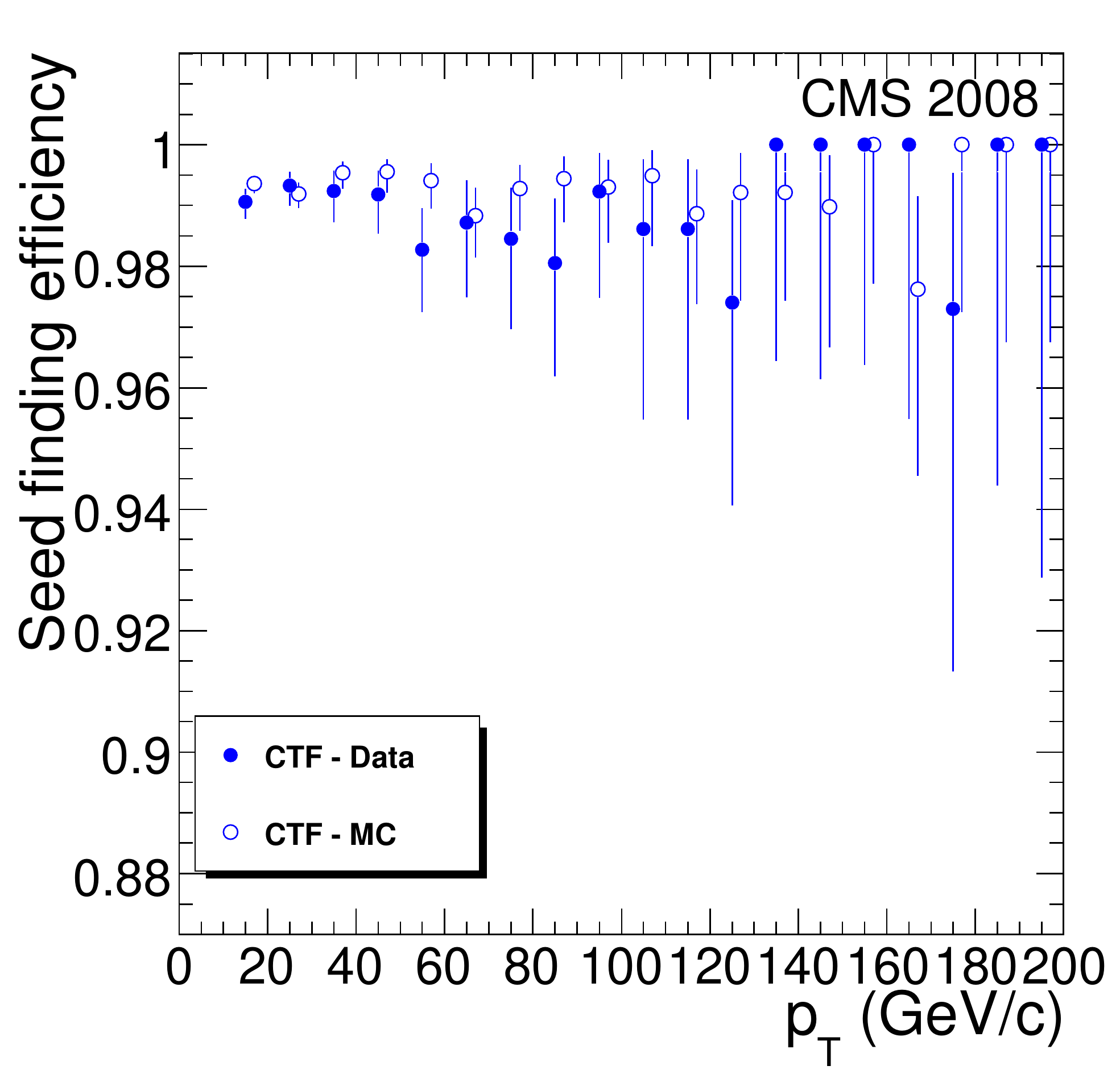}
 \includegraphics[width=5cm]{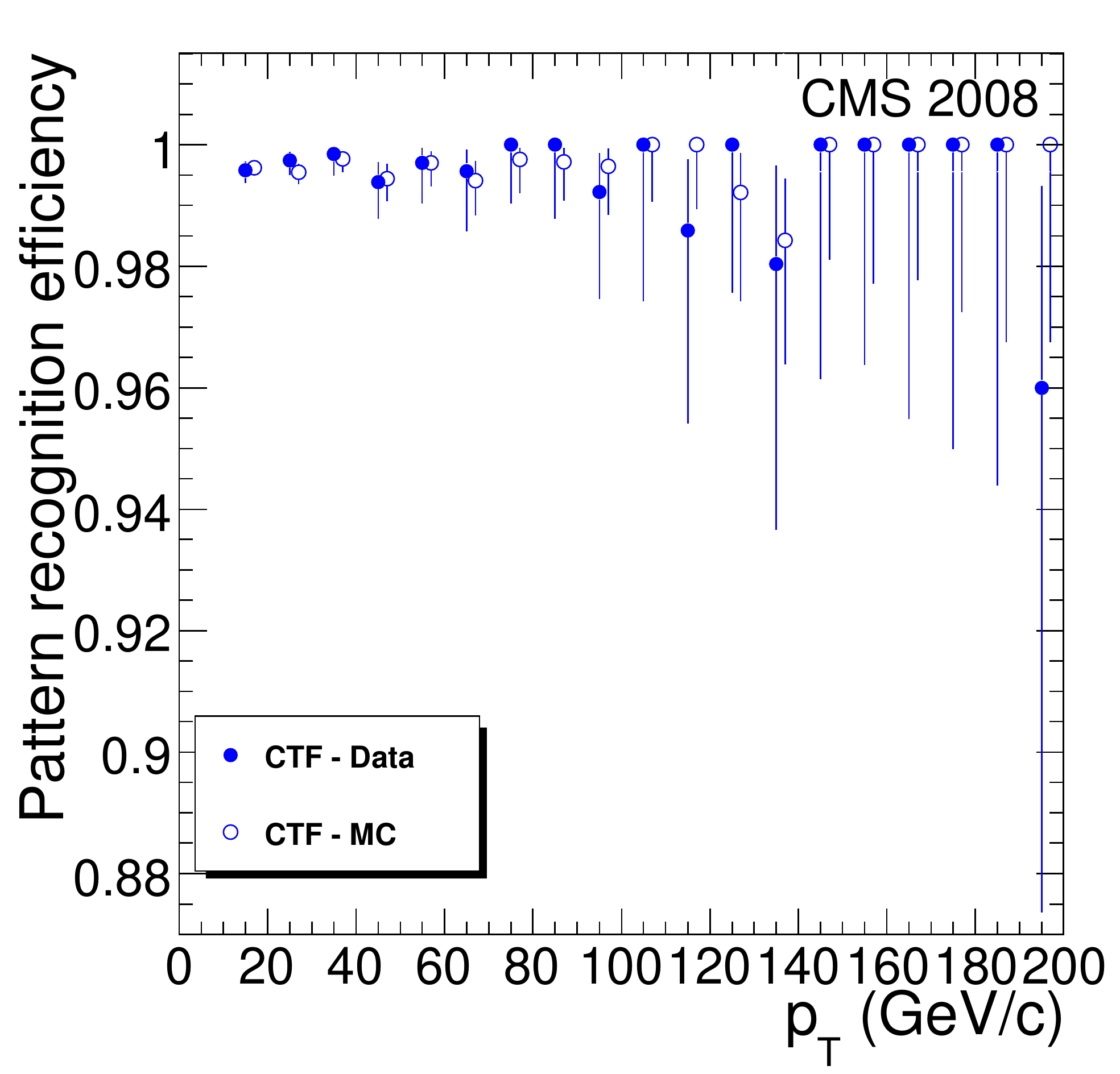}
 }
\caption{ %.
Track reconstruction efficiency (left), seed finding efficiency (middle), and
pattern recognition efficiency (right) as a function of the measured transverse
momentum of the reference track for inside-out tracking method. Note that the Monte Carlo points are shifted by $2\gevc$ so as to allow the uncertainties to be seen.}
\label{fig:trkEffIO}
\end{center}
\end{figure}

The efficiencies measured using this method are shown in Fig.~\ref{fig:trkEffIO} and in Table~\ref{tb:trkEffIO}.
These efficiencies can be further divided into a {\em seed finding} efficiency, which is the efficiency of building a seed for a given reference
track, and a {\em pattern recognition} efficiency, which is the efficiency of reconstructing a track once a seed has been found.
Inefficiencies affecting only a few detector channels have not been
taken into account when calculating the overall efficiency 
from the data.
The efficiencies measured in the Monte Carlo
simulation match those measured in the data to within $1$\%.

\begin{table}[tbh]
\caption{\label{tb:trkEffIO} %.
Reconstruction efficiency of the Inside-out tracking method.
}
\begin{center}
\begin{tabular}{|l|c|c|}
\hline
 & Data & MC \\
\hline
Seed finding efficiency (\%) & 99.17 $\pm$ 0.12 & 99.30 $\pm$ 0.08\\
Pattern recognition efficiency (\%) & 99.79 $\pm$ 0.06 & 99.64 $\pm$ 0.05 \\
Track reconstruction efficiency  (\%) & 98.96 $\pm$ 0.13 & 98.94 $\pm$ 0.09\\
\hline
\end{tabular}
\end{center}
\end{table}

\subsubsection{Summary of the track efficiency measurements}

The three methods of efficiency calculation presented in this section yield consistent results and indicate that a high track reconstruction
efficiency is attained for vertical tracks passing close to the nominal beam line, which is the topology most similar to the tracks
from proton-proton collisions.  Although the results are similar, some small differences were observed.  The main difference between the
efficiencies determined by the first and second methods arises from the fact that tracks are sought in only one half of the detector in the
second method, while in the first method, tracks may be found from seeds produced in both halves of the tracker.

The \ckf algorithm has been fully tested and is well understood, yielding a high quality performance.  The \costf algorithm, while not tuned
to the level of the \ckf, also achieves good performance and provides a fundamental cross-check.  The measurements of the ``Inside-out tracking method''
give confidence that the track reconstruction will perform well in proton-proton collisions.
Finally, the efficiencies measured in the Monte Carlo simulation agree very well with those measured in the data once the known detector inefficiencies
are accounted for in the simulation.  This indicates that the tracker and the reconstruction algorithms are well understood.

\subsection{Track parameter resolution}

The track reconstruction can be further validated using the CRAFT data sample by splitting the tracks into two separate parts.
A measure of the resolution of the track parameters can be determined by comparing the two legs of the split tracks.  To perform
this study, tracks are split at the point of closest approach to the nominal beam-line.  The top and bottom legs are treated as
two independent tracks and re-fitted accordingly.  The track parameters are then propagated to their respective points of closest
approach to the beam-line.  This method has been tested using Monte Carlo simulation and found to work well.
%A further validation of track reconstruction  cosmic ray data is done by splitting the tracks.
%By comparing the two legs of the split tracks, a measure of the resolution of the track parameters can be determined.
%For this purpose, a track is split at its point of closest approach to the nominal beam-line.
%The top and bottom legs are taken as two independent tracks and re-fit accordingly, and the track parameters propagated to their respective points
% of closest approach to the beam-line. A validation of the method itself is made with respect to simulation with good agreement.

For the purposes of this study, only events in which the \ckf reconstructed a single track whose point of closest approach to the beam-line is inside
the volume of the pixel barrel are considered.  The transverse
momentum of the track must be greater than $4\,\gevc$ and its \chisq
must satisfy the requirement $\chindf < 100$.  
In addition, the track must contain a minimum of 10 hits, with at least two hits being on double-sided strip modules.
There must also be six hits in the pixel barrel subsystem.  After splitting, each track segment is required to have at least six hits, three of which must
be in the pixel barrel.
%Only events in which the \ckf reconstructed a single track whose point of closest approach to the beam-line is inside the volume of the pixel barrel are used here.
%The transverse momentum of the track is required to be above $4\,\gevc$ and its normalised \chisq smaller than 100.
%In addition, the track is required to contain at least 10 hits, with at least two hits being on double-sided strip modules, and six hits in the pixel
% barrel sub-detector. After the splitting, each track segment is required to have at least six hits, three of which have to be in the pixel barrel sub-detector.

\begin{figure}[ht]
\begin{center}
 \centerline{
 \includegraphics[width=0.48\linewidth]{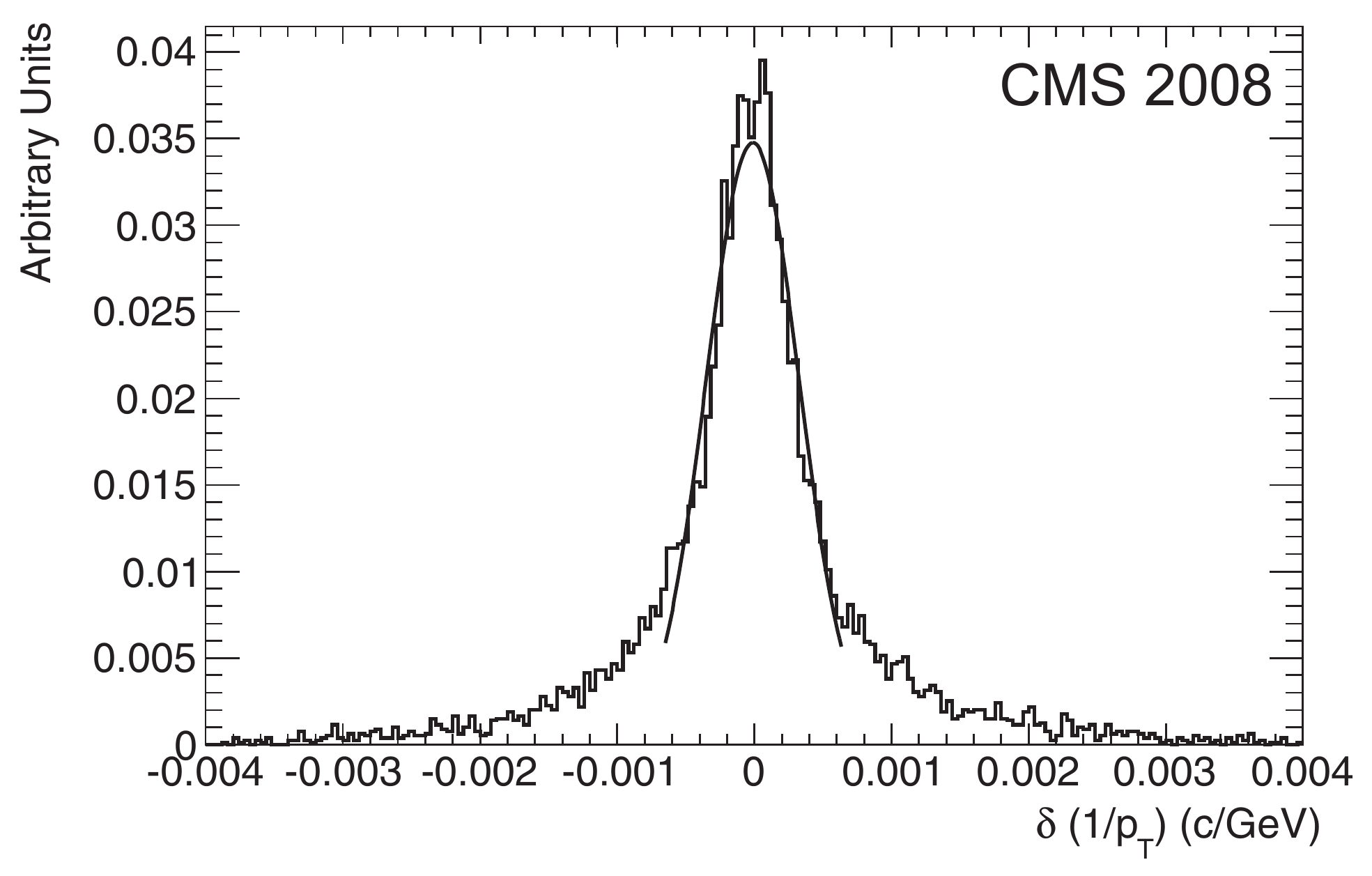}
 \includegraphics[width=0.48\linewidth]{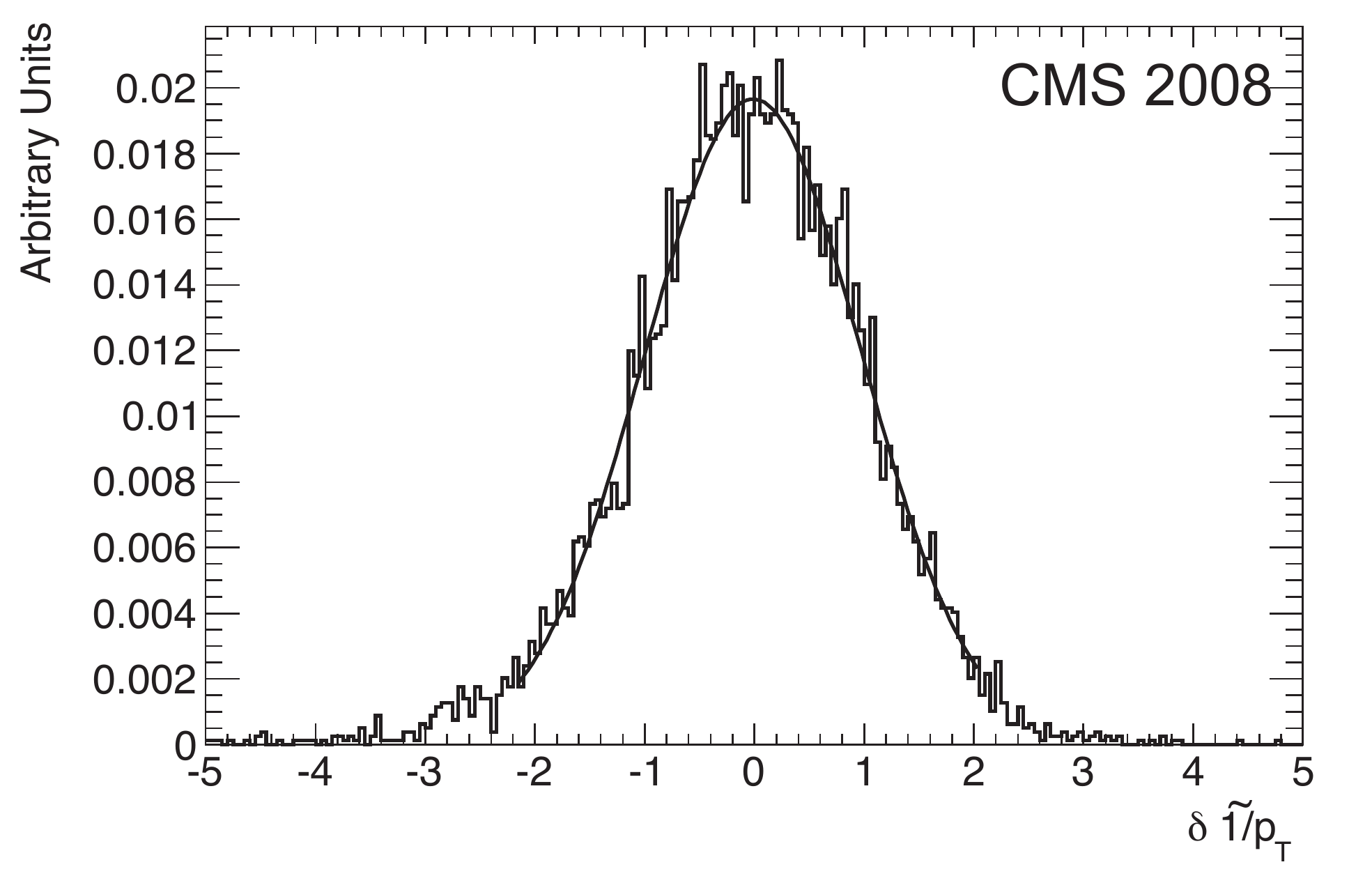}
 }
 \centerline{
 \includegraphics[width=0.48\linewidth]{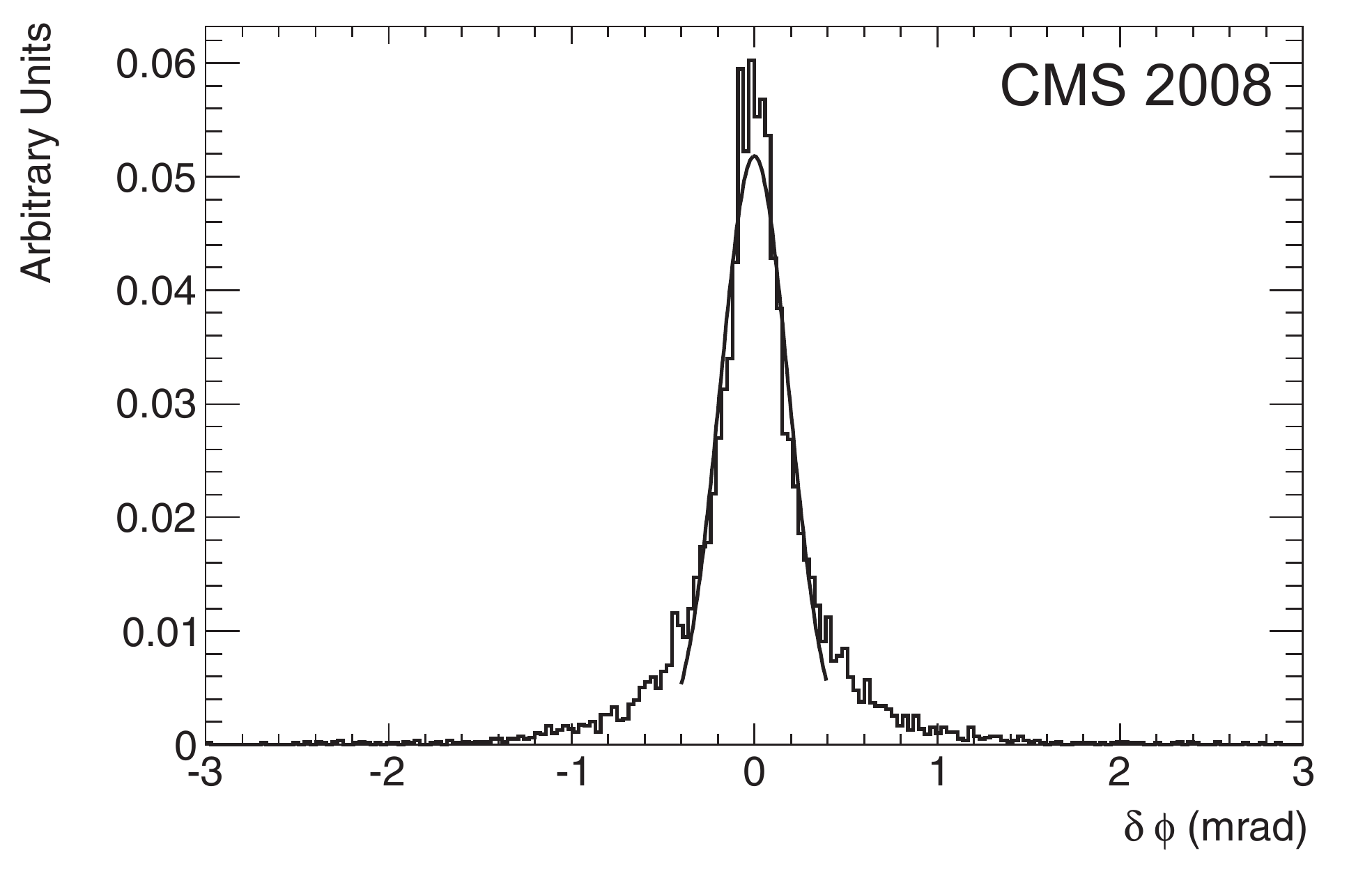}
 \includegraphics[width=0.48\linewidth]{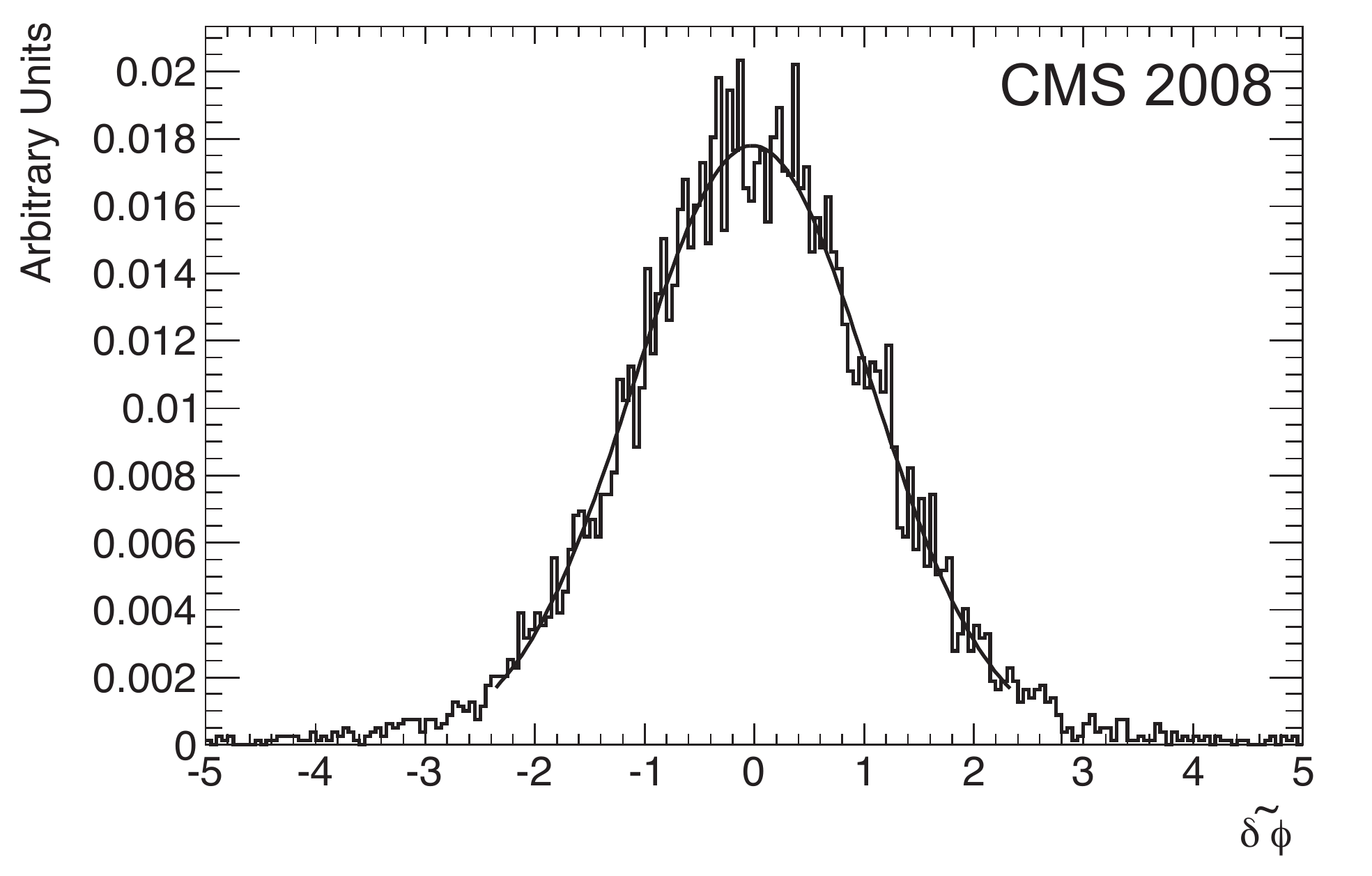}
 }
 \centerline{
 \includegraphics[width=0.48\linewidth]{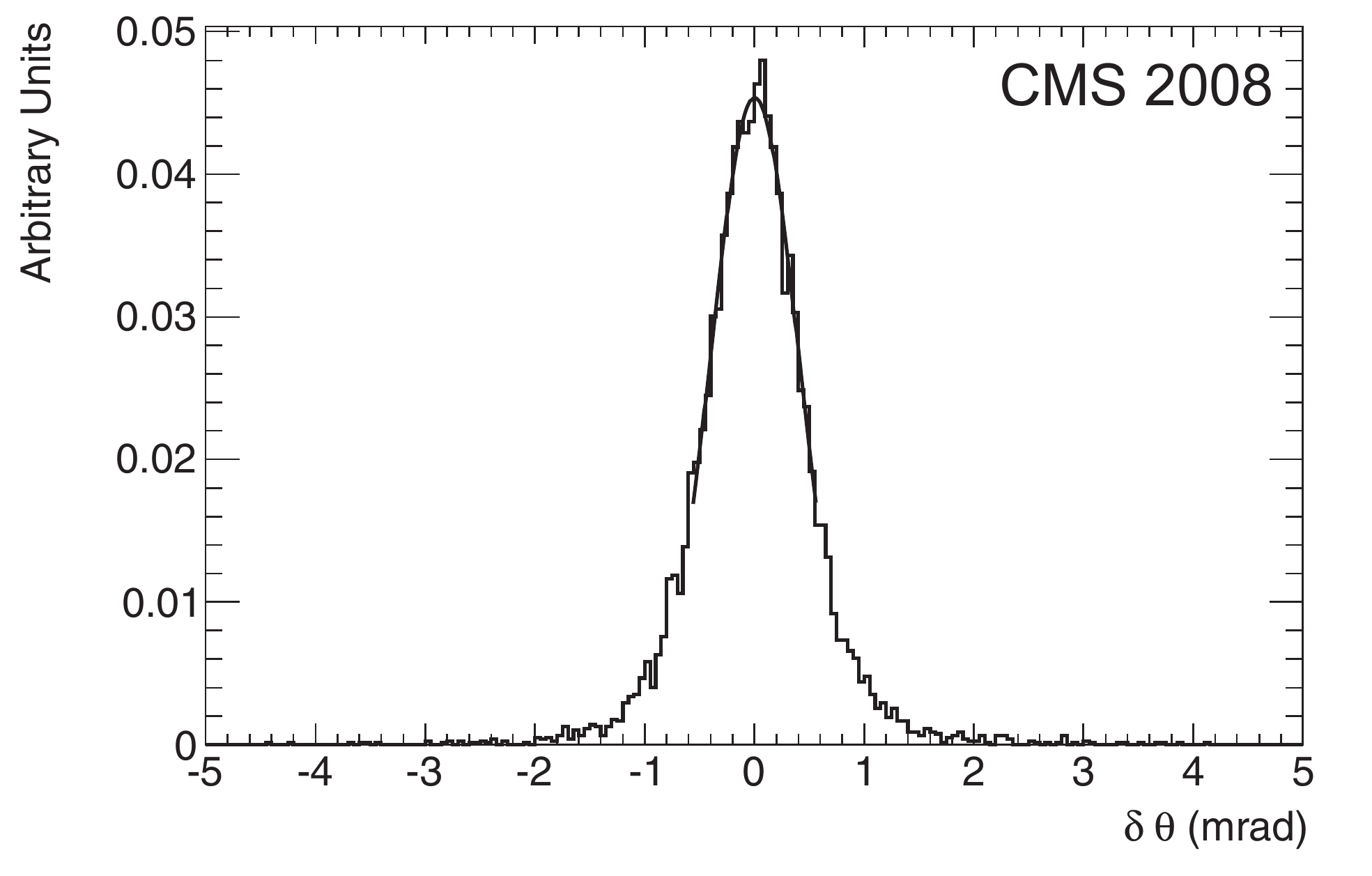}
 \includegraphics[width=0.48\linewidth]{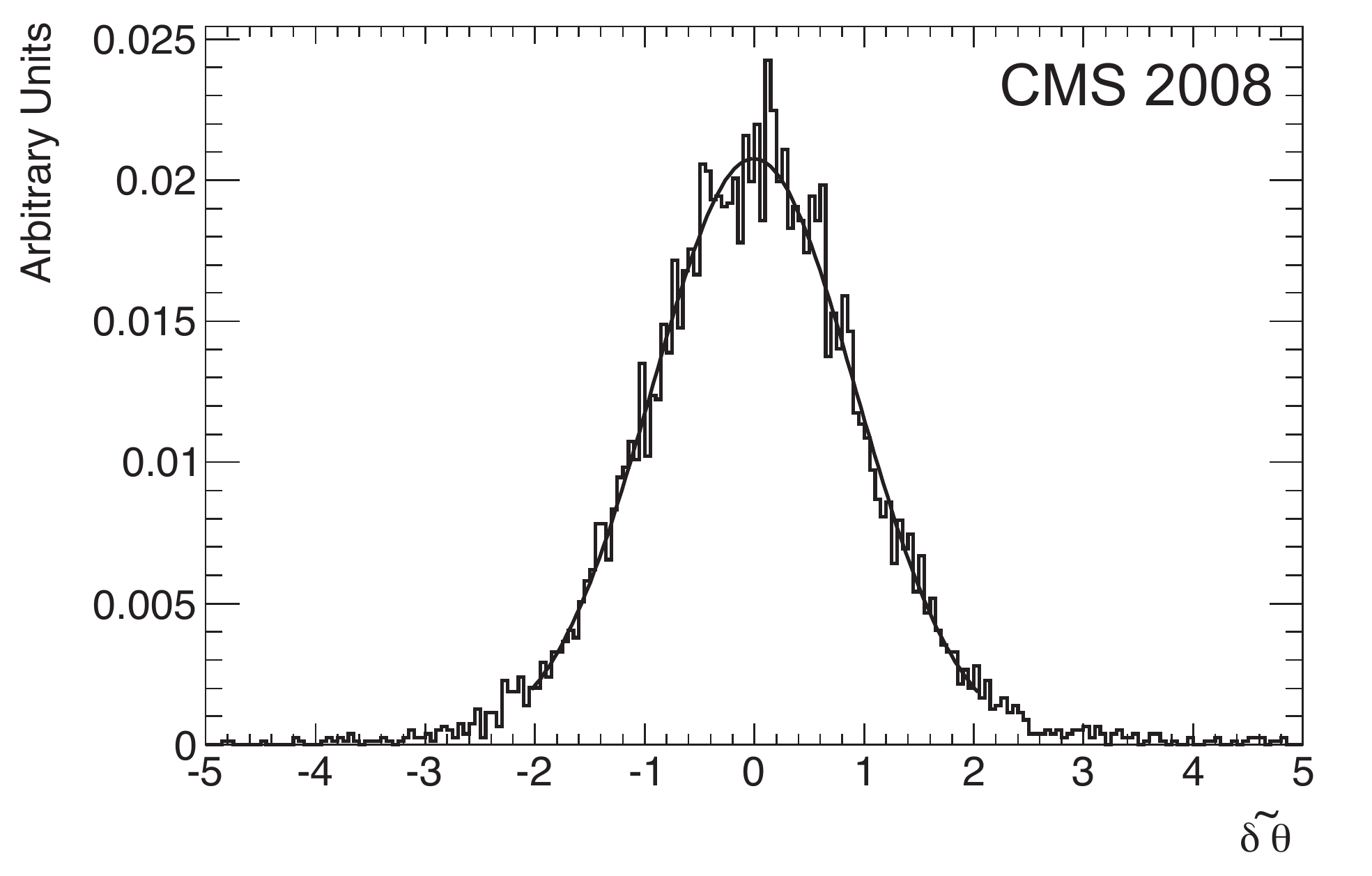}
 }
\caption{ %.
Residual distribution (left) and pull distribution (right) of the inverse
transverse momentum $1/\pt$ (top), azimuthal  $\phi$ (middle),  and polar 
$\theta$ angle (bottom).
}
\label{fig:trkParam}
\end{center}
\end{figure}

The results of this analysis are summarised in Table~\ref{tb:trkParam}, while the distributions of the residuals and pulls of the inverse transverse momentum
and the azimuthal ($\phi$) and polar ($\theta$) angles are shown in Fig.~\ref{fig:trkParam}.  The corresponding distributions for the transverse ($d_{xy}$) and longitudinal ($d_{z}$)  impact parameters
can be found elsewhere~\cite{craftPixel}.  For each track parameter, the residuals are defined as $\delta x = (x_1 - x_2)/\sqrt{2}$.  The factor of
 $\sqrt{2}$ is needed to account for the fact that the two legs are statistically independent.  The standardised residuals (or pulls) are defined by
${\widetilde{\delta x}} = (x_1 - x_2)/\sqrt{\sigma_{x1}^2 + \sigma_{x2}^2}$.  In Table~\ref{tb:trkParam} the mean and standard deviation (referred to as
the {\em resolution}) of a Gaussian fitted to the peak of the distributions are given.  In order to get an estimate of the tails of the distributions,
the half-widths of the symmetric intervals covering $95$\% of the distribution (also known as the {\em 95\% coverage}),
which, in the case of a Gaussian distribution, correspond to twice the standard deviation, are also given in Table~\ref{tb:trkParam}.

The same quantities are used to characterise the pull distributions.  In this case, the standard deviations of the fitted Gaussians are taken as the pull values.
It can be seen that the resolution of the angles and the impact parameters are well described by a Gaussian.
The resolution as a function of the momentum has been presented elsewhere~\cite{craftAlign}.

\begin{table}[bt]
\caption{\label{tb:trkParam} %.
Standard deviation, mean, and 95\% coverage of the residual and pull
distributions of the track parameters.
The units indicated pertain only to the residual distributions.
}
\begin{center}
\begin{tabular}{|l|c|c|c|c|c|c|} \hline
 Track parameter   & \multicolumn{3}{c|}{Residual distributions}
        & \multicolumn{3}{c|}{Pull distributions}\\ \cline{2-7}
     &Std. Dev.    & Mean    & 95\% Cov.    & Std. Dev.    & Mean    & 95\% Cov.\\
 \hline
$p_T$  (\gevc) 	& 0.083	& 0.000	& 1.92	& 0.99	& 0.01 & 2.1\\
Inverse  $p_T$  ($\gev^{-1}c$) 
					& 0.00035& 0.00003	& 0.00213& 0.99	&-0.01  & 2.1\\
$\phi$  	(mrad)		& 0.19	& 0.001	& 0.87	& 1.08	&-0.02	& 2.4\\
 $\theta$  (mrad)	& 0.40 	& 0.003	& 1.11	& 0.93	&-0.01	& 2.1\\
$d_{xy}$  ($\mu$m)	& 22	& 0.30	& 61	& 1.22	& 0.00	& 2.9\\
$d_{z}$ ($\mu$m)	& 39	& 0.28	& 94	& 0.94	&-0.01	& 2.1\\
\hline
\end{tabular}
\end{center}
\end{table}

\subsection{Hit resolution}

The hit resolution has been studied by measuring the track residuals, which are defined as the difference between the hit position and the track position.  The track is deliberately reconstructed excluding the
hit under study in order to avoid bias.  The uncertainty relating to the track position is much larger than the inherent hit resolution, so a single track residual is not sensitive to the resolution.
However, the track position difference between two nearby modules can be
measured with much greater precision.
A technique using tracks passing through overlapping modules from the same tracker layer is
employed to compare the difference in residual values for the two measurements in the overlapping modules~\cite{tifPaper}.  The difference in hit positions, $\Delta x_{hit}$, is compared to the difference in
the predicted positions, $\Delta x_{pred}$, and the width of the resulting distribution arises from the hit resolution and the uncertainty from the tracking predictions.  The hit resolution can therefore
be determined by subtracting the uncertainty from the tracking prediction.  This overlap technique also serves to reduce the uncertainty arising from multiple scattering, by limiting the track extrapolation to
short distances.  Any uncertainty from translational misalignment between the modules is also avoided by fitting a Gaussian to the distribution of the differences between the residuals.
%The hit resolution has been studied by measuring the track residuals defined as the difference between the hit position and the track position, where the track is reconstructed without 
%the hit under study to avoid bias. Nominally, the uncertainty from the track position is much larger than the inherent resolution, so a single track residual is not sensitive to the resolution. 
%The track position difference between two nearby modules, however, can be measured with greater accuracy. A technique using tracks passing through overlapping modules from the same tracker layer 
%is employed to compare the difference in residual values for the two measurements in the overlapping modules, as described in Ref.~\cite{tifPaper}. 
%The difference in hit positions ($\Delta x_{hit}$ in Ref.~\cite{tifPaper}) is compared to the difference in predicted positions ($\Delta x_{pred}$) and the width of the resulting distribution is 
%due to the resolution of hits and the uncertainty from the tracking predictions. Subtracting the uncertainty from the tracking prediction provides a measurement of the hit resolution.
%The overlap technique also serves to reduce the uncertainty due to multiple scattering by requiring the track extrapolation  over only a short distance.
%Any uncertainty from translational misalignment between the modules is also avoided by fitting a Gaussian to the distribution of the differences between the residuals.

For the purposes of this study, only events in which the \ckf reconstructed a single track are used, and only overlaps from barrel modules for which the residual rotational misalignment
is less than $5\mum$ are analysed.  The \chisq probability of the track is required to exceed $0.1$\% and the tracks must be reconstructed with at least 6 hits.  In addition, the track momenta are
required to be greater than $20\, \gevc$, ensuring that the uncertainty arising from multiple scattering is reduced to less than $3\mum$.  Remaining uncertainties from multiple
scattering and rotational misalignment between the overlapping modules are included as systematic uncertainties in the measurement.
%Only events in which the \ckf reconstructed a single track are used, and overlaps from the barrel modules for which the residual rotational
% misalignment is less than $5\mum$ are analysed. The $\chi^2$ probability of the track is required to be above 0.1\%,
% and the tracks are required to be reconstructed with a minimum of 6 hits. In addition, by requiring that the momentum of the track be above $20\, \gevc$, the
% uncertainty due to multiple scattering is reduced to less than $3\mum$. Remaining uncertainties from multiple scattering and rotational
% misalignment between the overlap modules are included as systematic uncertainties in the measurement.

The distribution of the differences between the residuals is fitted, with the width containing contributions from the hit resolutions and the uncertainty from the tracking predictions.
The latter is subtracted out in quadrature to leave the resolution on the difference of the hit positions between the two modules.  As the two overlapping modules are expected to have the 
same resolution, the resolution of a single sensor is determined by dividing by $\sqrt{2}$.
%The fitted width of the distribution of the differences between the residuals includes contributions from the inherent hit resolution and the
% difference of the predicted position of the track once the hits in the modules under study are removed ($\Delta x_{pred}$).
%The latter uncertainty is subtracted out in quadrature to calculate the resolution on the difference of the hit positions between the two modules.
%As the two overlapping modules should have the same resolution, the resolution of the position difference has equal contributions from each module, and the
% resolution of the single sensor is the position difference resolution divided by $\sqrt{2}$.

The sensor resolution is known to depend strongly on the angle of the track and the pitch of the sensor.  The results are therefore determined separately for different sensor pitches and
in 10 degree intervals for the track incidence angle.
%As expected, the sensor resolution depends strongly on the angle of the track and the pitch of the sensor. 
%The sample is thus divided into sub-samples based on the pitch of the sensors and on the angle of the tracks measured from normal incidence, in 10 degree increments. 
%% After the cuts, sufficient statistics remained to divide the sample into bins based on the different pitch values of the sensors and on track angle
%%  measured from normal in 10 degree increments. 
The results are shown in Table~\ref{tb:trkHitRes}, where they are compared to the predictions from Monte Carlo simulation.  The agreement between the data and the predictions
is very good for normally incident tracks, but suggests that the simulation may underestimate the resolution for larger track angles, as can be seen in the first two layers of
TIB.  
The resolutions vary from $20$ to $56\mum$ for the position difference,
 which corresponds to a variation between $14$ and $40\mum$ in the single
 sensor resolution.  
%The results are shown in Table~\ref{tb:trkHitRes}, and are compared to the values predicted by the model used in the simulation of the detector and used in the track reconstruction. 
%The comparison is very good for normally incident tracks, and suggests the model may underestimate the resolution for larger track angles, as seen in the first two TIB layers. 
%In total, the resolutions range from 20 to $46\mum$ for the position difference, or from 14 to $40\mum$ for the single sensor resolution with uncertainties less than 1 micron.

\begin{table}[bt]
\caption{\label{tb:trkHitRes} %.
Hit resolution measured on CRAFT data and predicted by the model in the Monte Carlo simulation, for the different local track angles. All values are in microns.}
\begin{center}
\begin{tabular}{|c|c|c|c|c|c|c|}
\hline
Sensor & Pitch & Resolution & \multicolumn{4}{|c|}{Track angle}\\ \cline{4-7}
&$(\mu \mathrm{m})$ & $(\mu \mathrm{m})$& $0\de-10\de$ & $10\de-20\de$ &
$20\de-30\de$ & $30\de-40\de$\\
\multirow{2}{*}{TIB 1-2} & \multirow{2}{*}{ 80} &
Measurement& $ 17.2 \pm  1.9 $ & $ 14.3 \pm  2.3 $ & $ 17.4 \pm  3.2$  & $ 25.7 \pm  6.0$             \\
% &Ratio& $1.03\pm0.03$ &$1.19\pm0.04$  &$1.39\pm0.04$  &$1.44\pm0.03$  \\
&& MC Prediction& $ 16.6 \pm  0.5 $ & $ 11.8 \pm  0.5 $ & $ 12.4 \pm  0.6$  & $ 17.9 \pm  1.5$             \\
\hline
\multirow{2}{*}{TIB 3-4} & \multirow{2}{*}{ 120} &
Measurement& $ 27.7 \pm  3.6 $ & $ 18.5 \pm  3.1 $ & $ 16.1 \pm  3.1$  & $ 24.1 \pm  6.7$             \\
%  &Ratio& $1.03\pm0.02$ &$0.94\pm0.02$  &$0.94\pm0.03$  &	 $1.08\pm0.02$\\
&& MC Prediction& $ 26.8 \pm  0.7 $ & $ 19.4 \pm  0.8 $ & $ 17.2 \pm  0.3$  & $ 21.4 \pm  2.0$	       \\
\hline
\multirow{2}{*}{TOB 1-4} & \multirow{2}{*}{ 183} &
Measurement& $ 39.6 \pm  5.7 $ & $ 28.0 \pm  5.8 $ & $ 24.8 \pm  6.5$  & $ 32.8 \pm  8.3$             \\
%  &Ratio& $1.01\pm0.01$ &$0.94\pm0.01$  &$0.94\pm0.01$  &$0.99\pm0.01$  \\
&& MC Prediction& $ 39.4 \pm  1.3 $ & $ 27.8 \pm  1.2 $ & $ 26.5 \pm  0.3$  & $ 32.5 \pm  2.1$	       \\
\hline
\multirow{2}{*}{TOB 5-6} & \multirow{2}{*}{ 122} &
Measurement& $ 23.2 \pm  3.6 $ & $ 19.5 \pm  3.6 $ & $ 20.9 \pm  6.1$  & $ 29.3 \pm  9.7$             \\
%  &Ratio& $0.96\pm0.02$ &$1.04\pm0.02$  &$1.05\pm0.02$  &$1.16\pm0.05$  \\
&& MC Prediction& $ 23.8 \pm  0.9 $ & $ 18.0 \pm  0.5 $ & $ 19.2 \pm  1.2$  & $ 25.4 \pm  1.6$	       \\
\hline
\end{tabular}
\end{center}
\end{table}

%% file: Conclusions.tex
\section{Summary} 

The Cosmic Run At Four Tesla  has been an important experience
for commissioning the tracker.
%All the main goals identified at the beginning of the
%CRAFT were achieved. 

The control and readout systems were successfully commissioned,
synchronised to the Level-1 Trigger, and operated in global runs with
all the other sub-detectors of the CMS experiment.  The total number
of modules used corresponds to 98.0\% of the total system. 
%This yield was
%consistent with the tracker acceptance relative to the Muon System.

About 15 million events with a muon in the tracker were collected. 
The hit and track reconstruction are seen to have an excellent
performance and the \ckf, which will be used in
proton-proton collisions as the default reconstruction algorithm, was tested successfully. The signal-to-noise
performance is in the range 25-30 for thin modules and 31-36 for
thick ones. The efficiency of hit reconstruction is above 99.5\%. 
In addition, with the collected data sample, it has been possible to 
calibrate the measurement of energy loss in silicon and to measure the
Lorentz angle. 

The track reconstruction efficiency has been measured with two
different methods: one using only muons reconstructed in the muon
chambers and one using only data from the tracker. The reconstruction
efficiency in data was found to be high and well described by the
Monte Carlo simulation. For tracks passing close to the centre of
the detector and having a direction close to the vertical axis,
the reconstruction efficiency was found to be higher than
99\%. The resolution on hit position and track parameters was also
consistent with expectations from Monte Carlo simulation.

CRAFT demonstrated the successful operation of
the tracker integrated with the other CMS subsystems. It was an important
milestone towards final commissioning with colliding beam data.
% and it
%showed that design performance can be achieved.

%% file: Acknowledgments.tex
\section*{Acknowledgments}

We thank the technical and administrative staff at CERN and other CMS
Institutes, and acknowledge support from: FMSR (Austria); FNRS and FWO
(Belgium); CNPq, CAPES, FAPERJ, and FAPESP (Brazil); MES (Bulgaria);
CERN; CAS, MoST, and NSFC (China); COLCIENCIAS (Colombia); MSES
(Croatia); RPF (Cyprus); Academy of Sciences and NICPB (Estonia);
Academy of Finland, ME, and HIP (Finland); CEA and CNRS/IN2P3
(France); BMBF, DFG, and HGF (Germany); GSRT (Greece); OTKA and NKTH
(Hungary); DAE and DST (India); IPM (Iran); SFI (Ireland); INFN
(Italy); NRF (Korea); LAS (Lithuania); CINVESTAV, CONACYT, SEP, and
UASLP-FAI (Mexico); PAEC (Pakistan); SCSR (Poland); FCT (Portugal);
JINR (Armenia, Belarus, Georgia, Ukraine, Uzbekistan); MST and MAE
(Russia); MSTDS (Serbia); MICINN and CPAN (Spain); Swiss Funding
Agencies (Switzerland); NSC (Taipei); TUBITAK and TAEK (Turkey); STFC
(United Kingdom); DOE and NSF (USA). Individuals have received support
from the Marie-Curie IEF program (European Union); the Leventis
Foundation; the A. P. Sloan Foundation; and the Alexander von Humboldt
Foundation.

%% file: CFT-09-002-authorlist.tex
\textbf{Yerevan Physics Institute,  Yerevan,  Armenia}\\*[0pt]
S.~Chatrchyan, V.~Khachatryan, A.M.~Sirunyan
\vskip\cmsinstskip
\textbf{Institut f\"{u}r Hochenergiephysik der OeAW,  Wien,  Austria}\\*[0pt]
W.~Adam, B.~Arnold, H.~Bergauer, T.~Bergauer, M.~Dragicevic, M.~Eichberger, J.~Er\"{o}
% ...
\vskip\cmsinstskip
% ...
